\pdfoutput=1
\documentclass[10pt,a4paper,twocolumn]{article}
\usepackage[T1]{fontenc}
\usepackage{lmodern}
\usepackage[a4paper,textwidth=522pt,textheight=712pt,columnsep=18pt,centering]{geometry}
\usepackage[numbers,sort&compress]{natbib}
\usepackage[font=small,labelfont=bf,labelsep=period]{caption}
\usepackage{adjustbox}
\usepackage{amsmath,amssymb}
\usepackage{bm}
\usepackage{graphicx}
\usepackage{booktabs}
\usepackage{tabularx}
\usepackage{algorithm}
\usepackage{algpseudocode}
\algrenewcommand\algorithmicrequire{\textbf{Input:}}
\algrenewcommand\algorithmicensure{\textbf{Output:}}
\usepackage{xcolor}
\usepackage{tikz}
\usetikzlibrary{arrows.meta,positioning,calc,backgrounds,fit}
\usepackage[hidelinks]{hyperref}

\newcommand{\wb}{\mathord{\circ}}   %
\newcommand{\bb}{\mathord{\bullet}} %
\newcommand{\bs}[1]{\bm{#1}}
\definecolor{muted}{cmyk}{0,0,0,0.55}

\makeatletter
\renewcommand\section{\@startsection{section}{1}{\z@}%
  {-3ex \@plus -1ex \@minus -.2ex}{1.5ex \@plus .2ex}{\normalfont\large\bfseries}}
\renewcommand\subsection{\@startsection{subsection}{2}{\z@}%
  {-2.5ex \@plus -1ex \@minus -.2ex}{1ex \@plus .2ex}{\normalfont\normalsize\bfseries}}
\renewcommand\subsubsection{\@startsection{subsubsection}{3}{\z@}%
  {-2.5ex \@plus -1ex \@minus -.2ex}{1ex \@plus .2ex}{\normalfont\normalsize\itshape}}
\renewcommand\paragraph{\@startsection{paragraph}{4}{\z@}%
  {1.5ex \@plus .5ex \@minus .2ex}{-1em}{\normalfont\normalsize\bfseries}}
\renewcommand\appendix{\par
  \setcounter{section}{0}\setcounter{subsection}{0}\setcounter{equation}{0}%
  \setcounter{figure}{0}\setcounter{table}{0}%
  \gdef\thesection{Appendix~\@Alph\c@section}%
  \@addtoreset{equation}{section}\@addtoreset{figure}{section}\@addtoreset{table}{section}%
  \gdef\theequation{\@Alph\c@section.\arabic{equation}}%
  \gdef\thefigure{\@Alph\c@section.\arabic{figure}}%
  \gdef\thetable{\@Alph\c@section.\arabic{table}}}
\makeatother

\begin{document}
	\hypersetup{pdftitle={Reinforcement Twinning with Twin-Guided Bayesian Policy Search and Replay-Driven Identification},pdfauthor={Miguel A. Mendez, Lorenzo Schena, Romain Poletti, Yannick Lecomte, Sebastiano~Randino, Jan van den Berghe}}

\twocolumn[{%
\centering
\vspace*{4pt}
{\LARGE\bfseries Reinforcement Twinning with Twin-Guided Bayesian Policy Search and Replay-Driven Identification\par}
\vskip 12pt
{\large Miguel A. Mendez\textsuperscript{1,2,3,$\ast$}, Lorenzo Schena\textsuperscript{2,4}, Romain Poletti\textsuperscript{2}, Yannick Lecomte\textsuperscript{2,1}, Sebastiano~Randino\textsuperscript{2,1}, Jan van den Berghe\textsuperscript{2}\par}
\vskip 8pt
{\small \textsuperscript{1}SAAS, Université libre de Bruxelles (ULB), Av. Franklin Roosevelt 50, 1050 Brussels, Belgium\\
\textsuperscript{2}Environmental and Applied Fluid Dynamics, von Karman Institute for Fluid Dynamics, 1640 Sint-Genesius-Rode, Belgium\\
\textsuperscript{3}Experimental Aerodynamics and Propulsion Lab, Universidad Carlos III de Madrid, 28911 Leganés, Spain\\
\textsuperscript{4}Department of Mechanical Engineering, Vrije Universiteit Brussel (VUB), 1050 Brussels, Belgium\\[3pt]
$^{\ast}$Corresponding author: mendez@vki.ac.be\par}
\vskip 12pt
\begin{minipage}{0.90\textwidth}
\small\setlength{\parindent}{1em}%
\centerline{\textbf{Abstract}}\par\smallskip
\noindent The feasibility of updating and synchronising digital twins of actuated systems from operational data depends on how the system is controlled and, therefore, on the information available for model identification. An imperfect twin, in turn, can guide control optimisation provided that inaccurate predictions do not constrain the search. This work presents a reinforcement-twinning framework that exploits this reciprocal relationship. A Gaussian-process surrogate of the cumulative reward over an episode combines the twin prediction, used as a calibrated trend, with a discrepancy learned from measured episode rewards. Candidate policies are proposed in a region centred on the policy the twin predicts to be optimal, and measured rewards can expand this region. Physical episodes are replayed on the twin after observer-based synchronisation with sparse measurements, and twin parameters are identified by combining Bayesian optimisation with gradient-based refinement. Information-oriented episodes deploy excitation policies designed on the twin to improve parameter identifiability.

The framework is demonstrated on a Burgers-flow analogue of dynamic-induction control in wind farms, with a deliberately misspecified model. Information episodes inform parameters poorly constrained by control-driven data. Despite substantially biased reward predictions, the twin provides a useful starting policy and search region, giving the guided search an early advantage over unguided Bayesian optimisation. This advantage persists when measured rewards can expand the region, whereas unguided alternatives eventually overtake searches confined to the twin-defined region. The results support reinforcement twinning for jointly adapting models and control policies when physical interaction is limited and model discrepancy cannot be neglected.\par\medskip
\end{minipage}\par
\vskip 14pt
}]

	\section{Introduction}
	\label{Sec:Intro}

Digital twins \cite{Jones2020DTReview,BokhtiarAlZami2025} have become a central paradigm for cyber-physical systems and Industry 4.0 \cite{Lee2008CPS,Lee2015CPS,Kagermann2013Industry4,Ficili2025}, linking a physical process to a continually updated virtual model through sensing, communication, and embedded computation. Unlike standalone models, twins remain connected to operational data and evolve with the systems they represent, supporting prediction, optimisation, and decision-making as new measurements become available \cite{Korenhof2021,Wagg2025,Torzoni2024digital}. In actuated systems, control and twinning are naturally coupled: the twin can guide control, while actuations determine the trajectories explored by the physical system and therefore the data available to update the twin. This creates the possibility of reciprocal learning between model and controller.

Exploiting this reciprocal coupling requires addressing two complementary problems. First, the state of the twin must be synchronised with that of the physical system, and the twin parameters identified or updated, from trajectories generated during operation, which may provide only sparse and unevenly informative data. Second, the controller must exploit useful model guidance without being confined to the policies favoured by imperfect predictions. Between these two lies informative actuation: actions can be chosen not only for control performance, but also to improve identifiability, reduce parameter uncertainty, or reveal model discrepancy. The literature spanning these problems is vast, and the following provides only a concise overview of the ideas and approaches most closely related to the present work.

\subsection{Model identification and assimilation}

Updating and synchronising a digital twin from sparse and noisy operational measurements can be a formidable challenge, especially when the twin is a high-dimensional model, such as a flow model governed by partial differential equations (PDEs). Parameter identification then requires estimating the full initial state, propagating it under the applied control actions and the exogenous forcing, and comparing the resulting predictions with measurements that typically observe only a small part of that state. The forcing itself is often measured only in part, or not at all. A point lidar reading, for instance, may stand in for the rotor-equivalent wind speed of a turbine.

State estimation becomes particularly challenging for PDE-governed systems, where data assimilation typically involves reconstructing high-dimensional fields from sparse measurements. The resulting inverse problem can be poorly conditioned, depending on sensor placement and the sensitivity of the dynamics to the model parameters \cite{ZakiWang2021,Zaki2025,Hoepffner2005,Colburn2011,Mons2016,Manohar2018,Fukami2021}. When reconstruction and prediction must also be performed fast enough for real-time forecasting, dimensionality reduction and mappings to low-dimensional latent representations can become necessary (see, e.g. \cite{Ozalp2026real}). An additional layer of complexity arises when state and parameter estimation are carried out jointly \cite{AschBocquetNodet2016,Moireau2008joint,Smith2025,Marques2024,Ozan2025}: errors in the reconstructed state can be absorbed into the inferred parameters, while model error can in turn bias the state estimate unless it is accounted for simultaneously \cite{Novoa2024inferring}.

Finally, for actuated systems, identifiability also depends on the trajectories generated by the applied control. This is the well-known problem of persistent excitation in adaptive control and more generally optimal experimental design \cite{GoodwinPayne1977,Ljung1999,Mehra1974,WalterPronzato1997,Gevers2005,Hjalmarsson2005,Chiuso2019}: parameters can only be identified if the system is driven through sufficiently informative dynamics. The control input therefore becomes part of the identification problem and can be chosen deliberately to improve identifiability, for example by reducing parameter uncertainty or exposing model discrepancy.

Recent studies design such experiments directly on the model \cite{Bhattacharya2026optimal,Yang2025active}. Sequential Bayesian experimental design has similarly been formulated as a Markov decision process and addressed through approximate dynamic programming, policy-gradient methods, and learned design policies based on expected information gain \cite{HuanMarzouk2016,ShenHuan2023,Shen2025variational,Foster2021DAD,Blau2022,Rainforth2024}. In learning-based control, active system identification selects inputs or exploration policies using Fisher-information or information-gain criteria \cite{WagenmakerJamieson2020,Mania2022,BuissonFenet2020,Treven2023,Mehta2022,Memmel2024ASID}, while reinforcement learning has recently been used to steer sensors for data acquisition in digital twins \cite{Ogbodo2026}. Broadly, these approaches focus on how controlled interaction can improve model identification.

\subsection{Model-based and model-free control}

Finding control actions that optimise a prescribed objective for a dynamical system for which a mathematical model is available is the classical problem of optimal control \cite{stengel1994optimal,Bertsekas2012,RawlingsMayne2017,gunzburger2002perspectives}. When the model is imperfect, however, the challenge is to exploit its predictions without making control performance overly dependent on its accuracy.

Self-tuning and adaptive control \cite{AstromWittenmark1973,astrom_adaptive_2008,Annaswamy2021,Matni2019} seek to reduce this dependence by updating the model and/or controller online as new measurements become available.
At the other extreme, model-free methods seek to solve the control problem directly from measured system performance, without relying on an explicit dynamical model \cite{Hjalmarsson1998IFT,SuttonBarto2018,Bucci2019,Pino2023,Pino2026,ParkLaw2016}. This avoids bias from an inaccurate model, but generally requires more interaction with the physical system and potentially lengthy, costly, or even unsafe trial-and-error experimentation.

Model-based and hybrid reinforcement-learning approaches bridge these extremes by using learned dynamics models to plan, generate synthetic experience, or guide control improvement \cite{sutton1991dyna,Deisenroth2011,nagabandi_neural_2018,moerland_model-based_2022,bansal_mbmf_2017,Johannink2019,Liu2021,Ozan2025}. For distributed systems, these models are often reduced-order representations learned from data, allowing feedback controllers to be trained offline and deployed in real time \cite{Tomasetto2025latent}. Such model-based guidance can substantially reduce the number of physical interactions compared with purely model-free approaches \cite{Werner2024,Werner2025system,Rai2019SimulationBO,Nobar2025,Nobar2026}, but it also introduces a dependence on model quality, since model errors can bias the resulting control strategy. The central challenge is therefore to exploit useful information from the model without allowing its inaccuracies to dominate the search.

\subsection{Dual control and reinforcement twinning}

Model identification and control meet when actions are valued both for the performance they produce and for the information they generate. This is the classical dual-control problem, originally formulated by Feldbaum \cite{Feldbaum1960,feldbaum_dual_1963}. Feldbaum showed that an optimal decision may sacrifice immediate performance for information that improves later decisions. This idea is closely related to the exploration--exploitation trade-off in reinforcement learning, where exploratory interactions can improve value estimates or policies used for subsequent control \cite{SuttonBarto2018,RummeryNiranjan1994,Mnih2015,LillicrapEtAl2016,HaarnojaEtAl2018}. In dual control, however, this exploration is explicitly valued through its effect on future uncertainty and performance. Exact dual control requires propagating both the system state and uncertainty about the model, which is tractable only for relatively small problems \cite{BarShalomTse1974,bernhardsson_dual_1989,astrom_dual_1986,Filatov2000}. Practical formulations rely on approximations, for example within model predictive control or Bayesian policy search \cite{Wittenmark1995,Mesbah2018,Klenske2016,Arcari2020,chen_dual_2021}.

Reinforcement Twinning (RT) \cite{Schena2024,Poletti2025} was introduced from a related perspective, coupling model-based and model-free learning through a continuously adapted digital twin. As in dual control, physical interactions have a dual role: they produce control performance while also generating information that can improve subsequent decisions. In RT, however, this reciprocity is organised explicitly around the twin. The aim is to improve the control of a physical system on which interactions are costly, using an imperfect twin, through two learning processes that share the same physical episodes. Reinforcement improves the control policy from the rewards measured on the physical system, with model-based guidance from the twin. Twinning synchronises the twin with the physical system and identifies its parameters from the recorded trajectories, including those of excitation policies that the twin itself proposes. The coupling of the two processes, rather than either one alone, defines RT.

Earlier realisations were primarily control-oriented. In the original formulation, an identified twin supported a model-based controller developed alongside a model-free actor--critic agent, with information transferred between the two when either learning process stagnated. Although the twin was continually updated from trajectories generated during operation, its main role remained to support control: A referee chose which of the two policies to deploy and, in \cite{Poletti2025}, also assessed the reliability of the twin. The present work replaces the two parallel policy searches and their referee with a unified Bayesian policy search that combines twin predictions with measured returns (Section~\ref{sec_A}).

\subsection{Contributions of this work}
The present work strengthens the reciprocal connection between model improvement and control in two ways. First, the controller-generated trajectories are replayed to synchronise and update the twin, which is then used to synthesise excitation policies for selected episodes to improve parameter identifiability. This brings the experimental-design perspective discussed above into the control loop. Second, model-based and model-free policy searches are integrated into a unified Bayesian optimisation framework. Measured control performance calibrates the twin's return prediction and helps learn its discrepancy with the physical system. The twin also informs the initial proposal region, while measured returns can progressively relax the search constraints. Both mechanisms let policy search benefit from the twin while retaining the freedom to depart from its guidance when physical evidence (data) supports it.

The return surrogate builds on the construction of Kennedy and O'Hagan, in which a high-fidelity output, that of an expensive computer code \cite{Kennedy2000,Kennedy2001} is modelled as the scaled prediction of a less accurate model plus a Gaussian-process discrepancy. The same construction has been used to calibrate flow and PDE models against measurements \cite{Guillas2014bayesian,Girolami2021statistical}, and underlies multi-fidelity surrogates in which a cheap model is corrected using a small number of expensive evaluations \cite{Forrester2007,LeGratiet2014,Bonfiglio2018multi,Peherstorfer2018,Foumani2023multi,Fiore2026predicting}, including approaches for correcting optima computed on a low-fidelity model \cite{Hart2023hyper}. A similar correction is used in modifier adaptation for real-time optimisation, where measurements on the physical process correct the cost predicted by a nominal process model \cite{Marchetti2009}. Its Gaussian-process variants learn this correction and restrict the next operating point to a trust region \cite{delRioChanona2019,delRioChanona2021}, and Bayesian optimisation of controllers has likewise combined simulations or digital twins with physical experiments \cite{Marco2017,Nobar2025,Nobar2026}. Here, the twin plays the role of the low-fidelity model and the measured returns that of the high-fidelity data, and its parameters are also identified from the recorded trajectories. The twin also sets the relative scales with which the surrogate compares policies, which are not fitted to the measured returns.

These developments are illustrated on a deliberately constructed benchmark, designed to examine the benefits and limitations of the different forms of twin guidance in the presence of structural model discrepancy. The benchmark is a one-dimensional Burgers-flow analogue of distributed energy extraction, motivated by dynamic-induction control in wind farms \cite{Munters2017,Munters2018}. The twin is misspecified through a persistent, policy-dependent structural discrepancy, and the physical system is observed only through sparse sensors.

The paper is organised as follows. Section~\ref{sec2} introduces the general RT setting and its main principles. Section~\ref{sec3} details the numerical implementation of each operation of the resulting algorithm: initialisation, synthesis of the model-based reference, the return surrogate, selection of performance and excitation policies, state synchronisation, and replay identification of the twin parameters. Section~\ref{sec4} defines the Burgers benchmark and its numerical set-up, and Section~\ref{sec5} presents the results. Section~\ref{sec6} summarises the work and discusses its main limitations and possible extensions.

	\section{Reinforcement twinning}	\label{sec2}

Reinforcement twinning targets actuated systems for which extensive trial-and-error experimentation on the physical system is impractical and the available model is imperfect both in its parameters and in its structure. This model depends on uncertain parameters that can be continuously refined from operational data, and is therefore referred to as a digital twin (hereinafter the twin, for brevity).

\begin{figure*}[t]
		\centering
		\begin{tikzpicture}[
			>=Latex,
			font=\footnotesize,
			box/.style={draw,fill=white,rounded corners=2pt,align=center,inner sep=4pt,text width=4.8cm,minimum height=1.3cm},
			panel/.style={rounded corners=5pt,inner xsep=7pt,inner ysep=6pt},
			arrow/.style={->,thick},
			upd/.style={->,thick,dashed,black!75},
			lab/.style={font=\scriptsize,align=center,inner sep=1pt}
			]
			\node[box] (sel)   at (0,0)
			{\textbf{Deployed policy} $\pi_{\bb}^{(n)}$\\
				$e_n=0$: performance $\pi_{\mathrm{perf}}(\cdot\mid\bs\theta_\pi^{(n)})$\\
				$e_n=1$: information $\pi_{\mathrm{info}}(\cdot\mid\bs\theta_\iota^{(n)})$};
			\node[box] (plant) at (0,-3.0)
			{\textbf{Physical episode} $n$\\
				hidden state $\bs s_{\bb}$, input $\bs z$,\\
				sparse noisy observations $\bs o_{\bb}$};
			\node[box] (rec)   at (0,-6.0)
			{\textbf{Record} $\mathcal E^{(n)}$\\
				every episode $\to$ replay buffer $\mathcal B$\\
				performance: $(\bs\theta_\pi^{(n)},\mathcal R_{\bb}^{(n)})\to\mathcal D_\pi$};
			\node[box] (rep)   at (7.5,-6.0)
			{\textbf{Factual replay} (recorded $\widehat{\bs z}$, $\bs a_{\bb}$)\\
				observer $\to$ replay states $\widehat{\mathcal I}^{\mathrm a}$\\
				free prediction vs.\ $\bs o_{\bb}$, $\mathcal R_{\bb}$ $\to\bs\theta_p^{(m+1)}$};
			\node[box] (cf)    at (7.5,-3.0)
			{\textbf{Counterfactual rollouts}\\
				virtual returns $\mathcal R_{\wb}^{(m)}(\bs\theta_\pi)$ on $\mathcal C^{+}$\\
				reference $\bs\theta_{\wb,\star}^{(n)}$ and $\mathcal J_{\mathrm{info}}(\bs\theta_\iota)$};
			\node[box] (next)  at (7.5,0)
			{\textbf{Choice of the next interaction}\\
				surrogate $\widehat{\mathcal R}_{\bb}$ ($\mathcal D_\pi$, $\mathcal R_{\wb}$) $\to\bs\theta_\pi^{(n+1)}$\\
				informativeness $\mathcal J_{\mathrm{info}}\to\bs\theta_\iota^{(n+1)}$};
			\coordinate (pt) at (0,0.95);   \coordinate (pb) at (0,-6.95);
			\coordinate (tt) at (7.5,0.95); \coordinate (tb) at (7.5,-6.95);
			\begin{scope}[on background layer]
				\node[panel,fill=black!7,fit=(sel)(plant)(rec)(pt)(pb)] (pp) {};
				\node[panel,draw=black!45,fit=(rep)(cf)(next)(tt)(tb)] (tp) {};
			\end{scope}
			\node[font=\footnotesize\bfseries,anchor=south] at (pp.north) {Physical environment ($\bb$)};
			\node[font=\footnotesize\bfseries,anchor=south] at (tp.north) {Digital twin ($\wb$), parameters $\bs\theta_p^{(m)}$};
			\draw[arrow] (sel)   -- node[lab,right=2pt] {deploy}  (plant);
			\draw[arrow] (plant) -- node[lab,right=2pt] {measure} (rec);
			\draw[arrow] (rec) -- node[lab,below=3pt] {replay\\buffer $\mathcal B$} (rep);
			\draw[arrow] (rep.north) -- node[lab,right=2pt] {$\bs\theta_p^{(m)}$} (cf.south);
			\draw[arrow] (cf.north)  -- node[lab,right=2pt] {$\mathcal R_{\wb}$, $\bs\theta_{\wb,\star}$, $\mathcal J_{\mathrm{info}}$} (next.south);
			\draw[arrow] (next) -- node[lab,above=3pt] {next interaction\\$\bs\theta_\pi^{(n+1)}$ or $\bs\theta_\iota^{(n+1)}$} (sel);
			\coordinate (bx) at ($(rec.east)+(1.0,0)$);
			\draw[arrow] ($(rec.east)+(0,0.4)$) -- ($(bx |- rec.east)+(0,0.4)$) -- (bx |- 0,-0.45) -- ($(next.west)+(0,-0.45)$);
			\node[lab,rotate=90,anchor=south] at ($(bx |- 0,-3.0)+(-0.1,0)$) {measured returns $\mathcal D_\pi$};
			\draw[upd] ($(rep.north)+(-1.75,0)$) -- node[lab,right=2pt,black!55] {twin update\\$m\to m{+}1$} ($(cf.south)+(-1.75,0)$);
			\foreach \i/\b in {1/sel,2/plant,3/rec,4/rep,5/cf,6/next}
			\node[circle,draw,fill=white,font=\scriptsize\bfseries,inner sep=0pt,minimum size=11pt] at (\b.north west) {\i};
		\end{tikzpicture}
		\caption{Reinforcement-twinning loop. Physical environment (left, $\bb$): the deployed policy (1) drives a physical episode (2), whose record (3) feeds the replay buffer and, for performance episodes, the archive of measured returns. Digital twin (right, $\wb$): factual replay (4) synchronises the twin with the data and refines its parameters; counterfactual rollouts (5) provide the virtual returns, the model-based reference and the informativeness of candidate excitation policies; with the measured returns, these determine the next interaction (6). All components are presented in Sections~\ref{sec2} and~\ref{sec3}.}
		\label{fig:rt_overview}
	\end{figure*}
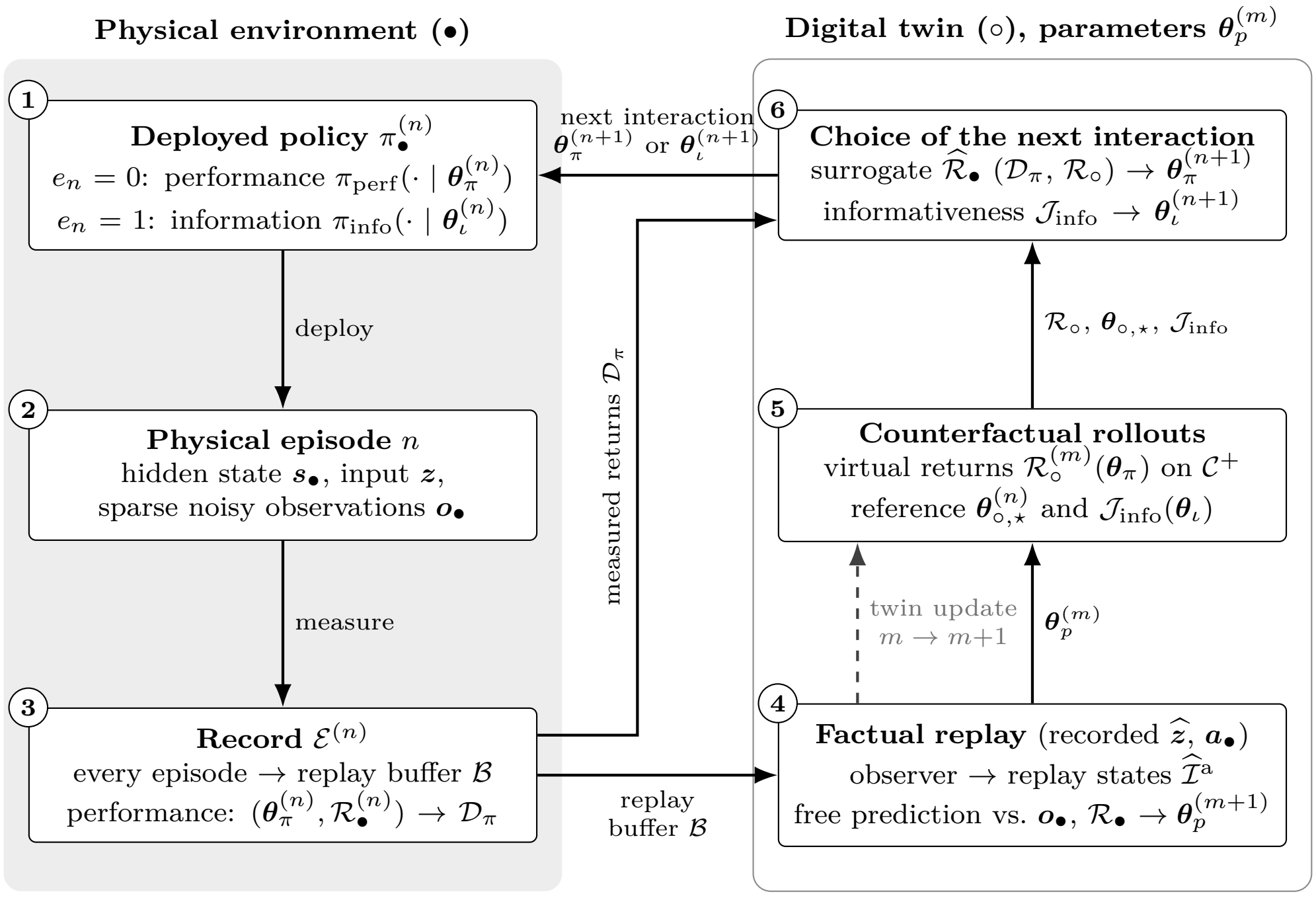

The simultaneous improvement of the control policy and the twin leverages data collected over sequences of interactions called episodes, following the loop in Figure~\ref{fig:rt_overview}. Physical episodes are replayed on the twin for state synchronisation and parameter identification, while counterfactual rollouts are used to evaluate candidate performance and excitation policies without interacting with the physical system. These virtual evaluations guide the next physical interaction, whose data then drive subsequent policy and twin updates. The remainder of this section introduces the physical and twin dynamics, defines the learning objectives, and assembles these components into the reinforcement-twinning loop. Throughout, subscript $\bb$ denotes physical quantities and $\wb$ twin quantities, physical episodes are indexed by $n$, and twin updates by $m$.

	\subsection{Physical system and digital twin}
	\label{sec2a}

	Each episode consists of $N$ interactions with the physical system, occurring at the sampling times $t_k=k\Delta t$, with $k=0,\dots,N-1$ and $\Delta t$ the sampling period. The physical system and its digital twin are treated as discrete dynamical systems evolving side by side as

	\begin{equation}
		\label{eq:rt_side_by_side}
		\left\{\;
		\begin{aligned}
			\bs s_{\bb,k+1}&=F_{\bb}(\bs s_{\bb,k},\bs z_k,\bs a_{\bb,k}),\\
			\bs s_{\wb,k+1}&=F_{\wb}(\bs s_{\wb,k},\widehat{\bs z}_k,\bs a_{\wb,k};\bs\theta_p),\\[0.45em]
			\bar{\bs o}_{\bb,k}&=h(\bs s_{\bb,k},\bs a_{\bb,k-1}),
			\qquad \bs o_{\bb,k}=\bar{\bs o}_{\bb,k}+\bs\varepsilon_{\bb,k},\\
			\bar{\bs o}_{\wb,k}&=h(\bs s_{\wb,k},\bs a_{\wb,k-1}),
			\qquad \bs o_{\wb,k}=\bar{\bs o}_{\wb,k}+\bs\varepsilon_{\wb,k},\\[0.45em]
			\bs s_{\bb,0}&\sim\mathcal I_0\;\;\text{(hidden)},\\
			\bs s_{\wb,0}&=\widehat{\bs s}_{\wb,0}^{\,\mathrm a}\;\;\text{(analysed)},\\[0.45em]
			\bs\chi_{\bb,k+1}&=G_\pi(\bs\chi_{\bb,k},\bs o_{\bb,k}),\\
			\bs\chi_{\wb,k+1}&=G_\pi(\bs\chi_{\wb,k},\bs o_{\wb,k}),\\[0.45em]
			\bs a_{\bb,k}&=\pi_{\bb}(\bs o_{\bb,k},\bs\chi_{\bb,k}),\\
			\bs a_{\wb,k}&=\pi_{\wb}(\bs o_{\wb,k},\bs\chi_{\wb,k}).
		\end{aligned}
		\right.
	\end{equation}

	Each pair of lines gives the physical and twin forms of the dynamics, the observation, the initial condition, the information state of the policy and the action. These are briefly described below to introduce the main components and their assembly in the algorithm, introduced in Section~\ref{sec2c}. Section~\ref{sec3} enters into the details of each.

	\paragraph{Dynamics} The physical system evolves from a hidden state $\bs s_{\bb,k}\in\mathbb R^{n_s}$ through an unknown transition operator $F_{\bb}$. At each time step $k$, it receives an exogenous forcing $\bs z_k\in\mathbb R^{n_z}$, a stochastic input realised from a process $\mathcal P_z$, whereas the twin receives its measurement $\widehat{\bs z}_k$. Each also receives a control action, $\bs a_{\bb,k}$ and $\bs a_{\wb,k}\in\mathbb R^{n_a}$ respectively, governed by a control policy. The measured input may be a component of $\bs o_{\bb,k}$, and its measurement error adds to the discrepancy between the two systems.

	Unmeasured inputs would require an input estimator \cite{GillijnsDeMoor2007}, not considered here. The action selected at $t_k$ acts over $[t_k,t_{k+1})$, whereas the exogenous input may vary within this interval. The transition from $t_k$ to $t_{k+1}$ may then use the input at both ends, and components of the action may evolve within the step in a way fixed at $t_k$, as in the benchmark (Sections~\ref{sec4p3} and~\ref{sec5p_solver}). The state is assumed or made Markovian, by augmenting it with any system memory such as actuator dynamics or delays, whereas the memory of the policy is kept in the separate information state $\bs\chi_k$, introduced below. The twin has a state $\bs s_{\wb,k}\in\mathbb R^{n_{s,\wb}}$, whose dimension generally differs from $n_s$, and evolves under the partially known operator $F_{\wb}$, with fixed structure and uncertain twin parameters $\bs\theta_p\in\Theta_p\subset\mathbb R^{n_{\theta_p}}$. It may be structurally misspecified, in which case no $\bs\theta_p$ makes the twin reproduce the noise-free physical observations for all inputs and actions; identification then yields the best in-family approximation, leaving a residual discrepancy that may depend on the policy. In the following, $\bs\theta_p^{(m)}$ denotes the twin parameters at the $m$-th scheduled twin update, while $\bs\theta_p$ without superscript refers to the same parameters as a free variable.

	\paragraph{Observations} The observation operator $h$ maps the state and the previously applied action to a noise-free output. On the physical system, this output, $\bar{\bs o}_{\bb,k}$, is latent: only the noisy measurement $\bs o_{\bb,k}\in\mathbb R^{n_o}$ is available to the controller, typically with $n_o\ll n_s$. The measurement error $\bs\varepsilon_{\bb,k}\in\mathbb R^{n_o}$ follows a known noise model, summarised by the covariance $\bs\Sigma_o\in\mathbb R^{n_o\times n_o}$ of the errors at one step. The noise model may also include correlations in time and the clipping of readings to their admissible range, after which the error need not be zero-mean or Gaussian (Section~\ref{sec4p2}). The dependence of $h$ on the previously applied action accounts for measurements that respond directly to the command, such as actuator states, while keeping the observation at $t_k$ available before the action at $t_k$ is chosen. When the operator depends on model parameters, as in the test case of Section~\ref{sec4p2}, the twin uses its own parameters $\bs\theta_p$.

	In factual replay, no synthetic observation noise is drawn: $\bar{\bs o}_{\wb,k}$ is compared directly with the recorded physical measurement $\bs o_{\bb,k}$. The resulting residual therefore contains both physical measurement noise and any discrepancy between the physical system and the twin. In counterfactual rollouts, synthetic noise $\bs\varepsilon_{\wb,k}$ is instead generated from the physical noise model to produced possible samples of $\bs o_{\wb,k}$. Both the candidate policy and the reward receive this observation, as do their physical counterparts. A complete noise trajectory is fixed for each entry of the scoring bank, ensuring that all candidate policies are evaluated under the same noise realisations. When the measured input $\widehat{\bs z}_k$ is also a component of the observation, its record, which already contains the sensor error, serves as both the twin input and that component of $\bs o_{\wb,k}$, with no synthetic noise added to it (Section~\ref{sec4p2}).

	\paragraph{Initial condition} The physical state at $k=0$ is hidden, drawn from an unknown distribution $\mathcal I_0$, and only partially observed, so the initial state of the twin cannot be set directly. It is instead reconstructed by an observer $\Psi_{\wb}$ from the measurements, inputs, and actions recorded over a window $\mathcal W_n^{-}$ of $L_{\mathrm a}$ steps preceding the episode, indexed by negative $k$ (Figure~\ref{fig:sync_free_forecast}),

	\begin{equation}
		\label{eq:rt_sync_intro}
		\widehat{\bs s}_{\wb,0}^{\,\mathrm a}
		=
		\Psi_{\wb}
		\!\left(
		\mathcal W_n^{-};
		\bs\theta_p^{(m)}
		\right),
	\end{equation}
		where the hat denotes quantities reconstructed from data. The states reconstructed for the stored episodes form the replay states $\widehat{\mathcal I}^{\mathrm a}$. This synchronisation is detailed in Section~\ref{sec_sync}. The time $k=0$, after which the twin is no longer corrected by measurements and runs freely in both factual replay and counterfactual rollouts, is hereafter called \emph{the branch}.

	\paragraph{Policy and information state} Each physical episode is driven by one of two policy families. The performance policy $\pi_{\mathrm{perf}}(\cdot\mid\bs\theta_\pi)$, with $\bs\theta_\pi\in\Theta_\pi\subset\mathbb R^{n_{\theta_\pi}}$, is the persistent target of the policy search. The excitation policy $\pi_{\mathrm{info}}(\cdot\mid\bs\theta_\iota)$, with $\bs\theta_\iota\in\Theta_\iota$, is deployed only in information-oriented episodes, to make the physical data informative about the twin parameters $\bs\theta_p$. The two parameterisations need not coincide. A binary schedule $e_n\in\{0,1\}$ determines which of them is the physical policy $\pi_{\bb}$ of \eqref{eq:rt_side_by_side} in episode $n$, that is

{\fontsize{9}{12}\selectfont%
	\begin{equation}
		\label{eq:rt_deployed_policy}
		\pi_{\bb}^{(n)}
		=
		\begin{cases}
			\pi_{\mathrm{perf}}(\,\cdot\mid\bs\theta_\pi^{(n)}),
			& e_n=0
			\quad\text{(performance-oriented)},\\[0.4em]
			\pi_{\mathrm{info}}(\,\cdot\mid\bs\theta_\iota^{(n)}),
			& e_n=1
			\quad\text{(information-oriented)}.
		\end{cases}
	\end{equation}
}%

	 Both policy families act on the current observation and on the information state $\bs\chi_k\in\mathbb R^{n_\chi}$, which carries the causal memory of the policy, such as a finite observation history or a low-pass filter. The information state is updated by the same operator $G_\pi$ of \eqref{eq:rt_side_by_side} on the physical system and on the twin, so that a candidate policy processes its observations in the same way in both. The twin's information state is initialised at the branch from the recorded value, $\bs\chi_{\wb,0}=\bs\chi_{\bb,0}$.

The twin policy $\pi_{\wb}$ of \eqref{eq:rt_side_by_side}, which gives the action $\bs a_{\wb,k}$ applied in the twin, takes one of two forms, depending on whether the twin operates in factual replay or in counterfactual rollout,
{\fontsize{9}{12}\selectfont%
	\begin{equation}
		\label{eq:twin_action_source}
		\pi_{\wb}(\bs o_{\wb,k},\bs\chi_{\wb,k})
		=
		\begin{cases}
			\bs a_{\bb,k}, & \text{factual replay},\\[0.3em]
			\pi(\bs o_{\wb,k},\bs\chi_{\wb,k}\mid\bs\theta), & \text{counterfactual rollout},
		\end{cases}
	\end{equation}
}%
	with $\pi\in\{\pi_{\mathrm{perf}},\pi_{\mathrm{info}}\}$. In factual replay, the twin policy thus returns the recorded action, whatever the twin observes. Factual replay is used for synchronisation and identification, whereas counterfactual rollouts are used to predict returns and to synthesise the model-based reference policy and the excitation policies. These operations are described in Section~\ref{sec3}.

	\subsection{Learning objectives}\label{sec2b}\label{virtual_experiments}

The proposed algorithm is guided by four learning criteria. Two rely on data from the physical system and define the learning targets: the physical return $\mathcal R_{\bb}(\bs\theta_\pi)$, whose expected value defines the policy optimum $\bs\theta_\pi^{\star}$, and the identification loss $\mathcal J_{\wb}(\bs\theta_p\mid\mathcal B)$, whose minimiser is the best-fitting twin parameter vector $\bs\theta_p^{\star}$. The remaining two are evaluated on the twin and support the learning process: the virtual return $\mathcal R_{\wb}(\bs\theta_\pi;\bs\theta_p,\mathcal C^{+})$, whose maximiser is the model-based reference $\bs\theta_{\wb,\star}^{(n)}$, and the information criterion $\mathcal J_{\mathrm{info}}(\bs\theta_\iota)$, whose minimiser $\bs\theta_\iota^{(n)}$ defines the excitation policy used in information episodes. These criteria are described next.

	\paragraph{Physical return} The control objective is the return of the performance policy over a physical episode. As in traditional reinforcement learning \cite{SuttonBarto2018}, it takes the standard discounted additive form

	\begin{equation}
		\label{R_o_b}
		\mathcal R_{\bb}
		(
		\bs{\theta}_{\pi}
		)
		=
		\sum_{k=0}^{N-1} \gamma^k r \!\left( \bs o_{\bb,k}, \widehat{\bs z}_k, \bs a_{\bb,k} \right)
		+
		\gamma^N r_N\!\left(\bs o_{\bb,N}\right),
	\end{equation}
	with $r$ the instantaneous reward, $r_N$ a reward on the final observation, which requires no further action, and $\gamma\in(0,1]$ a discount factor. The arguments $\widehat{\bs z}_k$ and $\bs a_{\bb,k}$ allow, for instance, a penalty on the actuation or a reward normalised by the measured input. The reward depends on the hidden state only through the observations, which therefore include the measured quantities it requires. The return \eqref{R_o_b} is a random variable, which changes from one episode to the next, even under the same policy, because of the stochastic exogenous input, the measurement noise and the different branch states. Episodes are assumed independent of each other, with the branch state being drawn from $\mathcal I_0$ in \eqref{eq:rt_side_by_side} regardless of the policies deployed before. The deployed policy therefore shapes only the data of its own episode, not the initial state of later ones. Ideally, the target of the policy search is the policy with the highest expected return,
	\begin{equation}
		\label{eq:rt_policy_target}
		\bs\theta_\pi^{\star}
		=
		\underset{\bs\theta_{\pi}\in\Theta_\pi}{\arg\max}
		\;
		\mathbb E
		\left[
		\mathcal R_{\bb}
		(
		\bs\theta_{\pi}
		)
		\right],
	\end{equation}
	with the expectation taken over the branch state, the exogenous input and the measurement noise. However, to avoid the need for repeating episodes with the same policy over many realisations of the stochastic variables on which the reward depends, the proposed algorithm trains a probabilistic surrogate $\widehat{\mathcal R}_{\bb}(\bs\theta_\pi)$ on the archive of measured returns $\mathcal D_\pi$, and the optimal policy $\bs\theta_\pi^{\star}$ is sought using an acquisition function as in traditional Bayesian optimisation \cite{Bo_1,Jones1998,Calandra2016bayesian}. The surrogate architecture is described in Section~\ref{sec_A}, while the selection of the next policy candidate is described in Section~\ref{sec_B}.

	\paragraph{Virtual return} The twin estimates the physical return of any policy without further interaction with the physical system. Let $\mathcal Z^{+}$ be a fixed set of $N_c$ measured records $\widehat{\bs z}^{+,[c]}$ of realisations of $\mathcal P_z$ over an episode. These are obtained from $N_c$ preparatory runs of the physical system before the policy search and are excluded from its interaction budget. The fixed \emph{scoring bank}
	\begin{equation}
		\label{eq:scoring_bank}
		\mathcal C^{+}
		=
		\left\{
		\left(
		\widehat{\bs s}_{\wb,0}^{\,\mathrm a,[c]},
		\bs\chi_{\wb,0}^{[c]},
		\widehat{\bs z}^{+,[c]},
		\bs\varepsilon_{\wb}^{+,[c]}
		\right)
		\right\}_{c=1}^{N_c}
	\end{equation}
	gives each scenario an initial state, the policy memory at the branch, the measured exogenous-input record, and a complete synthetic observation-noise trajectory. The superscript $+$ marks the inputs and noise realisations of the bank. The states are reconstructed once, with the initial twin, from synchronisation windows recorded in the same runs, whereas the noise trajectories are drawn from the physical noise model, with covariance $\bs\Sigma_o$. Both are then held fixed and kept separate from the replay states. Running the twin with parameters $\bs\theta_p$ in counterfactual mode from each entry under a candidate policy gives the \emph{virtual return} $\mathcal R_{\wb}(\bs\theta_\pi;\bs\theta_p,\mathcal C^+)$. Abbreviating the notation as $\mathcal R_{\wb}^{(m)}(\bs\theta_\pi)=\mathcal R_{\wb}(\bs\theta_\pi;\bs\theta_p^{(m)},\mathcal C^+)$ for the current twin, with $m$ the number of twin updates performed so far, this reads
	\begin{equation}
		\label{eq:operational_twin_return}
		\mathcal R_{\wb}^{(m)}(\bs\theta_\pi)
		=
		\mathbb E_{\mathcal C^+}
		\Bigg[
		\sum_{k=0}^{N-1}
		\gamma^k
		r\!\left(\bs o_{\wb,k},\widehat{\bs z}_k^{+},\bs a_{\wb,k}\right)
		+
		\gamma^N r_N\!\left(\bs o_{\wb,N}\right)
		\Bigg],
	\end{equation}
	where the expectation is the average over the $N_c$ entries of the bank, and the actions are those of the counterfactual rollout in \eqref{eq:twin_action_source} with $\pi=\pi_{\mathrm{perf}}$ and $\bs\theta=\bs\theta_\pi$. Both the policy and the reward receive the synthetic noisy observations $\bs o_{\wb,k}$, consistently with their physical counterparts in \eqref{R_o_b}.

	Unlike the physical return \eqref{R_o_b}, the virtual return is
	defined as an average because it approximates the expected return
	$\mathbb E[\mathcal R_{\bb}]$ targeted in
	\eqref{eq:rt_policy_target}, rather than the return of a single
	episode. The twin makes this averaging affordable because its
	scenarios require no additional physical interaction. The surrogate
	of Section~\ref{sec_A} builds on this distinction. Since all candidate
	policies are evaluated on the same bank $\mathcal C^{+}$, using the
	same noise realisations, $\mathcal R_{\wb}^{(m)}$ is a deterministic
	function of $\bs\theta_\pi$, as in the PEGASUS method of policy search
	\cite{NgJordan2000}. Nevertheless, structural misspecification of the
	twin can introduce systematic bias, while the finite bank can cause
	sampling error and favour policies that perform particularly well on
	its $N_c$ scenarios. Therefore, the maximiser
	\begin{equation}
		\label{eq:model_based_policy_synthesis_operator}
		\bs\theta_{\wb,\star}^{(n)}
		\approx
		\underset{\bs\theta_{\pi}\in\Theta_\pi}{\arg\max}
		\;
		\mathcal R_{\wb}^{(m)}
		(
		\bs\theta_{\pi}
		)
	\end{equation}
	is generally not optimal on the physical system (see also
	Abbeel et al.~\cite{Abbeel2006}). We refer to this maximiser as the
	\emph{model-based reference}: the policy that would be deployed if
	the current twin were trusted completely, corresponding to its
	certainty-equivalent policy \cite{BarShalomTse1974}. The superscript $(n)$ identifies the model-based reference used at physical episode $n$, synthesised with the twin parameters in use at its last synthesis; this reference is updated according to the schedule of Section~\ref{sec2c}. The symbol $\approx$ indicates that
	the reference is obtained through the finite numerical search
	described in Section~\ref{sec3_synth}.

	The model-based reference also defines the centre of the region of
	$\Theta_\pi$ from which new candidate policies are proposed. Its
	baseline extent is determined by the relevance scales of the virtual
	return (Section~\ref{sec_B}). Consequently, the physical optimum
	$\bs\theta_\pi^{\star}$ can be proposed only when it lies within the
	current region. The region is recentered and rescaled whenever the
	reference and relevance scales are recomputed and can expand when the evidence from measured returns drives the search to one of
	its faces lying in the interior of $\Theta_\pi$
	(Section~\ref{sec_B}).

	\paragraph{Identification loss} Identification seeks to make the twin reproduce the recorded physical episodes as closely as possible. Each episode is stored in the replay buffer $\mathcal B$ as a record $\mathcal E^{(i)}$ of its inputs, actions, measurements and measured return (Section~\ref{sec_sync}).

For a candidate $\bs\theta_p$, the twin starts from the analysed state
\eqref{eq:rt_sync_intro} of each episode and propagates its state without
measurement correction, using the recorded inputs and actions. The analysed
states, collected in $\widehat{\mathcal I}^{\mathrm a}$, are computed with the current twin parameters $\bs\theta_p^{(m)}$ and held fixed while candidates are compared. Conditional
on these states, the identification loss compares each free replay with its
recorded trajectory and return,
\begin{equation}
	\label{eq:rt_replay_loss}
	\mathcal J_{\wb}
	\big(
	\bs\theta_p
	\mid
	\mathcal B;
	\widehat{\mathcal I}^{\mathrm a}
	\big)
	=
	\mathcal J_{\mathrm{dyn}}
	\big(
	\bs\theta_p
	\mid
	\mathcal B;
	\widehat{\mathcal I}^{\mathrm a}
	\big)
	+
	c_{\mathrm{rew}}\,
	\mathcal J_{\mathrm{rew}}
	\big(
	\bs\theta_p
	\mid
	\mathcal B;
	\widehat{\mathcal I}^{\mathrm a}
	\big),
\end{equation}
	with
{\fontsize{9}{12}\selectfont%
	\begin{align}
\mathcal J_{\mathrm{dyn}}
\big(
\bs\theta_p
\mid
\mathcal B;
\widehat{\mathcal I}^{\mathrm a}
\big)
		&=
		\mathbb E_{\mathcal E\in\mathcal B}
		\Bigg[
		\frac{1}{N}
		\sum_{k=1}^{N}
		\ell_{\mathrm{dyn}}
		\Big(
		\bs o_{\bb,k}(\mathcal E),
		\bar{\bs o}_{\wb,k}(\bs\theta_p;\mathcal E)
		\Big)
		\Bigg],
		\label{J_o}
		\\
		\label{eq:rt_replay_rew_loss}
\mathcal J_{\mathrm{rew}}
\big(
\bs\theta_p
\mid
\mathcal B;
\widehat{\mathcal I}^{\mathrm a}
\big)
		&=
\mathbb E_{\mathcal E\in\mathcal B}
\left[
\left(
\frac{
	\mathcal R_{\wb}^{\mathrm{rep}}(\bs\theta_p;\mathcal E)
	-
	\mathcal R_{\bb}(\mathcal E)
}{
	s_R
}
\right)^2
\right],
	\end{align}
}%
where, for each record $\mathcal E$ sampled from $\mathcal B$,
$\bar{\bs o}_{\wb,k}(\bs\theta_p;\mathcal E)
=h(\bs s_{\wb,k}(\bs\theta_p;\mathcal E),\bs a_{\bb,k-1})$
is the noise-free prediction of the free replay. Its state evolves according to
$\bs s_{\wb,k+1}
=F_{\wb}(\bs s_{\wb,k},\widehat{\bs z}_k,\bs a_{\bb,k};\bs\theta_p)$
from the analysed initial state associated with that record in
$\widehat{\mathcal I}^{\mathrm a}$. The function $\ell_{\mathrm{dyn}}$ measures the discrepancy between a recorded measurement and this prediction. The sum in \eqref{J_o} starts at $k=1$, whereas the return \eqref{R_o_b} includes $k=0$, because the measurement at the branch is assimilated by the observer \eqref{eq:rt_sync_intro}.
The replay return $\mathcal R_{\wb}^{\mathrm{rep}}(\bs\theta_p;\mathcal E)$ is
the return predicted by the same free replay, with rewards evaluated on
$\bar{\bs o}_{\wb,k}(\bs\theta_p;\mathcal E)$, whereas
$\mathcal R_{\bb}(\mathcal E)$ is the measured return stored in the record. Since the reward can be nonlinear in the observations, this noise-free return can differ from the measured return in expectation, even for a twin that reproduces the physical system.
Unlike the virtual return \eqref{eq:operational_twin_return}, the replay return follows the recorded inputs and actions from the analysed state of the episode. The physical, virtual and replay returns thus apply the same reward and discounting \eqref{R_o_b} to the measured, counterfactual and replayed trajectories, respectively.
Finally, $s_R$ is a fixed scaling factor, and the expectations denote empirical
averages over the episode records selected from $\mathcal B$.

	 The first term, $\mathcal J_{\mathrm{dyn}}$, is an output error on the trajectory. The second, the return term $\mathcal J_{\mathrm{rew}}$, orients identification towards control, in the spirit of the identification for control of Gevers~\cite{Gevers2005} and Hjalmarsson~\cite{Hjalmarsson2005}, although here through the loss rather than through the excitation. Parameter vectors with similar trajectory errors are thus distinguished by the returns they predict for the recorded episodes. With $c_{\mathrm{rew}}=0$, the twin is identified from the trajectories alone, regardless of the use made of it for control. The form of $\ell_{\mathrm{dyn}}$, the choice of $s_R$ and $c_{\mathrm{rew}}$, and the split of the buffer into training and validation episodes are given in Section~\ref{sec_C}. The identification target is
\begin{equation}
	\label{eq:rt_identification_target}
	\bs\theta_p^{\star}
	\in
	\underset{\bs\theta_p\in\Theta_p}{\arg\min}
	\;
	\mathcal J_{\wb}
	\big(
	\bs\theta_p
	\mid
	\mathcal B;
	\widehat{\mathcal I}^{\mathrm a}(\bs\theta_p^{\star})
	\big),
\end{equation}
	where $\widehat{\mathcal I}^{\mathrm a}(\bs\theta_p^\star)$ are the replay states computed with $\bs\theta_p^\star$ itself. This fixed point is an idealised target, at which the alternation of synchronisation and identification would stop for a fixed buffer if each identification reached the minimum. The finite search of Section~\ref{sec_C}, accepted only if it lowers a validation loss, need not reach it. The fixed point generally differs from the minimiser obtained by recomputing the replay states for each candidate. Unlike $\bs\theta_\pi^{\star}$, this target depends on the replay buffer, hence on the policies that generated it.

	\paragraph{Information criterion} When $e_n=1$ in
	\eqref{eq:rt_deployed_policy} and an information-policy update is scheduled
	(Section~\ref{sec2c}), the current twin selects, without physical interaction,
	an excitation policy expected to be informative about $\bs\theta_p$, by solving

	\begin{equation}
		\label{eq:rt_nested_information_design}
		\bs\theta_\iota^{(n)}
		=
		\underset{\bs\theta_\iota\in\Theta_\iota}{\arg\min}
		\;\mathcal J_{\mathrm{info}}
		\!\left(
		\bs\theta_\iota
		\mid
		\bs\theta_p^{(m)},
		\widehat{\mathcal I}^{\mathrm a}
		\right).
	\end{equation}

	The criterion favours excitation policies in which the twin-predicted measurements respond strongly, relative to measurement noise, to changes in the twin's parameters. This sensitivity is evaluated by counterfactual rollouts under $\pi_{\mathrm{info}}(\cdot\mid\bs\theta_\iota)$ from the replay states of the $N_d$ most recent episodes, under their recorded exogenous inputs and policy memory. Each rollout gives, at the time steps $k=1,\ldots,N$ of the episode, the $n_{\mathrm{id}}\le n_o$ components of the noise-free prediction $\bar{\bs o}_{\wb,k}$ used for identification (Section~\ref{sec_C}), stacked in a vector $\bar{\bs O}_{\wb}(\bs\theta_\iota,\widetilde{\bs\theta}_p)\in\mathbb R^{Nn_{\mathrm{id}}}$, where $\widetilde{\bs\theta}_p\in[0,1]^{n_{\theta_p}}$ are the twin parameters normalised to the box $\Theta_p$ as in \eqref{eq:policy_whitening} (Section~\ref{sec_D}). Let $\bs S_{i,q}=\partial\bar{\bs O}_{\wb}/\partial\widetilde{\bs\theta}_p$ be the sensitivity of the rollout from episode $i$, evaluated at the $q$-th of $N_\theta$ values of the twin parameters, and $\bs\Sigma_{\mathrm{id},N}$ the covariance of the stacked measurement errors. Since the sensitivities depend on the parameters to be identified, the information is averaged over these $N_\theta$ values. The ensemble Fisher information
	\begin{equation}
		\label{eq:rt_info_ensemble_fisher}
		\overline{\bs F}
		(
		\bs\theta_\iota
		)
		=
		\frac{1}{N_\theta}
		\sum_{q=1}^{N_\theta}
		\sum_{i=n-N_d}^{n-1}
		\bs S_{i,q}^{\top}
		\bs\Sigma_{\mathrm{id},N}^{-1}
		\bs S_{i,q}
	\end{equation}
	then gives the E-optimal criterion
	\begin{equation}
		\label{eq:rt_e_optimal_policy}
		\mathcal J_{\mathrm{info}}
		\!\left(
		\bs\theta_\iota
		\mid
		\bs\theta_p^{(m)},
		\widehat{\mathcal I}^{\mathrm a}
		\right)
		=
		-\lambda_{\min}
		\left[
		\overline{\bs F}
		(
		\bs\theta_\iota
		)
		\right],
	\end{equation}
	with $\lambda_{\min}$ the smallest eigenvalue. Minimising $\mathcal J_{\mathrm{info}}$ thus selects the excitation policy whose least-informed parameter direction receives the most information. The criterion seeks information on every direction of $\bs\theta_p$, whether or not the return depends on it, unlike application-oriented designs \cite{Hjalmarsson2005,Gevers2005}. The selected excitation policy is used in information-oriented episodes until the next information-policy update. The records of these episodes enter $\mathcal B$ but not $\mathcal D_\pi$, since the return surrogate is defined on the performance-policy space $\Theta_\pi$.

	 The computation of the sensitivities, the error model behind $\bs\Sigma_{\mathrm{id},N}$, the choice of the $N_\theta$ values, and the use of $\overline{\bs F}$ as a ranking criterion are given in Section~\ref{sec_D}.

	\begin{algorithm*}[t]
		\caption{Reinforcement-twinning loop}
		\label{alg:RT_policy_search}
		\footnotesize
		\begin{algorithmic}[1]
			\Require initial twin parameters $\bs\theta_p^{(0)}$; for a run with performance episodes, the records of $N_c$ preparatory runs, made before the search and outside its budget; number $n_{\mathrm{init}}$ of initial episodes; search budget $M$, the number of physical episodes of the search, including the initial ones; schedules $e_n$, $\bs u_n$; observer, surrogate, synthesis, and identification configurations
			\Ensure performance policy maximising the posterior mean $\mu_{\bb}$ of \eqref{eq:gp_posterior_mean} among the policies of $\mathcal D_\pi$; $\bs\theta_p^{(m)}$; $\widehat{\mathcal R}_{\bb}$; $\bs\theta_{\wb,\star}$; $\mathcal B$; $\mathcal D_\pi$ (in a run with information episodes only: $\bs\theta_p^{(m)}$ and $\mathcal B$)
			\State $m\gets0$
			\If{the run has performance episodes}
			\State build the fixed scoring bank $\mathcal C^{+}$ from the $N_c$ preparatory runs and fixed synthetic noise trajectories (analysed states with $\bs\theta_p^{(0)}$)
			\State $\bs\theta_{\wb,\star}\gets\textsc{SynthesiseTwinPolicy}(\bs\theta_p^{(m)},\mathcal C^{+})$ \Comment{eq.~\eqref{eq:model_based_policy_synthesis_operator}}
			\EndIf
			\State run $n_{\mathrm{init}}$ initial episodes with performance policies or, in a run with information episodes only, excitation policies fixed in advance. The records are stored in $\mathcal B$ and the returns of performance policies in $\mathcal D_\pi$
			\State $\widehat{\mathcal I}^{\mathrm a}\gets\textsc{Synchronise}(\mathcal B,\bs\theta_p^{(m)})$ \Comment{eq.~\eqref{eq:rt_sync_intro}}
			\If{the run has performance episodes}
			\State $\widehat{\mathcal R}_{\bb}\gets\textsc{CalibrateReturnSurrogate}(\mathcal D_\pi,\bs\theta_p^{(m)},\mathcal C^{+})$ \Comment{Section~\ref{sec_A}}
			\EndIf
			\For{$n=n_{\mathrm{init}}+1,\ldots,M$}
			\State \textbf{if} $e_n=1$ and $u_{\iota,n}=0$: $\pi_{\bb}^{(n)}\gets$ the current excitation policy \Comment{no information-policy update}
			\State \textbf{else}: $\pi_{\bb}^{(n)}\gets\textsc{SelectInteraction}(e_n,\widehat{\mathcal R}_{\bb},\bs\theta_{\wb,\star},\bs\theta_p^{(m)},\mathcal C^{+},\widehat{\mathcal I}^{\mathrm a},\mathcal B)$ \Comment{eq.~\eqref{eq:rt_deployed_policy}}
			\State $\mathcal E^{(n)}\gets\textsc{DeployAndRecord}(\pi_{\bb}^{(n)})$; $\mathcal B\gets\mathcal B\cup\{\mathcal E^{(n)}\}$ \Comment{the only physical interaction}
			\State $\widehat{\mathcal I}^{\mathrm a}\gets$ the $N_{\mathrm s}$ most recent elements of $\widehat{\mathcal I}^{\mathrm a}\cup\textsc{Synchronise}(\mathcal E^{(n)},\bs\theta_p^{(m)})$ \Comment{eq.~\eqref{eq:branch_ensemble}}
			\State \textbf{if} $e_n=0$: $\mathcal D_\pi\gets\mathcal D_\pi\cup\{(\bs\theta_\pi^{(n)},\mathcal R_{\bb}^{(n)})\}$ and refit $\widehat{\mathcal R}_{\bb}$ on the enlarged archive
			\State \textbf{if} $u_{p,n}=1$: $\bs\theta_p^{(m+1)}\gets\textsc{IdentifyTwin}(\mathcal B,\widehat{\mathcal I}^{\mathrm a},\bs\theta_p^{(m)})$; $m\gets m+1$; resynchronise $\widehat{\mathcal I}^{\mathrm a}$ \Comment{eq.~\eqref{eq:rt_replay_loss}}
			\State \textbf{if} $u_{\wb\pi,n}=1$: $\bs\theta_{\wb,\star}\gets\textsc{SynthesiseTwinPolicy}(\bs\theta_p^{(m)},\mathcal C^{+})$
			\State \textbf{if} $u_{R,n}=1$: $\widehat{\mathcal R}_{\bb}\gets\textsc{CalibrateReturnSurrogate}(\mathcal D_\pi,\bs\theta_p^{(m)},\mathcal C^{+})$ \Comment{virtual returns of $\mathcal D_\pi$ and relevance scales of the current twin}
			\EndFor
		\end{algorithmic}
	\end{algorithm*}

	\begin{table}[ht!]
		\caption{Spaces, data sets, indices, schedules and evolving state used by Algorithm~\ref{alg:RT_policy_search}.}
		\label{tab:rt_loop_objects}
		\centering
		\footnotesize
		\begin{tabularx}{\columnwidth}{@{}l>{\raggedright\arraybackslash}X@{}}
			\toprule
			Object & Content \\
			\midrule
			$\Theta_p$, $\Theta_\pi$, $\Theta_\iota$ & Admissible spaces of the twin parameters, of the performance policy, and of the excitation policy. \\
			$\mathcal B$; $\mathcal B_{\mathrm{tr}},\mathcal B_{\mathrm{val}}\subset\mathcal B$ & Replay buffer of factual episode records $\mathcal E^{(i)}$ \eqref{eq:replay_record}, and its episode-disjoint training and validation views used for identification. \\
			$\mathcal D_\pi\subset\Theta_\pi\times\mathbb R$ & Performance-policy parameters paired with their measured physical returns; the data of the return surrogate. \\
			$\widehat{\mathcal I}^{\mathrm a}$, $\mathcal C^{+}$ & Ensemble of replay states, recomputed with the current twin; fixed scoring bank of analysed states reconstructed with the initial twin, exogenous-input scenarios $\mathcal Z^{+}$ recorded in the same preparatory runs, and synthetic observation-noise trajectories, on which counterfactual rollouts are evaluated. \\
			$\bs\theta_p^{(m)}$, $\bs\theta_{\wb,\star}^{(n)}$, $\widehat{\mathcal R}_{\bb}$ & Current twin parameters, model-based reference policy \eqref{eq:model_based_policy_synthesis_operator}, and surrogate of the physical return: the three model-dependent quantities refreshed by $u_{p,n}$, $u_{\wb\pi,n}$ and $u_{R,n}$. \\
			$\mathcal Q\subset\Theta_\pi$ & Finite pool of performance-policy candidates proposed around $\bs\theta_{\wb,\star}^{(n)}$ and ranked by the surrogate. \\
			$k$, $n$, $m$ & Time step within an episode (negative over the synchronisation window, zero at the branch), physical episode, and number of twin updates performed. \\
			$e_n$, $\bs u_n$ & Purpose of episode $n$, and the updates scheduled before it ($u_{\iota,n}$) or after it (Section~\ref{sec2c}). \\
			\bottomrule
		\end{tabularx}
	\end{table}

\subsection{The reinforcement-twinning loop}\label{sec2c}

The reinforcement-twinning loop couples policy search
\cite{SuttonBarto2018,Bertsekas2012,Deisenroth2013} with system identification
\cite{Ljung1999,GoodwinPayne1977,Chiuso2019}, while keeping their learning
targets distinct. The policy optimum $\bs\theta_\pi^{\star}$ is defined by
physical performance and does not depend on the twin. The twin nevertheless
influences which policies are tested through the surrogate
$\widehat{\mathcal R}_{\bb}$ and the proposal region. Conversely, the
identification target $\bs\theta_p^{\star}$ depends on the collected episodes
and therefore on the policies that generated them. Information-oriented
episodes are introduced to make these data more informative.

The present implementation does not combine performance and information into a
single dual-control objective. Instead, the binary schedule $e_n$
(Section~\ref{sec2a}) is fixed in advance and assigns each episode either a performance or an information purpose.  This separation is an implementation choice: $e_n$ could
instead be determined online by a criterion trading off expected return against
information gain, for example by combining the acquisition function of
Section~\ref{sec_B} with \eqref{eq:rt_e_optimal_policy}, or by valuing
information through its effect on future returns, as in approximate dual
control \cite{Klenske2016,Arcari2020}.

Algorithm~\ref{alg:RT_policy_search} implements this coupling as physical
episodes are collected following the loop in
Figure~\ref{fig:rt_overview}, with its main objects listed in
Table~\ref{tab:rt_loop_objects}. After $n_{\mathrm{init}}$ initial episodes,
each iteration selects a policy according to $e_n$,
deploys and records it, synchronises the twin to the new record, and performs
the scheduled model-dependent updates. The schedule
$\bs u_n=(u_{p,n},u_{\wb\pi,n},u_{R,n},u_{\iota,n})$ decouples these
updates from the rate of physical interaction.
The flag $u_{\iota,n}$ determines, before episode $n$, whether an
information-policy update is performed. If no update is scheduled, an
information-oriented episode reuses the most recently synthesised excitation
policy. An information-policy update is always scheduled for the first
information-oriented episode after the initial episodes. The remaining flags
act after episode $n$: $u_{p,n}$ triggers twin identification,
$u_{\wb\pi,n}$ the re-synthesis of the model-based reference, and $u_{R,n}$
the recalibration of the return surrogate.

Once $\mathcal D_\pi$ contains performance episodes, every twin update is
accompanied by re-synthesis and recalibration,
$u_{p,n}=1\Rightarrow u_{\wb\pi,n}=u_{R,n}=1$. Consequently, a single
surrogate fit never combines virtual returns computed with different versions of the twin. Every re-synthesis is likewise followed by recalibration, $u_{\wb\pi,n}=1\Rightarrow u_{R,n}=1$, because the recomputed relevance scales enter the surrogate. While the twin remains fixed, the surrogate is refitted after
every performance episode.

A scheduled twin update changes the parameters,
$\bs\theta_p^{(m)}\to\bs\theta_p^{(m+1)}$ (the dashed arrow in
Figure~\ref{fig:rt_overview}). To keep the policy-search components consistent
with the updated twin, the replay states are first re-synchronised. The virtual
returns of all policies in $\mathcal D_\pi$ are then recomputed on the unchanged
scoring bank $\mathcal C^{+}$, while their measured returns are retained. The
model-based reference \eqref{eq:model_based_policy_synthesis_operator} is
re-synthesised, thereby updating the proposal region, and the surrogate of
Section~\ref{sec_A} is finally recalibrated. None of these operations requires
further interaction with the physical system.

Each operation of Algorithm~\ref{alg:RT_policy_search} maps to
Figure~\ref{fig:rt_overview}. \textsc{Synchronise} and
\textsc{Identify\-Twin} form the factual-replay part of the loop, with
\textsc{Identify\-Twin} producing the dashed twin update.
\textsc{Synthesise\-Twin\-Policy},
\textsc{Calibrate\-Return\-Surrogate}, and
\textsc{Select\-Interaction} use counterfactual evaluations to construct the
model-based reference, calibrate the return surrogate, and select the next
interaction, respectively. \textsc{Deploy\-And\-Record}, in the left column,
is the only operation that interacts with the physical system.

	\section{Numerical implementation}
	\label{sec3}

	This section describes the operations of Algorithm~\ref{alg:RT_policy_search}
	in their order of execution. Benchmark-specific choices are given in
	Section~\ref{sec4}, and the budgets and schedules of the experiments are
	reported with the results.

	\subsection{Initialisation}
	\label{sec3_init}

	Algorithm~\ref{alg:RT_policy_search} starts from initial twin parameters $\bs\theta_p^{(0)}$, representing nominal or prior values. Its first $n_{\mathrm{init}}$ episodes, run with initial policies, count in the budget $M$, unlike the preparatory runs of the scoring bank. In a run with performance episodes, the
	fixed scoring bank $\mathcal C^{+}$ is first constructed from analysed states
	reconstructed with $\bs\theta_p^{(0)}$, the scenarios $\mathcal Z^{+}$ recorded
	in the preparatory runs, and synthetic noise trajectories drawn from the physical noise model. The initial model-based reference is then
	synthesised from this bank. The initial performance policies may form a
	space-filling set in $\Theta_\pi$ or be sampled around this reference.

	In a run with information episodes only, the initial excitation policies are
	instead fixed in advance, because the information criterion is evaluated from the replay states and inputs of recorded episodes (Section~\ref{sec2b}): scoring bank, model-based reference, and return surrogate are not needed in such a run.

	The $n_{\mathrm{init}}$ initial policies are deployed sequentially, one episode each, and their records stored in $\mathcal B$. The measured returns of performance policies also enter $\mathcal D_\pi$. Synchronising the twin to these records
	with $\bs\theta_p^{(0)}$ (Section~\ref{sec_sync}) gives the initial replay
	states $\widehat{\mathcal I}^{\mathrm a}$. In a run with performance episodes,
	the return surrogate is then calibrated from $\mathcal D_\pi$. Synchronisation,
	synthesis, and calibration are repeated according to the
	schedules of Algorithm~\ref{alg:RT_policy_search}.

	\subsection{Synthesis of the model-based reference}
	\label{sec3_synth}

	\textsc{SynthesiseTwinPolicy} computes the model-based reference
	$\bs\theta_{\wb,\star}^{(n)}$ of
	\eqref{eq:model_based_policy_synthesis_operator}. It maximises the virtual
	return $\mathcal R_{\wb}^{(m)}$ of the current twin,
	\eqref{eq:operational_twin_return}, over $\Theta_\pi$. Each evaluation requires
	one counterfactual rollout per element of $\mathcal C^{+}$, and the virtual return may have several local maxima that make the outcome of a gradient-based search strongly depend on its starting point. We therefore use a hybrid global--local optimiser:
	Bayesian optimisation proposes starting points, and local gradient-based search
	refines them. The same approach is used for identification in
	Section~\ref{sec_C}. Related methods combining Bayesian optimisation with local
	search can be found in \cite{GaoYuLi2020BOwLS,KyriakidisMendezBahr2024BOIPOPT}.

	The search is performed in the unit box of \eqref{eq:policy_whitening}. It begins with an archive of policies scored by their virtual return, initialised with a space-filling set and distinct from the archive $\mathcal D_\pi$ of measured returns. Each round fits a Mat\'ern-$5/2$ Gaussian process to this archive, proposes candidates by expected improvement,
	and refines them with projected AMSGrad \cite{ReddiKaleKumar2018}. The proposed
	and refined candidates are added to the archive. The search stops after a
	prescribed number of rounds or when the best score stagnates. To limit overfitting to the scoring bank, the local refinement evaluates the return on other realisations of the exogenous input and of the synthetic noise, while the candidates are compared, and the reference selected, under the inputs and noise of $\mathcal C^{+}$. The initial states of these rollouts may differ from the analysed states of $\mathcal C^{+}$ (Section~\ref{sec5p0b}).

Gradients are obtained by reverse-mode differentiation through the rollouts, which requires $F_{\wb}$, $h$, the policy and the reward to be differentiable. Each gradient costs only a small multiple of a forward evaluation and does not scale with the number of search variables \cite{Schena2024,Marques2024, GriewankWalther2008}. The virtual returns of the policies evaluated by this search, or of a separate space-filling set (Section~\ref{sec5p0b}), are also used to estimate the sensitivity of the return to each policy coordinate by fitting a Mat\'ern-$5/2$ Gaussian process with automatic relevance determination (ARD), whose length scales $\ell_j$ quantify the variation along the corresponding coordinates. These scales are used both to normalise the policy coordinates in the surrogate and to set the coordinate-wise widths of the region in $\Theta_\pi$ within which new policy candidates are proposed (Sections~\ref{sec_A} and~\ref{sec_B}). When $u_{\wb\pi,n}=1$, the reference policy and the ARD length scales are recomputed using the current twin, possibly by a simpler search, with the previous reference providing a warm start.

	\subsection{Return surrogate with a calibrated twin trend}\label{sec_A}

\textsc{CalibrateReturnSurrogate} builds the probabilistic surrogate
$\widehat{\mathcal R}_{\bb}$ of the physical return \eqref{R_o_b}, which is
used by \textsc{SelectInteraction}. The surrogate treats the virtual return \eqref{eq:operational_twin_return} as a biased predictor of the physical return, not as an objective of its own. It models the physical return using
the twin prediction as a calibrated trend,
\begin{equation}
	\label{eq:calibrated_return_model}
	\begin{split}
		\mathcal R_{\bb}^{(i)}
		&=
		\bs\varphi_{\wb}^{(m)}(\bs\theta_{\pi}^{(i)})^{\top}\bs b_{\wb}
		+
		\delta_{\wb}^{(m)}(\bs\theta_{\pi}^{(i)})
		+
		\varepsilon_{R}^{(i)},
		\\
		\bs\varphi_{\wb}^{(m)}(\bs\theta_{\pi})
		&=
		\begin{bmatrix}
			1\\
			\mathcal R_{\wb}^{(m)}(\bs\theta_{\pi})
		\end{bmatrix}.
	\end{split}
\end{equation}

Here, $\mathcal R_{\bb}^{(i)}$ is the measured return of archived episode $i$.
The vector
$\bs b_{\wb}=[b_{\wb,0},b_{\wb,1}]^{\top}\in\mathbb R^2$
contains the offset and slope of the affine calibration, and
$\bs\varphi_{\wb}^{(m)}(\bs\theta_\pi)\in\mathbb R^2$ is its trend basis. The
discrepancy
$\delta_{\wb}^{(m)}\sim\mathcal{GP}(0,k_\delta)$
is a zero-mean Gaussian process with covariance function $k_\delta$, specified
below. The episode noise
$\varepsilon_R^{(i)}\sim\mathcal N(0,\sigma_R^2)$ represents the variation in
the measured return between episodes run with the same policy.

The offset $b_{\wb,0}$ corrects an additive bias. The calibration slope
$b_{\wb,1}$ rescales the virtual return and, because it is unconstrained, can
also reverse its orientation. The discrepancy process
$\delta_{\wb}^{(m)}$ captures the policy-dependent differences that remain
after this affine calibration. This is the universal-kriging form of
multi-fidelity models
\cite{OHagan1978,Kennedy2000,Kennedy2001,Santner2018,Forrester2007,LeGratiet2014,Bonfiglio2018multi,Peherstorfer2018,Poloczek2017}.
It produces a single posterior and therefore a single acquisition function. Unlike in the calibration of Kennedy and O'Hagan \cite{Kennedy2001}, where the model parameters are estimated jointly with the discrepancy, the twin parameters are identified separately by replay (Section~\ref{sec_C}).
The influence of the twin is inferred from the fitted calibration
coefficients. These coefficients, the noise variance $\sigma_R^2$, and the
parameters of $k_\delta$ are estimated from the data as described below. The
index $m$ is suppressed from this point to simplify the notation.

The calibrated trend takes over the role of the policy referee of the earlier RT implementations \cite{Schena2024,Poletti2025}. There, a model-based and a model-free policy were trained in parallel, and the referee decided which of the two was deployed on the physical system by comparing their performance on the twin. In \cite{Poletti2025}, the referee also assessed the reliability of the twin through the ratio between the improvements measured on the physical system and those predicted by the twin over successive episodes, and gave control to the model-free policy when this ratio left a prescribed interval. The calibration slope $b_{\wb,1}$ plays the role of that ratio, but it is estimated from the whole archive of measured returns and weights the twin continuously rather than through thresholds. Because it is fitted together with the discrepancy, the slope depends on the kernel $k_\delta$, and it can exceed one, in which case the surrogate amplifies the differences predicted by the twin. A slope near one lets the virtual return drive the selection, whereas a slope near zero leaves it to the discrepancy process, which is learned from measured returns only. The twin then guides only where candidates are proposed and how the discrepancy measures distances between policies, through the proposal region of Section~\ref{sec_B} and the relevance scales of Section~\ref{sec3_synth}.

The surrogate \eqref{eq:calibrated_return_model} is built in a scaled policy
space. Each coordinate is first mapped to the unit interval using the sides
$[\theta_{\pi,j}^{\min},\theta_{\pi,j}^{\max}]$ of the box $\Theta_\pi$,
\begin{equation}
	\label{eq:policy_whitening}
	\widetilde{\theta}_{\pi,j}
	=
	\frac{
		\theta_{\pi,j}-\theta_{\pi,j}^{\min}
	}{
		\theta_{\pi,j}^{\max}-\theta_{\pi,j}^{\min}
	},
	\qquad
	j=1,\ldots,n_{\theta_\pi}.
\end{equation}

The distance between two policies is then
\[
d^2(\bs\theta_\pi,\bs\theta_\pi')
=
\sum_{j=1}^{n_{\theta_\pi}}
\left[
\frac{
\widetilde\theta_{\pi,j}-\widetilde\theta_{\pi,j}'
}{
\ell_j
}
\right]^2,
\]
where $\ell_j$ is the relevance scale of coordinate $j$ obtained in
Section~\ref{sec3_synth}. This scaling provides soft guidance: all coordinates
remain in the search, but directions along which the virtual return varies
slowly contribute less to the distance. The relevance scales are recomputed
with the model-based reference after a twin update. The discrepancy covariance is the Mat\'ern-$5/2$ kernel
\[
k_\delta(\bs\theta_\pi,\bs\theta_\pi')
=
\sigma_f^2
\left(
1+\sqrt{5}\,\frac{d}{\ell_\pi}+\frac{5}{3}\frac{d^2}{\ell_\pi^2}
\right)
\exp\!\left(-\sqrt{5}\,\frac{d}{\ell_\pi}\right),
\]
where $d=d(\bs\theta_\pi,\bs\theta_\pi')$, $\sigma_f^2$ is the discrepancy variance and $\ell_\pi$ is its length scale in the scaled policy space.

After episode $n$, the archive contains the $N_\pi\le n$ performance episodes
run so far, numbered in their order of execution,
\[
\mathcal D_{\pi}
=
\left\{
(\bs\theta_{\pi}^{(i)},\mathcal R_{\bb}^{(i)})
\right\}_{i=1}^{N_\pi}.
\]

Here $i$ counts performance episodes only and differs from the physical-episode index when information episodes are run. Let $\bs y=[\mathcal R_{\bb}^{(i)}]_{i=1}^{N_\pi}\in\mathbb R^{N_\pi}$
be the vector of measured returns and
$\bs H_{\wb}\in\mathbb R^{N_\pi\times 2}$ the trend matrix with rows
$[1,\ \mathcal R_{\wb}(\bs\theta_{\pi}^{(i)})]$. The covariance of the
measured returns is
$\bs V=\bs K_{\delta}+\sigma_R^2\bs I\in\mathbb R^{N_\pi\times N_\pi}$, where the discrepancy covariance matrix $\bs K_{\delta}$ has the entries
\[
[\bs K_{\delta}]_{ii'}
=
k_{\delta}
\left(
\bs\theta_{\pi}^{(i)},
\bs\theta_{\pi}^{(i')}
\right),
\]
and $\bs I$ is the identity matrix of the same size. Finally,
$\bs k_{\delta}(\bs\theta_{\pi})\in\mathbb R^{N_\pi}$ contains the
covariances between a candidate policy and the archived policies.

For fixed covariance hyperparameters, the calibration coefficients follow
from generalised least squares,
\begin{equation}
	\label{eq:gls_calibration}
	\widehat{\bs b}_{\wb}
	=
	\left(
	\bs H_{\wb}^{\top}
	\bs V^{-1}
	\bs H_{\wb}
	\right)^{-1}
	\bs H_{\wb}^{\top}
	\bs V^{-1}
	\bs y.
\end{equation}

This expression requires $\bs H_{\wb}$ to have full column rank, and therefore
requires $N_\pi\ge 2$ and variation among the archived virtual returns.

The surrogate $\widehat{\mathcal R}_{\bb}$ is the predictive distribution of
the return without episode noise. It is Gaussian,
\[
\widehat{\mathcal R}_{\bb}(\bs\theta_{\pi})
\sim
\mathcal N
\left(
\mu_{\bb}(\bs\theta_{\pi}),
\sigma_{\bb}^2(\bs\theta_{\pi})
\right),
\]
with mean
\begin{equation}
	\label{eq:gp_posterior_mean}
	\mu_{\bb}(\bs\theta_{\pi})
	=
	\bs\varphi_{\wb}(\bs\theta_{\pi})^{\top}
	\widehat{\bs b}_{\wb}
	+
	\bs k_{\delta}(\bs\theta_{\pi})^{\top}
	\bs V^{-1}
	\left(
	\bs y
	-
	\bs H_{\wb}
	\widehat{\bs b}_{\wb}
	\right).
\end{equation}

Its variance is
\begin{equation}
	\label{eq:gp_posterior_variance}
	\begin{split}
		\sigma_{\bb}^2(\bs\theta_{\pi})
		={}&
		k_{\delta}
		(
		\bs\theta_{\pi},
		\bs\theta_{\pi}
		)
		-
		\bs k_{\delta}(\bs\theta_{\pi})^{\top}
		\bs V^{-1}
		\bs k_{\delta}(\bs\theta_{\pi})
		\\
		&+
		\bs v_{\wb}(\bs\theta_{\pi})^{\top}
		\left(
		\bs H_{\wb}^{\top}
		\bs V^{-1}
		\bs H_{\wb}
		\right)^{-1}
		\bs v_{\wb}(\bs\theta_{\pi}),
	\end{split}
\end{equation}
where
\[
\bs v_{\wb}(\bs\theta_{\pi})
=
\bs\varphi_{\wb}(\bs\theta_{\pi})
-
\bs H_{\wb}^{\top}
\bs V^{-1}
\bs k_{\delta}(\bs\theta_{\pi})
\in\mathbb R^2.
\]

The mean combines the calibrated twin trend with a correlated residual
correction and estimates the expected return targeted in
\eqref{eq:rt_policy_target}. The first two terms of
\eqref{eq:gp_posterior_variance} form the usual kriging variance, while the
third propagates uncertainty in the calibration coefficients. Uncertainty in
the covariance hyperparameters is not propagated. Far from the evaluated
policies, $\bs k_{\delta}(\bs\theta_{\pi})\to\bs 0$, and the mean approaches
\(
\widehat b_{\wb,0}
+
\widehat b_{\wb,1}\mathcal R_{\wb}(\bs\theta_{\pi}).
\)
Thus, unexplored regions inherit the twin trend with an amplitude and
orientation learned from the physical returns.

The covariance hyperparameters $(\sigma_f,\ell_\pi,\sigma_R)$ and the
calibration coefficients are estimated by alternating two conditional steps.
For fixed hyperparameters, the coefficients are updated by generalised least
squares using \eqref{eq:gls_calibration}. For fixed coefficients
$\bs b_{\wb}$, the residual satisfies
\[
\bs y-\bs H_{\wb}\bs b_{\wb}
\sim
\mathcal N(\bs 0,\bs V),
\]
and the hyperparameters are estimated by standard maximum marginal likelihood,
\begin{equation}
	\label{eq:surrogate_likelihood}
	\begin{split}
		\mathcal L
		(
		\sigma_f,\ell_\pi,\sigma_R
		\mid
		\bs b_{\wb}
		)
		={}&
		-\tfrac12
		(\bs y-\bs H_{\wb}\bs b_{\wb})^{\top}
		\bs V^{-1}
		(\bs y-\bs H_{\wb}\bs b_{\wb})
		\\
		&-\tfrac12\log\det\bs V
		-\tfrac{N_\pi}{2}\log 2\pi.
	\end{split}
\end{equation}

The coefficients are initialised by ordinary least squares. The
maximum-marginal-likelihood update of the hyperparameters and the
generalised-least-squares update of the coefficients are then alternated a
fixed number of times.

The number of archived episodes used for fitting can be capped because
factorising $\bs V$ has a computational cost of order $N_\pi^3$. No cap is
needed for the number of episodes considered here. For longer deployments, the
factorisation could be updated when an episode is added, the hyperparameters
could be refitted only occasionally, and sparse Gaussian-process
approximations could be introduced. Such approximations are reviewed by
Rasmussen and Williams~\cite[ch.~8]{Rasmussen2006} and Cao et
al.~\cite{Cao2015}.

	\subsection{Selection of performance and excitation policies}\label{sec_B}\label{sec_D}

	\textsc{SelectInteraction} reads the schedule $e_n$ of \eqref{eq:rt_deployed_policy} and returns either a performance policy or an excitation policy.

	\paragraph{Performance episodes} At a performance episode,
	\textsc{SelectInteraction} selects the member of a finite candidate pool $\mathcal Q$, drawn from the proposal region \eqref{eq:proposal_region} defined below, that maximises the expected improvement under the posterior of
	\eqref{eq:calibrated_return_model},
	\begin{equation}
		\label{eq:RT_acquisition}
		\begin{split}
			\mathcal A_{\mathrm{RT}}(\bs\theta_{\pi})
			&=
			\mathbb E
			\left[
			\max
			\left(
			0,
			\widehat{\mathcal R}_{\bb}(\bs\theta_{\pi})
			-
			\mu_{\bb}^{\max}
			-
			\xi_{\mathrm{EI}}
			\right)
			\right]
			\\
			&=
			\Delta_{\bb}\,
			\Phi\!\left(\Delta_{\bb}/\sigma_{\bb}\right)
			+
			\sigma_{\bb}\,
			\phi\!\left(\Delta_{\bb}/\sigma_{\bb}\right).
		\end{split}
	\end{equation}

	Here,
	$\Delta_{\bb}
	=\mu_{\bb}(\bs\theta_{\pi})-\mu_{\bb}^{\max}-\xi_{\mathrm{EI}}$,
	$\sigma_{\bb}=\sigma_{\bb}(\bs\theta_\pi)$, and $\Phi$ and $\phi$ are the
	distribution and density of the standard normal distribution. The parameter
	$\xi_{\mathrm{EI}}\ge0$ is a user-set exploration margin.

	Because a measured return contains episode-to-episode variability, a single
	large observation is not taken as the current best value. Instead,
	\[
	\mu_{\bb}^{\max}
	=
	\max_{i\le N_\pi}
	\mu_{\bb}(\bs\theta_{\pi}^{(i)})
	\]
	is the largest posterior mean among the evaluated policies \cite{Picheny2013}. Through the episode-noise variance $\sigma_R^2$, the posterior mean at an evaluated policy weighs its measured return against the calibrated trend and the returns measured at nearby policies. Equation~\eqref{eq:RT_acquisition} is the expected
	improvement of Mo\v{c}kus~\cite{Bo_1}, in the closed form given by Jones et
	al.~\cite{Jones1998}.

	The acquisition can be strongly multimodal because the calibrated twin trend
	and the learned discrepancy vary nonlinearly across the policy space. It is
	therefore maximised approximately over a finite random pool rather than by a
	local gradient search. The pool size $N_{\mathcal Q}$ trades search resolution
	against computational cost: a larger pool is more likely to contain a
	candidate with high acquisition value, but requires more acquisition
	evaluations and, when the twin trend is used, more virtual-return evaluations.
	The influence of this choice is considered in Section~\ref{sec5_rt}.

	The twin also restricts the part of the policy space in which candidates are
	proposed. Let
	$\widetilde{\bs\theta}_{\wb,\star}^{(n)}
	\in[0,1]^{n_{\theta_\pi}}$
	be the normalised image of the model-based reference
	$\bs\theta_{\wb,\star}^{(n)}$ under \eqref{eq:policy_whitening}. At each
	performance episode, $N_{\mathcal Q}$ candidates are sampled uniformly from
	the proposal region
	\begin{equation}
		\label{eq:proposal_region}
		\begin{split}
			\mathcal P_n
			&=
			\left\{
			\bs\theta_{\pi}\in\Theta_\pi:
			\left|
			\widetilde\theta_{\pi,j}
			-
			\widetilde\theta_{\wb,\star,j}^{(n)}
			\right|
			\le \rho_j,\quad
			j=1,\ldots,n_{\theta_\pi}
			\right\},
			\\
			\mathcal Q
			&\subset\mathcal P_n,
			\qquad
			\rho_j
			=
			\min
			\left[
			g_j\min(\kappa\ell_j,\rho_{\max}),
			\,1
			\right].
		\end{split}
	\end{equation}

	The set $\mathcal P_n\subseteq\Theta_\pi$ is the part of the policy-parameter
	space from which candidates are drawn. Its centre is the model-based reference.
	The relevance scale $\ell_j$ sets the baseline half-width in coordinate $j$,
	$\kappa$ is a user-set multiplier, and $\rho_{\max}$ caps this baseline
	half-width. The expansion factor $g_j$ is initialised to one and allows the
	half-width to grow beyond its baseline value. The outer bound of one and the
	restriction $\bs\theta_\pi\in\Theta_\pi$ truncate the region at the physical
	bounds of the policy space.

	The expansion is guided by the best-evaluated policy according to the
	posterior mean,
	\[
	\bs\theta_{\pi,\mathrm{best}}=\bs\theta_\pi^{(i_{\mathrm{best}})},
	\qquad i_{\mathrm{best}}\in
	\underset{i\le N_\pi}{\arg\max}\;
	\mu_{\bb}(\bs\theta_\pi^{(i)}).
	\]

	Before sampling a candidate pool, $g_j$ is doubled if this policy lies
	within a distance $\tau\rho_j$ of a face of $\mathcal P_n$ that lies
	strictly inside $\Theta_\pi$, where $\tau$ is a user-defined fraction
	of the half-width. Expansion is also triggered if the policy already
	lies beyond such a face, which may occur when a twin update shifts
	the centre of the proposal region. Faces coinciding with the boundary
	of $\Theta_\pi$ do not trigger expansion, since the search cannot
	extend beyond them.

	The proposal region restricts the domain over which
	\eqref{eq:RT_acquisition} is maximised without altering the acquisition
	function itself. It thus plays a role analogous to that of a trust
	region in local Bayesian optimisation \cite{Eriksson2019}.
	Evidence from the physical system informs the adaptation of its size,
	but not its centre: between twin updates, $\mathcal P_n$ remains
	centred on the model-based reference. After a twin update, the updated
	reference and relevance scales define the new centre and baseline
	half-widths, while the expansion factors $g_j$ are retained.
	The adaptation therefore changes the extent of the region from which
	candidates are sampled, while keeping the pool size $N_{\mathcal Q}$
	constant. Section~\ref{sec5_rt} assesses the effect of this mechanism
	by comparing fixed and adaptive proposal regions.
	An alternative strategy, left for future work, would be to recentre
	the proposal region on the evaluated policy with the largest posterior
	mean, following the approach of \cite{Eriksson2019}.

\paragraph{Information episodes}
When $u_{\iota,n}=1$, the excitation policy
$\pi_{\mathrm{info}}(\cdot\mid\bs\theta_\iota)$ is synthesised on the
twin by minimising the E-optimal criterion
\eqref{eq:rt_e_optimal_policy}.

The sensitivities are computed with respect to the twin parameters
normalised to the unit box,
$\widetilde{\bs\theta}_p\in[0,1]^{n_{\theta_p}}$,
using the map \eqref{eq:policy_whitening} applied to the bounds
of $\Theta_p$. This normalisation accounts for differences in
units and ranges among the parameters. Since the eigenvalues
of $\overline{\bs F}$ depend on the parameter scaling, the
excitation policy selected by the E-optimal criterion depends
on the chosen normalisation, and hence on $\Theta_p$.

The sensitivities
$\bs S_{i,q}=\partial\bar{\bs O}_{\wb}/\partial\widetilde{\bs\theta}_p$
in \eqref{eq:rt_info_ensemble_fisher} are computed by differentiating
through the counterfactual rollouts. Each rollout is initialised
from the analysed state of a recorded episode, uses the corresponding
exogenous input, and predicts the quantities used for identification
(Section~\ref{sec_C}). Before the first twin update, the sensitivities
are evaluated at several parameter hypotheses distributed across
$\Theta_p$. After the first update, they are evaluated only at
the current parameter estimate $\bs\theta_p^{(m)}$.

The measurement errors of the $n_{\mathrm{id}}$ quantities used
for identification are assumed to be Gaussian, with covariance
$\bs\Sigma_{\mathrm{id}}$ at each time step, and temporally
independent. Accordingly,
$\bs\Sigma_{\mathrm{id},N}=\bs I_N\otimes\bs\Sigma_{\mathrm{id}}$
in \eqref{eq:rt_info_ensemble_fisher}, where $\otimes$ denotes
the Kronecker product.

$\overline{\bs F}$ is the Fisher information matrix
\cite{Mehra1974,WalterPronzato1997} under three conditions:
(i) the twin correctly describes the physical system,
(ii) the sensitivities are evaluated at a single parameter vector,
and (iii) the actions are prescribed in open loop.
If these conditions hold and $\overline{\bs F}$ is not singular, its inverse provides a
lower bound on the covariance of any unbiased estimator of
$\widetilde{\bs\theta}_p$ (Cram\'er--Rao bound).
The excitation policy selected by
\eqref{eq:rt_nested_information_design} then minimises the
largest eigenvalue of this covariance bound.

These conditions are not generally satisfied in this work.
Hence, $\overline{\bs F}$ is interpreted as an average
of the information associated with the parameter hypotheses and
is used solely to rank candidate excitation policies, but not to quantify parameter uncertainty.

	In this work, $\Theta_\iota$ is a finite library of excitation policies
	(Section~\ref{sec4p3}), screened via enumeration. For diagnostics, the information on parameter $j$ after accounting for the others is the Schur complement
	\begin{equation}
		\label{eq:rt_marginal_information}
		\overline F_{j\mid\sim j}
		=
		\overline F_{jj}
		-
		\overline{\bs F}_{j,\sim j}
		\,
		\overline{\bs F}_{\sim j,\sim j}^{+}
		\,
		\overline{\bs F}_{\sim j,j},
		\qquad
		j=1,\ldots,n_{\theta_p},
	\end{equation}
	with $\sim j$ all coordinates except $j$ and $(\cdot)^{+}$ the pseudo-inverse, needed when $\overline{\bs F}_{\sim j,\sim j}$ is singular; $\overline F_{j\mid\sim j}=0$ marks a parameter that is not identifiable to first order. When $\overline{\bs F}$ is invertible, $\overline F_{j\mid\sim j}=1/[\overline{\bs F}^{-1}]_{jj}$.

	\subsection{Deployment and state synchronisation}
	\label{sec_sync}

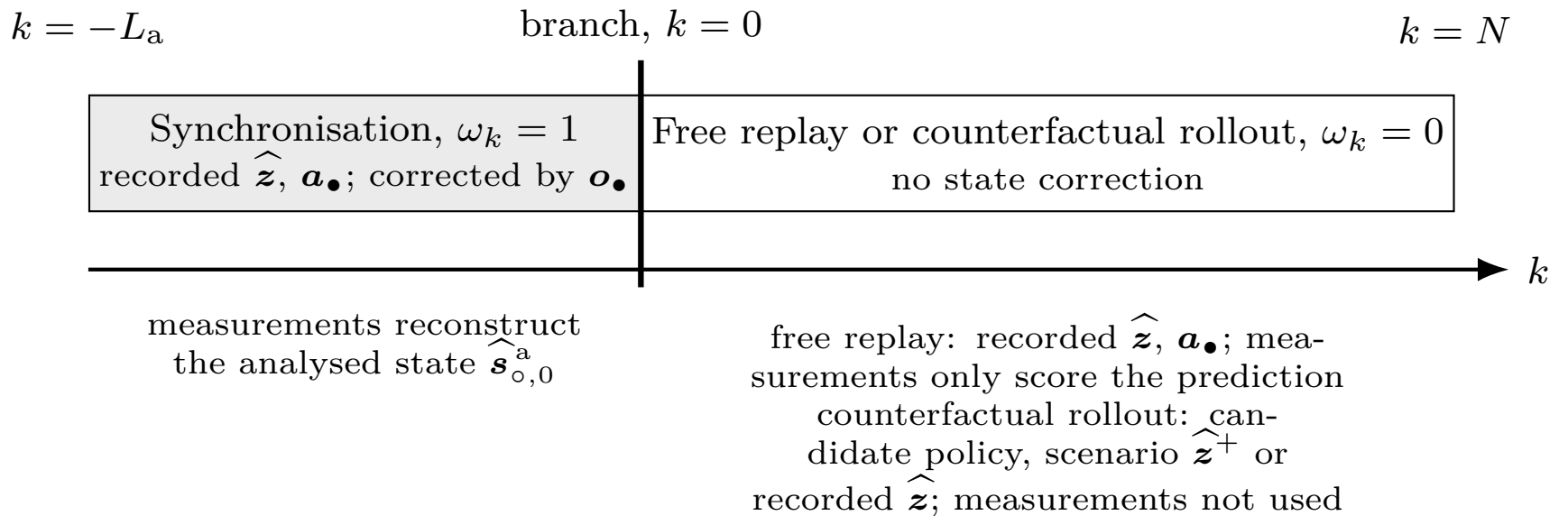
\begin{figure*}[t]
		\centering
		\begin{tikzpicture}[
			>=Latex,
			font=\footnotesize,
			cell/.style={draw,minimum height=0.85cm,align=center,inner sep=2pt},
			note/.style={font=\scriptsize,align=center,text width=3.4cm}
			]
			\node[cell,minimum width=3.8cm,anchor=west,fill=black!8] (sync) at (0,0)
			{Synchronisation, $\omega_k=1$\\{\scriptsize recorded $\widehat{\bs z}$, $\bs a_{\bb}$; corrected by $\bs o_{\bb}$}};
			\node[cell,minimum width=5.8cm,anchor=west] (free) at (sync.east)
			{Free replay or counterfactual rollout, $\omega_k=0$\\{\scriptsize no state correction}};
			\draw[very thick] ($(sync.north east)+(0,0.25)$) -- ($(sync.south east)+(0,-0.55)$);
			\node[anchor=south,font=\footnotesize] at ($(sync.north east)+(0,0.25)$) {branch, $k=0$};
			\node[anchor=south,font=\footnotesize] at ($(sync.north west)+(0,0.25)$) {$k=-L_{\mathrm a}$};
			\node[anchor=south,font=\footnotesize] at ($(free.north east)+(0,0.25)$) {$k=N$};
			\draw[->,thick] ($(sync.south west)+(0,-0.42)$) -- ($(free.south east)+(0.40,-0.42)$)
			node[right,font=\footnotesize] {$k$};
			\node[note,anchor=north] at ($(sync.south)+(0,-0.62)$)
			{measurements reconstruct the analysed state $\widehat{\bs s}_{\wb,0}^{\,\mathrm a}$};
			\node[note,anchor=north,text width=5.6cm] at ($(free.south)+(0,-0.62)$)
			{free replay: recorded $\widehat{\bs z}$, $\bs a_{\bb}$; measurements only score the prediction\\
				counterfactual rollout: candidate policy, scenario $\widehat{\bs z}^{+}$ or recorded $\widehat{\bs z}$; measurements not used};
		\end{tikzpicture}
		\caption{Time axis of one use of the twin. Before the branch, the recursion \eqref{eq:twin_recursion} is a sequential observer; after it, the twin runs freely, either along the recorded input and actions (free replay, used for identification and validation) or under a candidate policy (counterfactual rollout: scenarios of $\mathcal C^{+}$ for virtual returns, recorded inputs for excitation-policy synthesis). Interval lengths for the benchmark are given in Section~\ref{sec5p_solver}.}
		\label{fig:sync_free_forecast}
	\end{figure*}

	Figure~\ref{fig:sync_free_forecast} shows the two phases of a twin rollout.
	Before the branch $k=0$ (Section~\ref{sec2a}), measurements correct the twin state over the synchronisation window. For $k>0$, the twin runs
	without state correction, either under the recorded inputs and actions in a
	free replay or under a candidate policy in a counterfactual rollout.

	\textsc{DeployAndRecord} runs the selected policy for one physical episode and
	appends its record to $\mathcal B$. For a performance
	episode, $(\bs\theta_\pi^{(n)},\mathcal R_{\bb}^{(n)})$ also enters
	$\mathcal D_\pi$. Episode $i\le n$ is stored as
{\fontsize{9}{12}\selectfont%
	\begin{equation}
		\label{eq:replay_record}
		\mathcal E^{(i)}
		=
		\left\{
		\bs\theta^{(i)},\;
		\widehat{\bs z}_{-L_{\mathrm a}:N-1}^{(i)},\;
		\bs a_{\bb,-L_{\mathrm a}:N-1}^{(i)},\;
		\bs o_{\bb,-L_{\mathrm a}:N}^{(i)},\;\bs\chi_{\bb,0}^{(i)},\;
		\mathcal R_{\bb}^{(i)}
		\right\},
	\end{equation}
}%
	where $\bs\theta^{(i)}$ are the parameters of the deployed policy ($\bs\theta_\pi^{(i)}$ or $\bs\theta_\iota^{(i)}$ depending on $e_i$), the time axis starts at $k=-L_{\mathrm a}$, the beginning of the synchronisation window, $\bs\chi_{\bb,0}^{(i)}$ is the policy memory at the branch, which initialises that of the twin, and $\mathcal R_{\bb}^{(i)}$ is the measured return. The record contains the measurements, inputs, and actions
	needed to reconstruct the analysed state at the branch. The policy memory
	$\bs\chi_k$ is carried alongside this state. The buffer grows with $n$, and a bounded subset of its records may be supplied to each identification update. Given that the number of episodes run in this work is relatively small, no record is ever discarded. In a real deployment scenario however, memory management criteria would also be required to determine what data is worth keeping.

	Over a recorded episode, the twin is advanced by a forecast--analysis recursion with a switch $\omega_k\in\{0,1\}$ on the measurement correction,
	\begin{equation}
		\label{eq:twin_recursion}
		\begin{aligned}
			\bs s_{\wb,k+1}^{\mathrm f}
			&=
			F_{\wb}
			\!\left(
			\bs s_{\wb,k},
			\widehat{\bs z}_k,
			\bs a_{\bb,k};
			\bs\theta_p
			\right),
			\\
			\bs\nu_{k+1}
			&=
			\bs o_{\bb,k+1}
			-
			h\!\left(\bs s_{\wb,k+1}^{\mathrm f},\bs a_{\bb,k}\right),
			\\
			\bs s_{\wb,k+1} &= \Pi \!\left(
			\bs s_{\wb,k+1}^{\mathrm f}
			+
			\omega_k\,
			\bs K_{k+1}\bs\nu_{k+1}
			\right),
			\\
			\omega_k
			&=
			\begin{cases}
				1, & k<0 \quad\text{(synchronisation)},\\
				0, & k\ge 0 \quad\text{(free replay)}.
			\end{cases}
		\end{aligned}
	\end{equation}

	The first line produces the forecast
	$\bs s_{\wb,k+1}^{\mathrm f}$. The second gives the innovation
	$\bs\nu_{k+1}\in\mathbb R^{n_o}$, the difference between the physical
	measurement and the noise-free twin prediction. The gain
	$\bs K_{k+1}$ maps this innovation to a state correction, and $\Pi$ enforces the admissible state set. The gain may follow a filtering model
	\cite{AschBocquetNodet2016} or be prescribed, as in the nudging used here
	(Section~\ref{sec5p0}).

	Before the branch, $\omega_k=1$ and the recursion acts as an observer. Its
	value at $k=0$ is the analysed state
	$\widehat{\bs s}_{\wb,0}^{\,\mathrm a}$ of \eqref{eq:rt_sync_intro}. After the
	branch, $\omega_k=0$: innovations score the free prediction but do not correct
	it. Otherwise, an inaccurate twin could obtain a small identification loss
	because of the observer correction. The resulting output-error criterion
	\cite{Ljung1999} also matches the free propagation used in counterfactual
	rollouts.

	The observer propagates no covariance. Identification and excitation-policy
	synthesis therefore treat the analysed state as exact, so state errors may be
	absorbed by the identified parameters. Other state estimators could replace
	this observer without changing the replay protocol.

	\textsc{Synchronise} obtains
	$\widehat{\bs s}_{\wb,0}^{\,\mathrm a}$ by applying
	\eqref{eq:twin_recursion} to
	$\mathcal W_n^{-}=\{\bs o_{\bb,k}\}_{k=-L_{\mathrm a}}^{0}
	\cup\{\widehat{\bs z}_k,\bs a_{\bb,k}\}_{k=-L_{\mathrm a}}^{-1}$.
	This defines the operator $\Psi_{\wb}$ of \eqref{eq:rt_sync_intro}. The
	recursion starts from a state guess or from the state carried over from the
	previous episode. A sufficiently long window reduces the influence of this
	initial guess. Applying the operation to the $N_{\mathrm s}$ most recent
	episodes gives
	\begin{equation}
		\label{eq:branch_ensemble}
		\widehat{\mathcal I}^{\mathrm a}
		=
		\left\{
		\widehat{\bs s}_{\wb,0}^{\,\mathrm a,(i)}
		\right\}_{i=n-N_{\mathrm s}+1}^{n}.
	\end{equation}

	Each element is a single estimate of the branch state, not a sample from a
	state distribution. It may inherit bias from the twin, which further supports
	scoring identification on the free prediction. The replay states are
	recomputed after each twin update; the scoring-bank states, reconstructed once
	with the initial twin, remain fixed.

	\subsection{Replay-based identification}\label{sec_C}

	\textsc{IdentifyTwin} fits the twin parameters by replaying recorded episodes.
	It uses the $N_{\mathrm s}$ most recent episodes, for which replay states are
	available, and minimises \eqref{eq:rt_replay_loss} over a training subset
	$\mathcal B_{\mathrm{tr}}$. In the trajectory term \eqref{J_o}, the discrepancy
	between a measurement after the branch and the noise-free prediction from its
	free replay is
	\begin{equation}
		\label{eq:rt_replay_dyn_loss}
		\ell_{\mathrm{dyn}}
		\!\left(
		\bs o_{\bb,k},\;
		\bar{\bs o}_{\wb,k}
		\right)
		=
		\frac{1}{n_{\mathrm{id}}}
		\,
		\bs\nu_{k,\mathrm{id}}^{\top}
		\bs\Sigma_{\mathrm{id}}^{-1}
		\bs\nu_{k,\mathrm{id}}.
	\end{equation}

		Here $\bs\nu_{k,\mathrm{id}}\in\mathbb R^{n_{\mathrm{id}}}$ contains the
	components of the mismatch $\bs\nu_k$ used for identification, and
	$\bs\Sigma_{\mathrm{id}}$ is the corresponding block of $\bs\Sigma_o$.
	Measurements used only for synchronisation or return evaluation are excluded
	(Section~\ref{sec4p2}). Because $\omega_k=0$ after the branch, this term is a
	whitened mean squared output error. The observer influences it only through
	the analysed state at the branch.

	The return term \eqref{eq:rt_replay_rew_loss} is made dimensionless by the scale $s_R=\max\{s_{\min},\mathcal R_{\mathrm{rms}}\}$, with $\mathcal R_{\mathrm{rms}}$ the root-mean-square of the measured returns in the buffer at the first scheduled update and $s_{\min}$ a small positive floor set by the user. The return term constrains the returns of the deployed policies instead of the ranking of untried ones. The weight $c_{\mathrm{rew}}$ gives the return term a declared share $\eta_R\in[0,1)$ of the loss at the first update,
	\begin{equation}
		\label{eq:rt_reward_balance}
		c_{\mathrm{rew}}
		=
		\frac{\eta_R}{1-\eta_R}
		\frac{
			\mathcal J_{\mathrm{dyn}}
			(\bs\theta_p^{(0)}\mid\mathcal B_{\mathrm{tr}};
			\widehat{\mathcal I}^{\mathrm a})
		}{
			\max\!\left\{
			\mathcal J_{\mathrm{rew}}
			(\bs\theta_p^{(0)}\mid\mathcal B_{\mathrm{tr}};
			\widehat{\mathcal I}^{\mathrm a}),
			j_{\min}
			\right\}
		},
	\end{equation}
	evaluated on the training set of the first scheduled update, with
	$j_{\min}>0$ a user-set floor. While this floor is inactive,
	$\eta_R=1/2$ gives equal initial contributions while $\eta_R=0$ removes the return
	term. Both
	$c_{\mathrm{rew}}$ and $s_R$ are then held fixed, so later changes in the loss
	are not caused by rescaling. While the floor is inactive, the weighted return
	term is independent of $s_R$ because the same scale enters
	$\mathcal J_{\mathrm{rew}}$ and \eqref{eq:rt_reward_balance}.

The analysed states are computed with the current twin and held fixed while candidate parameters are compared. Thus, every candidate starts from states synchronised with $\bs\theta_p^{(m)}$, rather than from candidate-specific states. This avoids an observer pass at every loss evaluation and keeps the comparison consistent within an update. On the other hand, using states synchronised with $\bs\theta_p^{(m)}$ may mildly favour nearby candidate parameters, although this effect decreases when the twin rapidly forgets its initial state over the free interval. After identification, the replay states are recomputed with the updated twin before the next update.

	At each scheduled update, a bounded set of episodes is split into
	episode-disjoint training and validation sets $\mathcal B_{\mathrm{tr}}$ and
	$\mathcal B_{\mathrm{val}}$. The hybrid global--local strategy of
	Section~\ref{sec3_synth} is applied in the parameter space normalised to the unit box (Section~\ref{sec_D}). A Gaussian-process surrogate of the
	validation loss proposes candidates by expected improvement. Projected
	AMSGrad then refines each candidate on the training loss, using reverse-mode
	gradients through the free replay. The refined candidates are scored on the
	fixed validation set. The best replaces the current parameters only if it
	lowers the validation loss; otherwise the twin remains unchanged.

		\section{Selected benchmark}\label{sec4}

	The proposed benchmark is a one-dimensional Burgers flow from which $N_a$ devices extract a transported resource. It retains features that make the control of advection-dominated flows difficult: nonlinear advection, partial observability, delayed action effects, and upstream--downstream competition. The dynamics also depend strongly on the actuation. The problem is a one-dimensional analogue of distributed energy-extraction systems such as wind farms, where upstream actuation affects both local power extraction and the flow available to downstream devices. Varying upstream turbine thrust over time, known as dynamic-induction control, can enhance wake recovery, as shown by the large-eddy simulations \cite{Munters2017,Munters2018,Yilmaz2018} and wind-tunnel experiments \cite{Frederik2020Periodic}. Dynamic individual pitch control pursues the same aim through enhanced wake mixing \cite{Frederik2020Helix}.

	Section~\ref{sec4p1} presents the physical system and the digital twin, Section~\ref{sec4p2} the observations, the reward and the return, and Section~\ref{sec4p3} the policy. Sections~\ref{sec5p_solver}--\ref{sec5p0b} give the numerical set-up and the episode protocol, the observer, and the settings of the identification and the policy search.

	\subsection{Physical system and digital twin}
	\label{sec4p1}

	In the notation of Section~\ref{sec2a}, the digital twin state includes the velocity at the nodes of the numerical grid and the actuator states, while the physical system state also includes hidden interaction states absent from the twin. The action contains the device commands and the amplitude, frequency, and phase of the upstream harmonic, which also enter the recovery law (Section~\ref{sec4p3}). The exogenous input is the inflow, which the twin and the policy measure at the inlet sensor. The observations are point velocities, actuator states, and metered powers and energies (Section~\ref{sec4p2}), and the twin parameters are those of the recovery law and the footprint widths \eqref{eq:unified_theta_p}.

	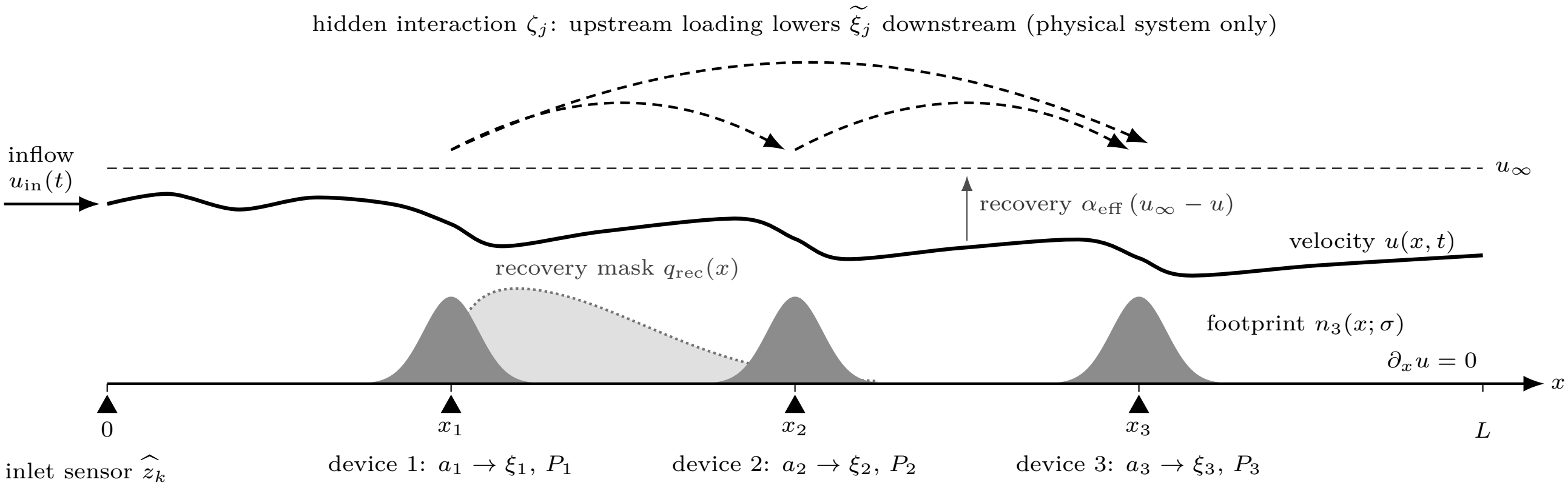
\begin{figure*}[t]
		\centering
		\begin{tikzpicture}[
			x=1.5cm,y=1cm,
			>=Latex,
			font=\footnotesize,
			lab/.style={font=\scriptsize,align=center,inner sep=1.5pt}
			]
			\draw[densely dashed] (0,2.35) -- (10,2.35);
			\node[lab,anchor=west] at (10.05,2.35) {$u_\infty$};
			\fill[black!12] plot coordinates {(2.050,0.0080) (2.100,0.0137) (2.150,0.0232) (2.200,0.0390) (2.250,0.0645) (2.300,0.1045) (2.350,0.1647) (2.400,0.2492) (2.450,0.3582) (2.500,0.4846) (2.550,0.6148) (2.600,0.7338) (2.650,0.8320) (2.700,0.9065) (2.750,0.9595) (2.800,0.9952) (2.850,1.0179) (2.900,1.0311) (2.950,1.0372) (3.000,1.0378) (3.050,1.0340) (3.100,1.0267) (3.150,1.0162) (3.200,1.0029) (3.250,0.9871) (3.300,0.9689) (3.350,0.9487) (3.400,0.9265) (3.450,0.9025) (3.500,0.8770) (3.550,0.8500) (3.600,0.8218) (3.650,0.7926) (3.700,0.7625) (3.750,0.7316) (3.800,0.7003) (3.850,0.6687) (3.900,0.6369) (3.950,0.6050) (4.000,0.5734) (4.050,0.5420) (4.100,0.5111) (4.150,0.4807) (4.200,0.4510) (4.250,0.4221) (4.300,0.3941) (4.350,0.3670) (4.400,0.3409) (4.450,0.3159) (4.500,0.2919) (4.550,0.2692) (4.600,0.2475) (4.650,0.2271) (4.700,0.2078) (4.750,0.1897) (4.800,0.1727) (4.850,0.1569) (4.900,0.1421) (4.950,0.1284) (5.000,0.1158) (5.050,0.1041) (5.100,0.0934) (5.150,0.0835) (5.200,0.0746) (5.250,0.0664) (5.300,0.0589) (5.350,0.0522) (5.400,0.0461) (5.450,0.0407) (5.500,0.0357) (5.550,0.0314) (5.600,0.0274)} -- (5.6,0) -- (2.05,0) -- cycle;
			\draw[black!55,densely dotted,thick] plot coordinates {(2.050,0.0080) (2.100,0.0137) (2.150,0.0232) (2.200,0.0390) (2.250,0.0645) (2.300,0.1045) (2.350,0.1647) (2.400,0.2492) (2.450,0.3582) (2.500,0.4846) (2.550,0.6148) (2.600,0.7338) (2.650,0.8320) (2.700,0.9065) (2.750,0.9595) (2.800,0.9952) (2.850,1.0179) (2.900,1.0311) (2.950,1.0372) (3.000,1.0378) (3.050,1.0340) (3.100,1.0267) (3.150,1.0162) (3.200,1.0029) (3.250,0.9871) (3.300,0.9689) (3.350,0.9487) (3.400,0.9265) (3.450,0.9025) (3.500,0.8770) (3.550,0.8500) (3.600,0.8218) (3.650,0.7926) (3.700,0.7625) (3.750,0.7316) (3.800,0.7003) (3.850,0.6687) (3.900,0.6369) (3.950,0.6050) (4.000,0.5734) (4.050,0.5420) (4.100,0.5111) (4.150,0.4807) (4.200,0.4510) (4.250,0.4221) (4.300,0.3941) (4.350,0.3670) (4.400,0.3409) (4.450,0.3159) (4.500,0.2919) (4.550,0.2692) (4.600,0.2475) (4.650,0.2271) (4.700,0.2078) (4.750,0.1897) (4.800,0.1727) (4.850,0.1569) (4.900,0.1421) (4.950,0.1284) (5.000,0.1158) (5.050,0.1041) (5.100,0.0934) (5.150,0.0835) (5.200,0.0746) (5.250,0.0664) (5.300,0.0589) (5.350,0.0522) (5.400,0.0461) (5.450,0.0407) (5.500,0.0357) (5.550,0.0314) (5.600,0.0274)};
			\node[lab,anchor=south west,text=black!75] at (2.78,1.07) {recovery mask $q_{\mathrm{rec}}(x)$};
			\foreach \xj in {2.5,5.0,7.5}{
				\fill[black!45] plot[domain=\xj-0.75:\xj+0.75,samples=60] (\x,{0.95*exp(-(\x-\xj)^2/(2*0.2^2))}) -- (\xj+0.75,0) -- (\xj-0.75,0) -- cycle;
			}
			\node[lab,anchor=west] at (7.95,0.64) {footprint $n_3(x;\sigma)$};
			\draw[very thick] plot[smooth,tension=0.55] coordinates {(0,1.96) (0.45,2.07) (0.95,1.90) (1.5,2.03) (2.1,1.95) (2.5,1.74) (2.85,1.50) (3.6,1.66) (4.6,1.80) (5.0,1.58) (5.35,1.36) (6.2,1.48) (7.1,1.57) (7.5,1.37) (7.85,1.18) (8.8,1.29) (10,1.40)};
			\node[lab,anchor=south west] at (8.55,1.36) {velocity $u(x,t)$};
			\draw[->,black!70] (6.25,1.56) -- (6.25,2.27);
			\node[lab,anchor=west,text=black!80] at (6.3,1.95) {recovery $\alpha_{\mathrm{eff}}\,(u_\infty-u)$};
			\draw[thick,->] (0,0) -- (10.45,0) node[lab,anchor=west] {$x$};
			\foreach \xt/\lt in {0/{$0$},2.5/{$x_1$},5.0/{$x_2$},7.5/{$x_3$},10/{$L$}}{
				\draw (\xt,0) -- (\xt,-0.09);
				\node[lab,anchor=north] at (\xt,-0.36) {\lt};
			}
			\foreach \xs in {0,2.5,5.0,7.5}{
				\fill (\xs,-0.12) -- ($(\xs,-0.12)+(-0.075,-0.2)$) -- ($(\xs,-0.12)+(0.075,-0.2)$) -- cycle;
			}
			\foreach \xj/\j in {2.5/1,5.0/2,7.5/3}{
				\node[lab,anchor=north] at (\xj,-0.72) {device \j: $a_\j\rightarrow\xi_\j$, $P_\j$};
			}
			\draw[->,thick] (-0.75,1.96) -- (-0.05,1.96);
			\node[lab,anchor=south] at (-0.48,2.02) {inflow\\$u_{\mathrm{in}}(t)$};
			\node[lab,anchor=north] at (-0.15,-0.72) {inlet sensor $\widehat z_k$};
			\node[lab,anchor=south east] at (10.0,0.08) {$\partial_x u=0$};
			\draw[->,densely dashed,thick] (2.5,2.55) to[bend left=28] (4.93,2.55);
			\draw[->,densely dashed,thick] (5.0,2.55) to[bend left=28] (7.43,2.55);
			\draw[->,densely dashed,thick] (2.5,2.55) to[bend left=24] (7.57,2.62);
			\node[lab,anchor=south] at (5.0,3.72) {hidden interaction $\zeta_j$: upstream loading lowers $\widetilde\xi_j$ downstream (physical system only)};
		\end{tikzpicture}
		\caption{Sketch of the benchmark. The inflow $u_{\mathrm{in}}(t)$ enters at $x=0$ and is advected past the $N_a=3$ devices, which extract over the footprints $n_j(x;\sigma)$ with the effective intensities $\widetilde\xi_j$, while the velocity recovers towards $u_\infty$ at the rate $\alpha_{\mathrm{eff}}$. Oscillating the command of the first device raises this rate over the mask $q_{\mathrm{rec}}(x)$. Triangles mark the velocity sensors. The dashed arrows show the interaction that is hidden from the digital twin.}
		\label{fig:benchmark_sketch}
	\end{figure*}

	The resource to be extracted is a scalar velocity field $u(x,t)$ on $x\in[0,L]$ (Figure~\ref{fig:benchmark_sketch}), governed by
	\begin{equation}
		\label{env}
		\begin{split}
		\partial_t u + u\,\partial_x u
		={}&
		\nu_0\,\partial_{xx}u
		+
		\alpha_{\mathrm{eff}}(x,t)\,\bigl(u_\infty-u\bigr)
		\\
		&-
		\sum_{j=1}^{N_a}\widetilde\xi_j(t)\,n_j(x;\sigma)\,u_+(x,t),
		\end{split}
	\end{equation}
	where $\nu_0$ is a constant diffusivity, $\alpha_{\mathrm{eff}}$, defined in \eqref{eq:effective_coefficients}, is the rate of recovery towards the constant level $u_\infty$, and the last term is the extraction by the devices, with the footprints $n_j$ and the effective intensities $\widetilde\xi_j$ defined below. Extraction acts only on positive values, $u_+=\max(u,0)$. The inflow $u(0,t)=u_{\mathrm{in}}(t)$ is a stochastic exogenous input (Section~\ref{sec5p_solver}), and the outflow satisfies $\partial_x u(L,t)=0$. The recovery level $u_\infty$ is instead a constant, close to the mean inflow. In the wind-farm analogy, it stands for the undisturbed wind outside the wakes, from which the wakes recover momentum. Being constant, it carries none of the inflow fluctuations, which reach the devices only by advection from the inlet. All quantities are dimensionless.

	The recovery rate has a constant part and a dynamic increment, which the policy triggers by oscillating the command of the upstream device ($j=1$),
	\begin{equation}
		\label{eq:effective_coefficients}
		\begin{aligned}
		\alpha_{\mathrm{eff}}(x,t)
		&=
		\alpha_0
		+
		q_{\mathrm{rec}}(x)\,\delta\alpha(t)
		\left[1+\mu_\alpha\sin\!\left(\phi(t)-\phi_\alpha\right)\right],
		\\
		\delta\alpha(t)
		&=
		\Delta\alpha\,g_\alpha\!\left(A_1(t),f_1(t)\right),
		\end{aligned}
	\end{equation}
	where $A_1$, $f_1$ and $\phi$ are the amplitude, frequency and phase that the policy assigns to the upstream harmonic (Section~\ref{sec4p3}); the phase carries no device index because the downstream harmonics are locked to it. The mask $q_{\mathrm{rec}}(x)$ concentrates the increment just downstream of the first device. The gain $g_\alpha\in[0,1]$ grows steeply with the amplitude $A_1$ up to $A_{\mathrm{ref}}$. As a function of the frequency $f_1$, it peaks near $f^\star$ with bandwidth $s_f$ and decreases towards zero below the roll-on frequency $f_\alpha$ (see \eqref{eq:dynamic_induction_gains} and Figure~\ref{fig:appendix_recovery_gain}). The scale $\Delta\alpha$ sets the size of the increment, which is modulated at the frequency of the upstream harmonic, with modulation depth $\mu_\alpha$ and phase lag $\phi_\alpha$ (\ref{app:benchmark_details}). Since $g_\alpha$ vanishes at zero amplitude or frequency, $\alpha_{\mathrm{eff}}=\alpha_0$ under steady commands. The increment is made deliberately strong and selective: dynamic operation pays off only in a narrow band of amplitude and frequency, which steady operation does not reveal and which the policy search has to find.

	Each device $j$ extracts over the Gaussian footprint
	\begin{equation}
		n_j(x;\sigma)=\exp\!\left[-\frac{(x-x_j)^2}{2\sigma^2}\right],
	\end{equation}
	centred at $x_j$, with a width $\sigma$ common to all devices. Its nominal extraction intensity, the counterpart of the thrust coefficient of a turbine, is the actuator state $\xi_j(t)\ge0$, which follows the command $a_j(t)$ with a known time constant $\tau_\xi$,
	\begin{equation}
		\label{eq:actuator_dynamics}
		\tau_\xi\dot\xi_j+\xi_j=a_j(t),
		\qquad
		a_j(t)\in[0,\xi_{\max}].
	\end{equation}

	The commands $a_{j,k}$ are selected at $t_k$ and held constant over $[t_k,t_{k+1})$ (zero-order hold).

	To reproduce the structural errors that arise when a digital twin is an imperfect representation of the physical system, the simulated physical system and the digital twin are deliberately given different model structures. These differ in the effective extraction intensity $\widetilde\xi_j$ entering \eqref{env}. In the digital twin, each device extracts with its own actuator state, $\widetilde\xi_j=\xi_j$, and the devices interact only through the flow. In the physical system, the effectiveness of each device is also reduced by the recent loading of the devices upstream of it, adding a hidden interaction to that already mediated by the flow. This additional interaction is represented through hidden states $\zeta_j$ satisfying
	\begin{equation}
		\label{eq:plant_interaction}
		\tau_\zeta\dot\zeta_j
		=
		-\zeta_j+\sum_{i<j}w_{ij}\,\frac{\xi_i(t)}{\xi_{\max}},
		\qquad \zeta_1=0,
	\end{equation}
	which modify the effective intensity as
	\begin{equation}
		\label{eq:effective_actuation}
		\widetilde\xi_j(t)=\xi_j(t)\,\bigl[1-c_\zeta\,\zeta_j(t)\bigr].
	\end{equation}

The weights $w_{ij}$ decay exponentially with the distance between devices over a characteristic length $\ell_w$ and are normalised to sum to one for each $j$ (\ref{app:benchmark_details}). Each $\zeta_j$ tracks a weighted history of the normalised upstream loading, with $\tau_\zeta$ setting its response time and $c_\zeta$ the interaction strength. Crucially, because this mechanism is absent from the digital twin, neither the hidden states $\bs\zeta$ nor the parameters $(c_\zeta,\tau_\zeta,\ell_w)$ belong to its model structure. The resulting discrepancy is a structural error whose effect depends on the deployed policy.

The diffusivity $\nu_0$, the recovery level $u_\infty$, the actuator constants $\tau_\xi$ and $\xi_{\max}$, and all constants not included in $\bs\theta_p$ are assumed known. The twin uses the same recovery-law structure as the physical system, but its five recovery parameters and the footprint width are unknown. The resulting parameter vector is
	\begin{equation}
		\label{eq:unified_theta_p}
		\bs\theta_p
		=
		[\alpha_0,\,\Delta\alpha,\,f^\star,\,s_f,\,f_\alpha,\,\sigma]^\top ,
	\end{equation}
	identified in a box $\Theta_p$; the box and the values of $\bs\theta_p$ in the physical system are given in \eqref{eq:theta_p_box} and \eqref{eq:theta_p_physical} of \ref{app:benchmark_details}, and Table~\ref{tab:burgers_settings} lists the main constants of the benchmark. Since no $\bs\theta_p$ reproduces \eqref{eq:plant_interaction}--\eqref{eq:effective_actuation}, identification yields the best in-family approximation of Section~\ref{sec2a}; the corresponding parameter values used in the physical-system simulation serve for comparison only and are not the target of identification. With the field on the $n_x$ nodes of the numerical grid (Section~\ref{sec5p_solver}), $\bs u_k\in\mathbb R^{n_x}$, the two states are $\bs s_{\wb,k}=[\bs u_k,\bs\xi_k]\in\mathbb R^{n_x+N_a}$ and $\bs s_{\bb,k}=[\bs u_k,\bs\xi_k,\bs\zeta_k]\in\mathbb R^{n_x+2N_a}$. Each side has its own field and actuator states; the extracted energies of Section~\ref{sec4p2}, zero at the branch $t=0$, are accumulated alongside.

	\begin{table}[t]
		\centering
		\caption{Main constants of the Burgers benchmark. Additional constants,
			search bounds and physical parameter values are given in
			\ref{app:benchmark_details}.}
		\label{tab:burgers_settings}
		\footnotesize
		\begin{tabularx}{\columnwidth}{@{}>{\raggedright\arraybackslash}Xll@{}}
			\toprule
			Quantity & Symbol & Value \\
			\midrule
			Domain length, grid points & $L,\ n_x$ & $10,\ 150$ \\
			Numerical and control time steps & $\Delta t_{\mathrm{num}},\ \Delta t$ & $2\times 10^{-3},\ 0.1$ \\
			Number and locations of devices & $N_a,\ x_j$ & $3,\ (2.5,5.0,7.5)$ \\
			Maximum intensity, actuator time constant & $\xi_{\max},\ \tau_\xi$ & $8,\ 0.25$ \\
			Diffusivity, recovery level & $\nu_0,\ u_\infty$ & $0.05,\ 6.0$ \\
			Interaction strength, time scale, length & $c_\zeta,\ \tau_\zeta,\ \ell_w$ & $0.55,\ 4.0,\ 3.5$ \\
			Sensor noise (velocity, actuator state, power) & $\sigma_u,\ \sigma_\xi,\ \sigma_P$ & $0.175,\ 0.15,\ 1.0$ \\ %
			Balance weight, instantaneous share & $w_{\mathrm{std}},\ \lambda$ & $2,\ 0.10$ \\
			Washout, synchronisation, free interval & $T_{\mathrm{wash}},\ T_{\mathrm{sync}},\ T_{\mathrm{eval}}$ & $5,\ 4,\ 20$ \\
			\bottomrule
		\end{tabularx}
	\end{table}

	\subsection{Observations, reward and return}
	\label{sec4p2}

	\paragraph{Observations} At each decision time $t_k=k\Delta t$, the physical system is observed through point sensors of the velocity at the inlet and at the device locations, sensors of the actuator states, and the metered power and energy of each device,
	\begin{equation}
		\label{observation_definition}
		\begin{aligned}
			\bs o_{\bb,k}
			&=
			\bigl[\,\bs o_{\bb,k}^{u},\ \bs o_{\bb,k}^{\xi},\ \bs o_{\bb,k}^{P},\ \bs o_{\bb,k}^{E}\,\bigr],
			\\
			\bs o_{\bb,k}^{u}
			&=
			\bigl[u(0,t_k),u(x_1,t_k),\ldots,u(x_{N_a},t_k)\bigr]+\bs\varepsilon_{u,k},
			\\
			\bs o_{\bb,k}^{\xi}
			&=
			\bs\xi_k+\bs\varepsilon_{\xi,k},
			\qquad
			\bs o_{\bb,k}^{P}
			=
			\bs P_k+\bs\varepsilon_{P,k},
			\\
			\bs o_{\bb,k}^{E}
			&=
			\sum_{k'=0}^{k-1}\bs o_{\bb,k'}^{P},
		\end{aligned}
	\end{equation}
	with the powers $\bs P_k=[P_{1,k},\ldots,P_{N_a,k}]$ defined below in \eqref{eq:instantaneous_power}, so that $\bs o_{\bb,k}^{u}\in\mathbb R^{N_a+1}$, $\bs o_{\bb,k}^{\xi},\bs o_{\bb,k}^{P},\bs o_{\bb,k}^{E}\in\mathbb R^{N_a}$ and $n_o=4N_a+1$ (Section~\ref{sec2a}). The energy channel is formed from the earlier power measurements rather than by the operator $h$. Before clipping, the sensor errors are Gaussian and independent across sensors and time, $\bs\varepsilon_{u,k}\sim\mathcal N(\bs 0,\sigma_u^2\bs I)$, $\bs\varepsilon_{\xi,k}\sim\mathcal N(\bs 0,\sigma_\xi^2\bs I)$ and $\bs\varepsilon_{P,k}\sim\mathcal N(\bs 0,\sigma_P^2\bs I)$ (Table~\ref{tab:burgers_settings}). Noisy velocity measurements are clipped to $[0,\infty)$ and actuator measurements to $[0,\xi_{\max}]$. The physical system is driven by the inflow, $z_k=u_{\mathrm{in}}(t_k)$, a scalar ($n_z=1$), and the digital twin and the policy receive its measurement $\widehat z_k=[\bs o_{\bb,k}^{u}]_1$ (Section~\ref{sec2a}).

	Counterfactual rollouts use the same measurement construction. Synthetic velocity, actuator-state, and power errors are drawn from the distributions above and added to the corresponding noise-free twin outputs; the velocity and actuator-state measurements are then clipped to the same bounds as their physical counterparts. In particular,
	\begin{equation}
		\bs o_{\wb,k}^{P}
		=
		\bar{\bs o}_{\wb,k}^{P}+\bs\varepsilon_{P,k}^{+},
		\qquad
		\bs o_{\wb,k}^{E}
		=
		\sum_{k'=0}^{k-1}\bs o_{\wb,k'}^{P}.
		\label{eq:twin_noisy_power_energy}
	\end{equation}

	Thus the error in the energy channel is the accumulated power-meter error and is temporally correlated; no independent noise is added to energy. The measured inlet record $\widehat z_k^{+}$ already contains the inlet-sensor error and is used directly as both the twin input and the corresponding policy observation, without perturbing it a second time. The complete synthetic error sequences are fixed in $\mathcal C^{+}$, so every candidate policy sees the same measurement realisations.

	The observer of Section~\ref{sec5p0} assimilates the velocities measured at the three devices and the three actuator states. In the physical system and in the digital twin alike, the velocity at a device is obtained from the discretised field $\bs u_k\in\mathbb R^{n_x}$ by linear interpolation. The discrepancy $\ell_{\mathrm{dyn}}$ of \eqref{eq:rt_replay_dyn_loss} uses only these velocities ($n_{\mathrm{id}}=N_a$), with $\bs\Sigma_{\mathrm{id}}=\sigma_u^2\bs I$. The velocity field between the sensors is never measured. The simulated field and the interaction states $\bs\zeta$ of the physical system serve only as diagnostics, to quantify what the observer cannot see, and never enter the observer, the identification, the surrogate or the policy search.

	The power extracted by device $j$, that is, the rate at which its extraction term in \eqref{env} removes the kinetic energy $\int_0^L u^2/2\,\mathrm dx$ from the flow, is
	\begin{equation}
		\label{eq:instantaneous_power}
		P_{j,k}
		=
		\widetilde\xi_{j,k}
		\int_0^L
		n_j(x;\sigma)\,u_+(x,t_k)^2\,\mathrm dx ,
	\end{equation}
	with the effective intensity $\widetilde\xi_{j,k}=\widetilde\xi_j(t_k)$ of \eqref{eq:effective_actuation}; in the digital twin, $\widetilde\xi_j=\xi_j$ and the footprint has the twin's width $\sigma$.
	Power depends on the field over each device footprint and cannot be reconstructed from the pointwise velocity measurements alone. We therefore assume it is directly measured and include it in $\bs o_{\bb,k}$, consistent with the reward definition in Section~\ref{sec2b}. During twin identification, however, the power measurements are not used as additional observations of the system dynamics; these enter only through the return term \eqref{eq:rt_replay_rew_loss}. The trajectory term is thus kept to the measured velocities, while the powers, which depend directly on the effective intensity \eqref{eq:effective_actuation} that the twin cannot represent, enter in the scalar form that matters for control.

	\paragraph{Reward and return} The reward favours a large total power and penalises unbalanced extraction among the devices, both instantaneous and cumulative. The balance penalty is a simple proxy for an even sharing of the loads among the devices, not a fatigue model. The instantaneous reward reads
	\begin{equation}
		\label{erre}
		\begin{split}
		r(\bs o_k)
		={}&
		\tfrac{1}{N+1}
		\biggl\{
		\sum_{j=1}^{N_a} P_{j,k}
		-
		w_{\mathrm{std}}
		\Bigl[
		\lambda\operatorname{std}_j P_{j,k}
		\\
		&+
		(1-\lambda)(\operatorname{std}_j E_{j,k+1}-\operatorname{std}_j E_{j,k})
		\Bigr]
		\biggr\},
		\end{split}
	\end{equation}
	where $\operatorname{std}_j$ is the population standard deviation across the $N_a$ devices and $E_{j,k}=\sum_{k'=0}^{k-1}P_{j,k'}$ is the cumulative extracted energy scaled by $1/\Delta t$, so that $E_{j,k+1}=E_{j,k}+P_{j,k}$. In the physical return, $P_{j,k}$ and $E_{j,k}$ in \eqref{erre} are the metered components of $\bs o_{\bb,k}^{P}$ and $\bs o_{\bb,k}^{E}$; in the virtual return, they are the corresponding synthetic measurements $\bs o_{\wb,k}^{P}$ and $\bs o_{\wb,k}^{E}$ of \eqref{eq:twin_noisy_power_energy}. The weight $w_{\mathrm{std}}\ge0$ sets the cost of the imbalance, and $\lambda\in[0,1]$ splits it between the instantaneous powers and the cumulative energies: the last term is the growth, over one step, of the imbalance of the energies extracted so far. %

	With $\gamma=1$ and $r_N=r$ in \eqref{R_o_b}, the return sums the reward over the $N+1$ observation instants $k=0,\ldots,N$ of the free interval (Section~\ref{sec5p_solver}); the final reward needs no action, since the reward depends on the observations only. The successive differences in the last term of \eqref{erre} then cancel, leaving only the imbalance of the energies extracted over the whole episode, and the return of an episode is
	\begin{equation}
		\label{eq:ground_truth_dynamic_objective}
		\begin{split}
		\mathcal R_{\bb}^{(n)}
		&=
		\sum_{k=0}^{N} r(\bs o_{\bb,k})
		\\
		&=
		\sum_{j=1}^{N_a}\overline{P}_{j}
		-
		w_{\mathrm{std}}
		\left[
		(1-\lambda)\operatorname{std}_{j}\!\left(\overline{P}_{j}\right)
		+
		\lambda\,
		\overline{\operatorname{std}_{j}\!\left(P_{j,k}\right)}
		\right],
		\end{split}
	\end{equation}
	where overbars denote averages over these instants. The factor $1/(N+1)$ in \eqref{erre} makes the return a time average over the free interval. The return thus rewards the mean total power and penalises the imbalance of the mean powers and of the instantaneous powers, with weights $1-\lambda$ and $\lambda$. A physical episode gives one sample of \eqref{eq:ground_truth_dynamic_objective}, for its inflow and measurement-noise realisations. The virtual return \eqref{eq:operational_twin_return} applies the same reward to the synthetic noisy observations and averages it over the $N_c$ fixed scenarios of the scoring bank $\mathcal C^{+}$ (Section~\ref{sec5p_solver}). It therefore represents a sample-average approximation of the expected measured physical return, subject to the dynamical and observational discrepancies of the twin.

	\subsection{Policy parameterisation}
	\label{sec4p3}

	All performance policies used in this work belong to the same family, with parameters $\bs\theta_\pi$ in a fixed box. The policy neither contains twin parameters nor uses quantities from the identified model. A given $\bs\theta_\pi$ therefore denotes the same deployed controller before and after a twin update, as required by the surrogate of the physical return (Section~\ref{sec_A}). Because the policy reads only the inlet measurement, through its memory $\bs\chi_k$, its commands are identical on the digital twin and on the physical system for a given inlet record.

	The policy information state $\bs\chi_k=(u_{\mathrm f,k},\phi_k)$ (Section~\ref{sec2a}) stores a low-pass filtered inlet velocity and the phase of an oscillator ($n_\chi=2$). The filter update \eqref{in_FIR} and the phase update in \eqref{action_1} together define the operator $G_\pi$ of that section. Through the phase update, $G_\pi$ depends on the frequency $f_0$ of the performance policy and, for the excitation policy \eqref{eq:information_policy}, on its block frequencies and hence on time. The filter
	\begin{equation}
		\label{in_FIR}
		u_{\mathrm f,k+1}=u_{\mathrm f,k}+\frac{\Delta t}{\tau_{\mathrm f}}\left(\widehat z_k-u_{\mathrm f,k}\right),
	\end{equation}
	is initialised with the first inlet sample of the washout (Section~\ref{sec5p_solver}). It smooths inflow fluctuations on time scales shorter than $\tau_{\mathrm f}$, so the policy responds only to the slower variations of the inflow. The mean allocation defined below reads the filtered inflow through the bounded input
	\begin{equation}
		\label{eq:policy_context}
		\eta_k
		=
		\tanh\!\left(\frac{u_{\mathrm f,k}-u_{\mathrm{ref}}}{s_{\mathrm{in}}}\right),
	\end{equation}
	with a reference inflow $u_{\mathrm{ref}}$ and a scale $s_{\mathrm{in}}$ (Table~\ref{tab:burgers_full_settings}).

	The commands $\bs a_k^{\mathrm c}=[a_{1,k},\ldots,a_{N_a,k}]\in\mathbb R^{N_a}$ are the sum of a mean allocation and a harmonic part,
	\begin{equation}
		\label{eq:unified_policy}
		\bs a_k^{\mathrm c}
		=
		\mathrm{sat}_{[0,\xi_{\max}]}
		\!\left[
		\pi_{\mathrm{qs}}(\eta_k;\bs\theta_{\mathrm{qs}})
		+
		\bs d_k
		\right],
	\end{equation}
	where $\mathrm{sat}_{[0,\xi_{\max}]}$ clips each component. The mean allocation,
	\begin{equation}
		\label{eq:learned_quasi_policy}
		\begin{aligned}
		\left[\pi_{\mathrm{qs}}(\eta_k;\bs\theta_{\mathrm{qs}})\right]_j
		&=
		\xi_{\max}\,
		\operatorname{sigm}\!\left(\beta_{j,0}+\beta_{j,1}\eta_k\right),
		\\
		\bs\theta_{\mathrm{qs}}
		&=
		\{\beta_{j,0},\beta_{j,1}\}_{j=1}^{N_a},
		\end{aligned}
	\end{equation}
	with $\operatorname{sigm}(q)=(1+e^{-q})^{-1}$, distributes the extraction among the devices according to the filtered inflow.

	The upstream device carries the harmonic that can trigger the dynamic increment of \eqref{eq:effective_coefficients},
	\begin{equation}
		\label{action_1}
		\begin{aligned}
			d_{1,k}
			&=
			d_{1,\max}\tanh\!\left(A_{1,k}\sin\phi_k/d_{1,\max}\right),
			\\
			A_{1,k}
			&=
			\mathrm{sat}_{[0,A_{\mathrm{ref}}]}
			\!\left[
			A_0+k_A\left|u_{\mathrm f,k}-u_{\mathrm{ref}}\right|
			\right],
			\\
			\phi_{k+1}
			&=
			\phi_k+2\pi f_{1,k}\Delta t,
			\qquad
			f_{1,k}
			=
			f_0 .
			\end{aligned}
	\end{equation}

	The hyperbolic tangent smoothly limits the harmonic to $\pm d_{1,\max}$, while $A_{\mathrm{ref}}$ caps the amplitude at the value at which the recovery gain \eqref{eq:dynamic_induction_gains} saturates. Through $k_A$, the amplitude grows with $|u_{\mathrm f,k}-u_{\mathrm{ref}}|$, allowing the upstream device to strengthen the recovery when the inflow departs from its reference value; $A_0$, $k_A$ and $f_0$ are policy parameters. Because $A_{1,k}$, $f_{1,k}$ and $\phi_k$ enter the recovery law \eqref{eq:effective_coefficients} directly, the complete action of Section~\ref{sec2a} is $\bs a_k=[\bs a_k^{\mathrm c},A_{1,k},f_{1,k},\phi_k]\in\mathbb R^{n_a}$, with $n_a=N_a+3$. This vector is recorded and imposed in a factual replay. The recovery law responds directly to these commanded quantities, not to the actuator-state oscillation, whose amplitude is reduced by the limiter and the actuator lag \eqref{eq:actuator_dynamics}. At $A_1=A_{\mathrm{ref}}$ and $f_0\approx f^\star$, the oscillation of $\xi_1$ is about two thirds of $A_{\mathrm{ref}}$, so a law driven by the actuator state would receive only about a fifth of the gain. When the mean allocation approaches $\xi_{\max}$, the saturation in \eqref{eq:unified_policy} also clips the command without changing the recovery. In \eqref{eq:effective_coefficients}, $A_1$ and $f_1$ are held over each control step, whereas $\phi$ is interpolated linearly between control instants. Before the branch, $A_1=0$; at the branch, $\phi_0=0$ for the deployed policy and every candidate alike.

	The downstream devices add harmonics phase-locked to the upstream one,
	\begin{equation}
		\label{eq:downstream_dynamic_policy}
		d_{j,k}
		=
		A_j\sin(\phi_k-\psi_j),
		\qquad j=2,\ldots,N_a,
	\end{equation}
	with amplitudes $A_j$ and phase lags $\psi_j$; the lags set the timing of the downstream extraction relative to the upstream harmonic, for instance to match the advection delay between the devices. These amplitudes are constant, which keeps the parameterisation low-dimensional; only the upstream amplitude, which drives the recovery, adapts to the inflow. Beyond $\bs\chi_k$, the policy is memoryless, its commands are bounded by construction, and the extraction intensities stay non-negative. The controller is a periodic forcing whose mean and upstream amplitude are scheduled by the measured inflow, and, since it reads only the inlet, model error can change only the effect of its commands, not the commands themselves.

	The policy parameters of Section~\ref{sec2a} are the thirteen coordinates
	\begin{equation}
		\label{eq:unified_theta_pi}
		\bs\theta_{\pi}
		=
		\bigl[
		\{\beta_{j,0},\beta_{j,1}\}_{j=1}^{N_a},\,
		A_0,\,
		k_A,\,
		f_0,\,
		\{A_j,\psi_j\}_{j=2}^{N_a}
		\bigr]^\top ,
	\end{equation}
	six for the mean allocation and seven for the harmonic part, searched jointly in the box given in Table~\ref{tab:burgers_full_settings}. With $A_0=k_A=A_j=0$, the harmonic part vanishes and the command reduces to the mean allocation.

	Within a performance episode the frequency of the upstream harmonic is constant, $f_{1,k}=f_0$, so that $(\Delta\alpha,f^\star,s_f,f_\alpha)$ enter the recovery law \eqref{eq:effective_coefficients} only through the product of $\Delta\alpha$ and the frequency factors of the gain \eqref{eq:dynamic_induction_gains} at $f_0$. These four parameters can therefore be separated only from data at several frequencies at which the gain is not negligible, recorded in different episodes or in one episode whose frequency changes.

	Information episodes instead use an open-loop command whose frequency changes within the episode. It is built from the harmonic part of the performance policy without its limiter. The excitation policy $\pi_{\mathrm{info}}$ of Section~\ref{sec2a} is
	\begin{equation}
		\label{eq:information_policy}
		a_{j,k}
		=
		7+\sin\!\left(\phi_k-\psi_j\right),
		\qquad
		j=1,\ldots,N_a,
		\qquad
		\psi_1=0.
	\end{equation}

	The phase $\phi_k$ follows the update in \eqref{action_1}, with the frequency held constant over each of four blocks of five time units. %
	The mean command is $7$ rather than the $7.98$ of the baseline controller of Section~\ref{sec5_baseline}, because a unit-amplitude oscillation about $7.98$ would be clipped at $\xi_{\max}$ (Table~\ref{tab:burgers_settings}) for half of each cycle. %
	The unit amplitude equals $A_{\mathrm{ref}}$, at which the amplitude factor of the gain \eqref{eq:dynamic_induction_gains} is one. The recovery law \eqref{eq:effective_coefficients} therefore receives $A_1=1$, the frequency of the current block as $f_1$, and $\phi_k$. %
	The parameters $\bs\theta_\iota$ of an excitation policy are its four block frequencies and the phase lags $\psi_2$ and $\psi_3$. The library $\Theta_\iota$ of Section~\ref{sec_D} contains 16 excitation policies, formed by combining one of four block-frequency sequences, $(0.06,0.30,0.40,0.55)$, $(0.08,0.24,0.40,0.60)$, $(0.20,0.32,0.45,0.58)$ and $(0.70,0.58,0.33,0.40)$, %
	with one of four phase-lag pairs $(\psi_2,\psi_3)$, $(0,0)$, $(\pi/2,\pi)$, $(\pi/2,3\pi/2)$ and $(\pi,\pi/2)$. %

	\subsection{Numerical set-up and episode protocol}
	\label{sec5p_solver}

	The inflow $u_{\mathrm{in}}(t)$ is a realisation of a Gaussian process whose mean is drawn uniformly from $[6.0,6.1]$. Its covariance is the sum of two squared-exponential kernels, with standard deviations $(0.30,0.15)$ and correlation times $(2.0,0.5)$, and an uncorrelated component of standard deviation $0.01$. This defines the process $\mathcal P_z$ of Section~\ref{sec2a}, measured by the inlet sensor. The inflow is sampled on the control grid, clipped to non-negative values, and linearly interpolated between control instants. The digital twin treats its measured input $\widehat z_k$ in the same way: the step from $t_k$ to $t_{k+1}$ uses $\widehat z_k$ and $\widehat z_{k+1}$, with the final sample $\widehat z_N$ stored in the observation $\bs o_{\bb,N}$ of \eqref{eq:replay_record}. Each physical episode draws a new inflow and measurement-noise realisation.

	Fixed seeds define the $N_c=8$ input and synthetic-noise scenarios of the scoring bank (Section~\ref{sec2b}). A separate fixed set of eight realisations is used only for reporting and is hereafter called the reporting inflows. Unless stated otherwise, the return reported for a policy in Section~\ref{sec5} is the mean measured return over these eight inflows and their measurement-noise realisations. These post-processing evaluations are not supplied to the algorithm and are therefore excluded from its interaction budget; they enter neither $\mathcal D_\pi$ nor $\mathcal B$.

	The physical system and the digital twin are advanced with the same semi-implicit finite-difference scheme, on $n_x$ uniformly spaced nodes from $x=0$ to $x=L$ and with time step $\Delta t_{\mathrm{num}}$ (Table~\ref{tab:burgers_settings}). Convection, discretised by central differences, and the recovery and extraction terms are treated explicitly. Diffusion is treated implicitly through a tridiagonal solve, with the zero-gradient outlet condition imposed through a ghost node. The field is clipped to non-negative values after each step. Within each step, the actuator states \eqref{eq:actuator_dynamics} are advanced first by explicit Euler and clipped to $[0,\xi_{\max}]$. In the physical system, the interaction states \eqref{eq:plant_interaction} are also advanced first by explicit Euler. The integral in \eqref{eq:instantaneous_power} is evaluated by the rectangle rule on the nodes, and the record stores the measured inlet at the control rate. Since the physical system and the twin share this discretisation, it adds no discrepancy between them; the identified width $\sigma$, resolved by two to three grid spacings, and the reported returns refer to this grid.

	Each episode follows the time axis of Figure~\ref{fig:sync_free_forecast} and is preceded by a washout. It therefore has three phases, whose durations are given in Table~\ref{tab:burgers_settings}. (i) During the \emph{washout}, of duration $T_{\mathrm{wash}}$, the physical system starts from the uniform field $u=1$ and zero actuator and interaction states, and evolves until its flow no longer depends on this initial field. (ii) During \emph{synchronisation}, of duration $T_{\mathrm{sync}}$ ($L_{\mathrm a}=T_{\mathrm{sync}}/\Delta t=40$ control steps; see \eqref{eq:replay_record}), the physical system continues to evolve while the twin is initialised as described in Section~\ref{sec5p0}. The observer assimilates the measurements, allowing the influence of the twin's initial field to decay. (iii) The \emph{free interval} lasts $T_{\mathrm{eval}}$ ($N=T_{\mathrm{eval}}/\Delta t=200$ control steps); the return and the training and validation losses \eqref{eq:rt_replay_loss} are computed over this interval. The branch at $t=0$ separates synchronisation from the free interval. Throughout the first two phases, every device is held at the constant command $a_{j,k}=7$, so the state at the branch is independent of the policy. The policy under test acts from the branch onwards.

	Both washout and synchronisation exploit the advection-dominated flow and the measured inflow. The P\'eclet number based on the mean inflow, $\overline{u}_{\mathrm{in}}L/\nu_0$, is about $1200$. Under the constant command, the flow crosses the domain in about three time units, less than the duration of either phase, so the initial field is carried out through the outlet. A system with a longer memory would require a longer window or more sensors.

	For each scenario of $\mathcal Z^{+}$, the physical system is run once under the constant command, from the start of the washout to the end of the free interval. Using the initial twin parameters $\bs\theta_p^{(0)}$, the observer reconstructs the state at the branch. This analysed state enters $\mathcal C^{+}$ (Section~\ref{sec3_init}) together with the policy memory at the branch (the filtered inlet velocity and $\phi_0=0$), the measured inlet record over the free interval, and a fixed synthetic measurement-noise trajectory generated as described in Section~\ref{sec4p2}. These eight preparatory runs are outside the search budget; drawing the synthetic noise requires no additional physical interaction.

	\subsection{Observer definition}
	\label{sec5p0}

	The state observer $\Psi_{\wb}$ of \eqref{eq:rt_sync_intro} nudges the digital twin towards the measurements. Nudging is a simple form of sequential data assimilation that has also been used for turbulent flows \cite{ClarkDiLeoni2020,Zauner2022}. Each velocity innovation is spread over the field by a Gaussian kernel centred at the corresponding device,
	\begin{equation}
		\label{eq:burgers_nudging_gain}
		(\bs K_u)_{ij}
		=
		\kappa_u
		\exp\!\left[
		-\frac{\left(X_i-x_j\right)^2}{2\ell_K^2}
		\right],
	\end{equation}
	where $X_i$, $i=1,\ldots,n_x$, are the grid nodes and $x_j$ are the device locations. The gain $\kappa_u$ sets the magnitude of the correction, while the length scale $\ell_K$ determines how far a point measurement informs the surrounding field. The field update is $\widehat{\bs u}_k^{\,\mathrm a}=\bs u_k^{\mathrm f}+\bs K_u\bs\nu_k^{u}$, where $\bs K_u\in\mathbb R^{n_x\times N_a}$ is constant in time and $\bs\nu_k^{u}\in\mathbb R^{N_a}$ contains the velocity innovations at the devices. The metered powers and energies are not assimilated. The actuator states are corrected separately by
	\begin{equation}
		\label{eq:burgers_actuator_nudging}
		\widehat{\bs\xi}_k^{\,\mathrm a}
		=
		\bs\xi_k^{\mathrm f}
		+
		\kappa_{\xi}
		\left(
		\bs o_{\bb,k}^{\xi}-\bs\xi_k^{\mathrm f}
		\right).
	\end{equation}

	After each analysis, the operator $\Pi$ in \eqref{eq:twin_recursion} re-imposes the measured inlet value, restores the zero-gradient outlet condition, and clips the field to non-negative values and the actuator states to $[0,\xi_{\max}]$. The gains and length scale are $(\kappa_u,\ell_K,\kappa_\xi)=(0.25,0.20,1)$. With $\kappa_\xi=1$, each measured actuator state replaces its forecast.

	The observer is initialised at $t=-T_{\mathrm{sync}}$ with a uniform field equal to the measured inlet velocity and with zero actuator states. Its forecast model is the digital twin with $\alpha_{\mathrm{eff}}=\alpha_0$. This reduction is exact for the twin during the constant-command synchronisation window (Section~\ref{sec5p_solver}), because the dynamic increment in \eqref{eq:effective_coefficients} then vanishes.

	Because the filtered inlet velocity depends only on the measured inflow and $\phi_0=0$ for every policy (Section~\ref{sec4p3}), all policies have the same memory at the branch for a given inflow. The analysed state itself (Section~\ref{sec_sync}) contains only the field and the actuator states.

	\subsection{Identification and policy-search settings}
	\label{sec5p0b}

	The identifications from the baseline data and in the information phase (Sections~\ref{sec5_baseline} and~\ref{sec5_excitation}) fit all the episodes recorded so far, the first $80\%$ for training and the last $20\%$ for validation, with $\eta_R=0$ in \eqref{eq:rt_reward_balance}, that is, with the trajectory term \eqref{J_o} of the identification loss alone. The analysed states are first computed with the twin at the centre of the box \eqref{eq:theta_p_box}, and all recorded episodes are then kept synchronised with the current twin ($N_{\mathrm s}=n$). The hybrid optimiser starts every time from the same archive of 64 normalised parameter vectors, a scrambled Sobol sequence that includes the centre of the box. Six rounds of one expected-improvement proposal follow, each refined by 20 AMSGrad steps with learning rate $0.025$. %

	At a scheduled twin update of the policy search (Section~\ref{sec5_rt}), 60 recorded episodes are drawn at random from the buffer and split into 40 for training and 20 for validation. The hybrid optimiser (Section~\ref{sec_C}) starts from an archive of eight normalised parameter vectors (Section~\ref{sec_D}): the current parameters, four Gaussian perturbations of these with standard deviation $0.08$, and three uniform draws in the box \eqref{eq:theta_p_box}. Up to four rounds of two expected-improvement proposals follow, stopped early if the best loss improves by less than 1\% over two rounds, and each candidate is refined by 50 AMSGrad steps with learning rate $0.02$. The return share of the identification loss is $\eta_R=1/2$ in \eqref{eq:rt_reward_balance}. All recorded episodes are kept synchronised with the current twin parameters ($N_{\mathrm s}=n$ in \eqref{eq:branch_ensemble}). The floors $s_{\min}=1$ and $j_{\min}=10^{-12}$ of the identification loss, in the scale $s_R$ of Section~\ref{sec_C} and in \eqref{eq:rt_reward_balance}, are numerical safeguards that were never activated in the numerical experiments presented in this work.

	At each information-policy update, the excitation-policy synthesis problem \eqref{eq:rt_nested_information_design}, with the E-optimal criterion \eqref{eq:rt_e_optimal_policy}, is solved by enumeration: the criterion is evaluated for each of the 16 excitation policies of the library $\Theta_\iota$ (Section~\ref{sec4p3}), and the best one is deployed. The twin is rolled out under each candidate, from the analysed states \eqref{eq:branch_ensemble} and under the measured inflows of the $N_d=2$ most recent recorded episodes; the sensitivities of the velocities that it predicts at the devices give the Fisher information \eqref{eq:rt_info_ensemble_fisher}. Before the first twin update of the information phase (Section~\ref{sec5_excitation}), this information is averaged over $N_\theta=4$ parameter vectors of $\Theta_p$ \eqref{eq:theta_p_box}, the centre of the box and three points of a scrambled Sobol sequence; afterwards it is evaluated at the current twin parameters. %

	The initial model-based reference \eqref{eq:model_based_policy_synthesis_operator} is synthesised on the initial twin before the first episode of the search; it uses no physical data beyond the episodes that identified this twin and the preparatory runs of the scoring bank (Section~\ref{sec5p_solver}). Its search starts from a scrambled Sobol archive of 64 policies, followed by 40 rounds of one expected-improvement proposal, each refined by up to 20 AMSGrad steps with learning rate $0.03$. Every rollout of this search starts from the state that the digital twin itself reaches at the branch under the constant command, not from an analysed state. The gradient steps use two further inflow records, with synthetic measurement noise drawn with fixed seeds, while the archive is scored, and the reference selected, on the complete input-and-noise scenarios of $\mathcal C^{+}$. In operation, these records could be taken from the episodes under the baseline controller (Section~\ref{sec5_baseline}). In each run, the relevance scales of Section~\ref{sec3_synth} are fitted to the virtual returns of a separate set of 512 Sobol policies. After a twin update, a new set of 512 Sobol policies is scored with the new twin and the relevance scales are refitted. The reference is then re-synthesised more simply than in Section~\ref{sec3_synth}, without expected improvement or gradients. The best among the current reference and this set is refined by 256 policies drawn uniformly around it, in a box of half-width $0.05$ in the coordinates of \eqref{eq:policy_whitening}. The model-based strategy of the results (Section~\ref{sec5_modelbased}) deploys the initial reference without retuning it on physical returns.

	On the physical system, the expected improvement \eqref{eq:RT_acquisition}, with exploration margin $\xi_{\mathrm{EI}}=0.01$ in units of the return, is maximised over $N_{\mathcal Q}=128$ candidates drawn uniformly in the proposal region \eqref{eq:proposal_region}, with $\kappa=0.5$, $\rho_{\max}=0.5$ and $\tau=0.1$. The $n_{\mathrm{init}}$ initial policies of the policy search (Section~\ref{sec3_init}) are drawn uniformly in the same region around the initial reference, which is one of these. The surrogate of Section~\ref{sec_A} is refitted before every selection, on returns standardised by the mean and the standard deviation of the archive. Each fit makes two alternations between \eqref{eq:surrogate_likelihood} and \eqref{eq:gls_calibration}, with the likelihood maximised from four starting points within $\ell_\pi\in[0.3,5]$, $\sigma_f^2\in[10^{-2},10^{2}]$ and $\sigma_R^2\in[10^{-4},10^{2}]$, in standardised units.

	The digital twin and the policy are implemented in JAX, which provides the reverse-mode gradients of Sections~\ref{sec3_synth} and~\ref{sec_C}; the inlet filter \eqref{in_FIR}, which depends only on the measured inflow, is computed beforehand.

	\section{Results}\label{sec5}
	\setcounter{dbltopnumber}{2}\setcounter{topnumber}{3}\setcounter{totalnumber}{5}%
	\renewcommand{\dbltopfraction}{0.9}\renewcommand{\topfraction}{0.9}%
	\renewcommand{\textfraction}{0.1}\renewcommand{\floatpagefraction}{0.8}%
	\renewcommand{\dblfloatpagefraction}{0.8}%

	\begin{table*}[t]
		\centering
		\caption{Performance on the physical system at each stage of Section~\ref{sec5}, as means over the eight reporting inflows (Section~\ref{sec5p_solver}): mean power of each device $\overline P_j$, their sum, the standard deviation $\operatorname{std}_j$ across the devices of the mean powers of each episode, averaged in the same way, and the return $\mathcal R_{\bb}$ of \eqref{eq:ground_truth_dynamic_objective}. Physical episodes: episodes that the stage adds on the physical system; the model-based reference and the strategies guided by the twin also rely on the 52 episodes of the information phase. The 50 episodes of Section~\ref{sec5_baseline} are not counted, since a system under its baseline controller runs them anyway, and neither are the eight preparatory runs of the scoring bank and the evaluations on the reporting inflows (Section~\ref{sec5p_solver}). Runs: independent runs of the search, over which the rows of the searches are also averaged; their return is given with one standard error over the runs. Excitation policy: the excitation policy selected in Section~\ref{sec5_excitation}, with the devices in phase. Searches: the policy of $\mathcal D_\pi$ with the largest posterior mean after 500 episodes, for the strategies of Table~\ref{tab:arms} (Section~\ref{sec5_rt}). The unguided searches draw their candidates in the whole box $\Theta_\pi$; the strategies guided by the twin draw them in the fixed region ($g_j=1$) or, in the last three rows, in the adaptive region \eqref{eq:proposal_region}.} %
		\label{tab:scoreboard}
		\footnotesize
		\begin{adjustbox}{max width=\linewidth}
\begin{tabular}{@{}l@{\hspace{5.5pt}}r@{\hspace{5.5pt}}r@{\hspace{5.5pt}}r@{\hspace{5.5pt}}r@{\hspace{5.5pt}}r@{\hspace{5.5pt}}r@{\hspace{5.5pt}}r@{\hspace{5.5pt}}r@{}}
\toprule
 & & physical & & & & & & \\
Policy & runs & episodes & $\overline P_1$ & $\overline P_2$ & $\overline P_3$ & $\sum_j \overline P_j$ & $\operatorname{std}_j$ & $\mathcal R_{\bullet}$ \\
\midrule
constant command & -- & 0 & 70.4 & 12.7 & 11.1 & 94.3 & 27.6 & 39.14 \\
excitation policy & -- & 52 & 70.1 & 26.2 & 16.7 & 113.1 & 23.3 & 66.48 \\
model-based reference & -- & 0 & 76.7 & 31.8 & 29.0 & 137.5 & 21.9 & 93.71 \\
\midrule
unguided search & 50 & 500 & 59.1 & 40.9 & 26.2 & 126.2 & 14.0 & $97.94\pm0.27$ \\
unguided search, 2048 candidates & 50 & 500 & 53.3 & 38.0 & 29.8 & 121.0 & 10.5 & $99.60\pm0.23$ \\
constant-trend variant & 50 & 500 & 70.9 & 38.9 & 25.0 & 134.7 & 19.4 & $95.75\pm0.15$ \\
fixed-twin variant & 50 & 500 & 71.8 & 37.3 & 26.0 & 135.0 & 19.6 & $95.81\pm0.09$ \\
updated-twin variant & 50 & 500 & 70.1 & 36.7 & 26.9 & 133.7 & 18.6 & $96.40\pm0.15$ \\
\midrule
constant-trend variant, adaptive region & 50 & 500 & 53.0 & 38.8 & 29.3 & 121.2 & 10.3 & $100.08\pm0.28$ \\
fixed-twin variant, adaptive region & 50 & 500 & 49.3 & 36.4 & 31.9 & 117.6 & 7.6 & $102.33\pm0.19$ \\
updated-twin variant, adaptive region & 50 & 500 & 51.8 & 36.5 & 31.3 & 119.6 & 8.9 & $101.56\pm0.19$ \\
\bottomrule
\end{tabular}
\end{adjustbox}
	\end{table*}

	\begin{figure*}[t]
		\centering
		\includegraphics[width=\textwidth]{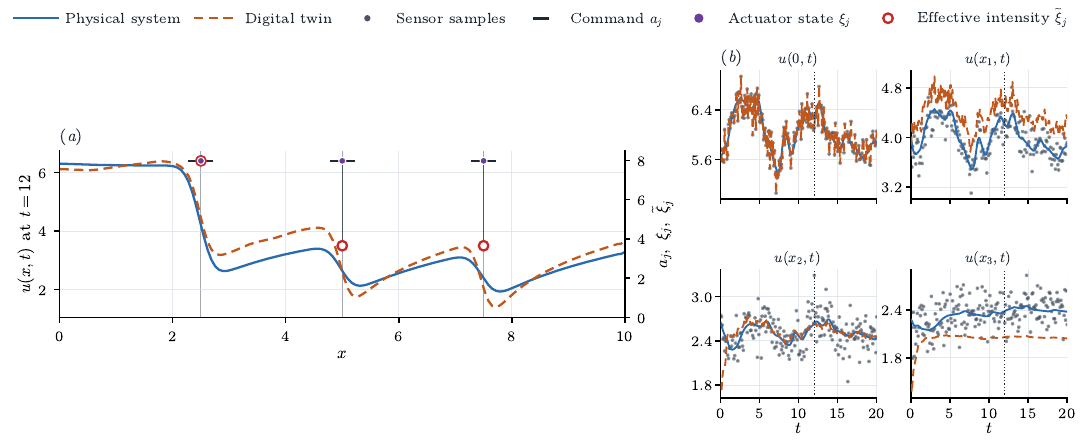}
		\caption{One of the 50 episodes run with the constant command $a_j=7.98$, used for validation: physical system (solid) and digital twin identified from the 50 episodes (dashed; first block of Table~\ref{tab:twin_parameters}) in free prediction from the branch.
			(a) Velocity at $t=12$; at each device, the right axis shows $a_j$ (black bar), actuator state $\xi_j$ of \eqref{eq:actuator_dynamics} (filled purple circle), and, for the physical system, effective intensity $\widetilde\xi_j$ in \eqref{eq:effective_actuation} (open red circle).
			(b) Velocity at the inlet and the three devices over the free interval; dots denote sensor samples \eqref{observation_definition} and the dotted line marks $t=12$. At the inlet, the twin follows the measured velocity as boundary condition (Section~\ref{sec4p2}). At $t=0$, the twin is initialised from the analysed state (Sections~\ref{sec5p_solver} and~\ref{sec5p0}). As in the validation loss of the identification (Sections~\ref{sec_C} and~\ref{sec5p0b}), this state is reconstructed by the observer with the twin parameters at the centre of the admissible box \eqref{eq:theta_p_box}, not with the identified ones.} %
		\label{fig:baseline_episode}
	\end{figure*}

This section presents the implementation of reinforcement twinning in successive stages, as if introduced on an existing system operating under a baseline controller. The information phase, model-based synthesis, and policy search could be combined in a single run of Algorithm~\ref{alg:RT_policy_search}, with schedules $e_n$ and $\bs u_n$ switching from information to performance episodes and synthesising the model-based policy at the switch. Here they are run separately so each can be analysed on its own. Algorithm~\ref{alg:RT_policy_search} is run in two stages: first with information episodes only, then with performance episodes only, starting from the twin identified in the first run.

Section~\ref{sec5_baseline} first evaluates the baseline controller, here a constant maximal command applied to all devices, both in terms of control performance and model identification.

The RT implementation begins with an information phase, consisting of two initial episodes under a fixed excitation policy followed by 50 episodes with $e_n=1$ in Algorithm~\ref{alg:RT_policy_search}, to collect informative data for twin identification. This phase is described in Section~\ref{sec5_excitation}. The identified twin is then used, before any policy search on the physical system, to analyse the sensitivity of the return to the policy parameters and to synthesise the model-based reference of Section~\ref{sec3_synth}; this step is reported in Section~\ref{sec5_modelbased}. Finally, Section~\ref{sec5_rt} considers the coupled RT phase, in which the twin guides policy search on the physical system while the resulting episodes are used in turn to refine the twin.

All performance policies belong to the family introduced in Section~\ref{sec4p3}, and all episodes are evaluated using the setting described in Section~\ref{sec5p_solver}. Tables~\ref{tab:scoreboard} and~\ref{tab:twin_parameters} collect the key results across these stages: the physical-system performance of the policy obtained at each stage, and the twin parameters identified under the baseline controller, during the information phase, and through reinforcement twinning. The policy searches compare the five strategies defined in Table~\ref{tab:arms}.

	\begin{table*}[t]
		\centering
		\caption{Twin parameters \eqref{eq:unified_theta_p} in the physical system \eqref{eq:theta_p_physical}, identified in Sections~\ref{sec5_baseline} and~\ref{sec5_excitation}, and at the end of the searches of Section~\ref{sec5_rt}. Identified: lowest validation loss. Within $5\%$: range of the archived fits whose validation loss is within $5\%$ of the lowest. End of the search: median and range over the 50 runs of the updated-twin variant with the adaptive region (Section~\ref{sec5_rt}). Grey: practically no information on the parameter, $F_{j\mid\sim j}\approx0$ (Section~\ref{sec5_baseline}).} %
		\label{tab:twin_parameters}
		\footnotesize
		\begin{adjustbox}{max width=\linewidth}
\begin{tabular}{@{}l@{\hspace{6pt}}r@{\hspace{10pt}}r@{\hspace{6pt}}c@{\hspace{10pt}}r@{\hspace{6pt}}c@{\hspace{10pt}}r@{\hspace{6pt}}c@{}}
\toprule
 & physical & \multicolumn{2}{c}{constant command} & \multicolumn{2}{c}{excitation policy} & \multicolumn{2}{c}{end of the search} \\
 & system & \multicolumn{2}{c}{50 episodes} & \multicolumn{2}{c}{52 episodes} & \multicolumn{2}{c}{50 runs} \\
\cmidrule(lr){3-4}\cmidrule(lr){5-6}\cmidrule(l){7-8}
 & & identified & within 5\,\% & identified & within 5\,\% & median & lowest--highest \\
\midrule
$\alpha_0$ & 0.5 & 0.800 & 0.800--0.800 & 0.800 & 0.800--0.800 & 0.645 & 0.507--0.751 \\
$\Delta\alpha$ & 6 & \textcolor{muted}{6.89} & \textcolor{muted}{4.80--7.05} & 10.0 & 8.44--10.0 & 5.96 & 5.02--7.61 \\
$f^\star$ & 0.4 & \textcolor{muted}{0.549} & \textcolor{muted}{0.400--0.549} & 0.399 & 0.393--0.399 & 0.399 & 0.349--0.424 \\
$s_f$ & 0.1 & \textcolor{muted}{0.154} & \textcolor{muted}{0.154--0.195} & 0.111 & 0.111--0.119 & 0.0892 & 0.0774--0.133 \\
$f_\alpha$ & 0.08 & \textcolor{muted}{0.104} & \textcolor{muted}{0.0689--0.118} & 0.157 & 0.103--0.157 & 0.197 & 0.102--0.200 \\
$\sigma$ & 0.2 & 0.162 & 0.162--0.166 & 0.161 & 0.160--0.161 & 0.136 & 0.124--0.143 \\
\bottomrule
\end{tabular}
\end{adjustbox}
	\end{table*}

	\begin{figure*}[t]
		\centering
		\includegraphics[width=\textwidth]{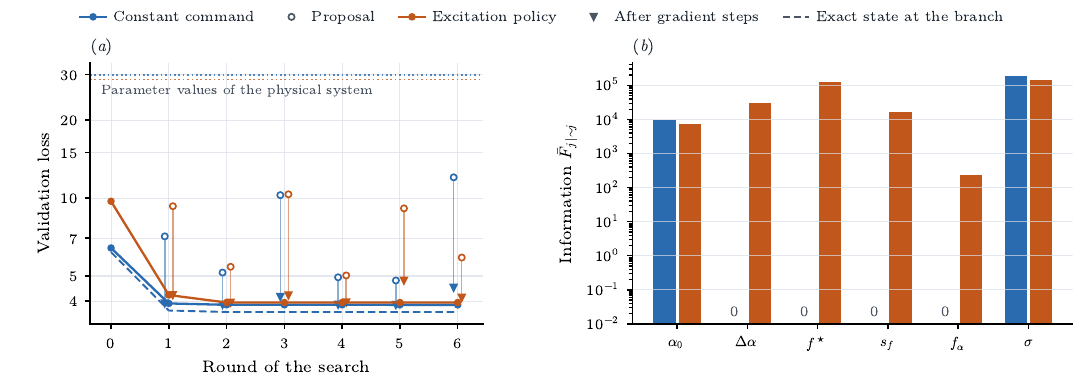}
		\caption{Identification under the constant command of Section~\ref{sec5_baseline} and under the excitation policy selected in Section~\ref{sec5_excitation} (`excitation policy'; last identification of the information phase). (a) Validation trajectory loss \eqref{J_o} against the round of the search. Lines with filled circles: best loss after each round. Open circles: the parameter vector proposed by expected improvement in that round; filled triangles: the same vector after the gradient-based refinement (Section~\ref{sec5p0b}), joined to it by a vertical segment. Dashed: fit under the constant command started from the exact state at the branch; dotted: loss at the parameter values \eqref{eq:theta_p_physical} of the physical system. (b) Information coefficients $\overline F_{j\mid\sim j}$ from two episodes in each case (the first two under the constant command, the last two of the information phase), averaged over $N_\theta=4$ parameter vectors (Section~\ref{sec5p0b}) of the admissible box; `0' marks a parameter without information. Figure~\ref{fig:information_growth} gives the information of all the recorded episodes, evaluated under the identified twin.}
		\label{fig:identification_curves}
	\end{figure*}

	\subsection{Baseline controller and data}
	\label{sec5_baseline}

	The constant baseline policy is a member of the family in Section~\ref{sec4p3} with $A_0=k_A=A_j=0$ and $\beta_{j,1}=0$, so that the command \eqref{eq:unified_policy} reduces to $a_j=\xi_{\max}\operatorname{sigm}(\beta_{j,0})$ (see \eqref{eq:learned_quasi_policy}),
	which is $7.98$ at the upper bound of $\beta_{j,0}$ (Table~\ref{tab:burgers_full_settings}; $\xi_{\max}$ in Table~\ref{tab:burgers_settings}). %

	On the eight reporting inflows of Section~\ref{sec5p_solver}, the return \eqref{eq:ground_truth_dynamic_objective} of the constant command is $\mathcal R_{\bb}=39.14$ (Table~\ref{tab:scoreboard}, first row). %
	Although the three devices receive the same command, the upstream device extracts more than five times the mean power of each of the other two. %
	Two mechanisms combine to produce this imbalance. The upstream extraction depletes the flow that reaches the other two devices,
	and the interaction \eqref{eq:plant_interaction}--\eqref{eq:effective_actuation} lowers their effective intensity to about half of the nominal one. %
	The balance penalty of \eqref{eq:ground_truth_dynamic_objective}, driven by $\operatorname{std}_j\overline P_j=27.6$ (weights $w_{\mathrm{std}}$ and $\lambda$ in Table~\ref{tab:burgers_settings}), substantially lowers the return of this controller. %
	Commanding maximal extraction at every device is therefore far from optimal. No optimised steady command is used as a baseline here, so the gains reported below over the constant command include those of a better mean allocation, not only of the dynamic increment. Figure~\ref{fig:baseline_episode} shows both mechanisms in one of the 50 episodes run under this command, for the physical system and for the twin identified below.

This operation also provides the first data for identifying the twin.

Replay identification (Section~\ref{sec_C}) uses the records \eqref{eq:replay_record} from 50 episodes on the physical system. Because no policy is searched at this stage, the twin is fitted to sensor measurements, not returns. The identification loss \eqref{eq:rt_replay_loss} omits its return term ($\eta_R=0$ in \eqref{eq:rt_reward_balance}) and therefore reduces to the trajectory term \eqref{J_o}. The first 40 episodes are used for fitting and the remaining 10 for validation.

The parameter search uses the hybrid global–local strategy described in Sections~\ref{sec3_synth} and~\ref{sec_C}, with the settings of Section~\ref{sec5p0b}. Figure~\ref{fig:identification_curves} compares this identification with that in Section~\ref{sec5_excitation}. Panel (a) plots the validation trajectory loss \eqref{J_o} during the search; panel (b) reports the information available for each parameter, discussed below. The validation loss converges to $3.87$. Since the loss is whitened by the sensor noise \eqref{eq:rt_replay_dyn_loss}, the noise-limited value would be close to one. The observed loss is therefore about four times that level.

In every round of both identifications, gradient refinement brings the loss of the parameter vector proposed by expected improvement (open circles) to within about $20\%$ of the best value. Figure~\ref{fig:identification_curves}a does not establish whether expected improvement finds better starting points than random or Sobol sampling. The first block of Table~\ref{tab:twin_parameters} lists the identified parameters alongside those of the physical system.

In the validation episode shown in Figure~\ref{fig:baseline_episode}, the twin’s predicted velocity differs from the physical system’s by approximately $10\%$ in root-mean-square error, evaluated over the domain and free interval. The largest discrepancy occurs between devices 1 and 2, with over predictions also at sensor $x_1$. %

Four of the six parameters in \eqref{eq:unified_theta_p} are unidentifiable under the data considered here. Under a steady command, the recovery rate \eqref{eq:effective_coefficients} reduces to $\alpha_0$ (Section~\ref{sec4p1}); hence, $\Delta\alpha$, $f^\star$, $s_f$, and $f_\alpha$ do not affect the measurements. For each of these parameters, the information coefficient $F_{j\mid\sim j}$ (\eqref{eq:rt_marginal_information}), evaluated for the 50 episodes using the recorded actions and identified twin, is practically zero (grey entries in Table~\ref{tab:twin_parameters}). Additional episodes under the same command therefore provide no information about these parameters.

Figure~\ref{fig:identification_curves}b reports the ensemble Fisher information coefficients $\overline F_{j\mid\sim j}$ defined in \eqref{eq:rt_info_ensemble_fisher}. The coefficients are evaluated using the first two episodes ($N_d=2$) and the $N_\theta=4$ parameter vectors specified in Section~\ref{sec5p0b}. The coefficients are essentially zero for the four unidentifiable parameters and approximately $10^{4}$ for $\alpha_0$ and $2\times10^{5}$ for $\sigma$.

	The values returned by the fit for these four unidentifiable parameters are therefore rather arbitrary: Among the fits within $5\%$ of the best validation loss, $\Delta\alpha$ ranges from $4.80$ to $7.05$ and $f^\star$ from $0.400$ to $0.549$ (Table~\ref{tab:twin_parameters}), while the velocity field of the twin changes by at most $0.07$, or $3.0\%$ of the local velocity. %

	The two informed parameters, $\alpha_0$ and $\sigma$, are biased by the model error (Section~\ref{sec4p1}). Because the twin does not represent the reduction in effective intensity at downstream devices \eqref{eq:effective_actuation}, these parameters partly compensate for that discrepancy. The fitted $\alpha_0$ reaches the upper bound of its admissible interval \eqref{eq:theta_p_box}: $0.8$, compared with the physical value of $0.5$. The fitted $\sigma$ is $0.162$, compared with $0.2$ for the physical system (Table~\ref{tab:twin_parameters}). Thus, the fit is constrained by the upper bound on $\alpha_0$; the effect of widening its admissible interval on the validation loss was not tested. This kind of compensation is a known consequence of fitting models with structural deficiencies \cite{Brynjarsdottir2014}. Accordingly, the true physical parameters in \eqref{eq:theta_p_physical} are not optimal for this twin. At those values, the validation loss is almost eight times that of the identified twin. %

Most of the loss, beyond sensor noise, arises from the twin’s structure rather than from the reconstructed state used to initialise each replay. To assess the observer’s contribution (Sections~\ref{sec_sync} and~\ref{sec5p0}), the fit was repeated with each replay initialised from the exact physical-system state at the branch point. The resulting validation loss, shown by the dashed curve in Figure~\ref{fig:identification_curves}a, is only about $6\%$ lower.

The twin is therefore limited in two respects: the constant command provides no information about four parameters, and the twin’s structure cannot represent the device interactions described by \eqref{eq:plant_interaction}–\eqref{eq:effective_actuation}.

	\subsection{Identification with informative excitation}
	\label{sec5_excitation}

	\begin{figure*}[t]
		\centering
		\includegraphics[width=\textwidth]{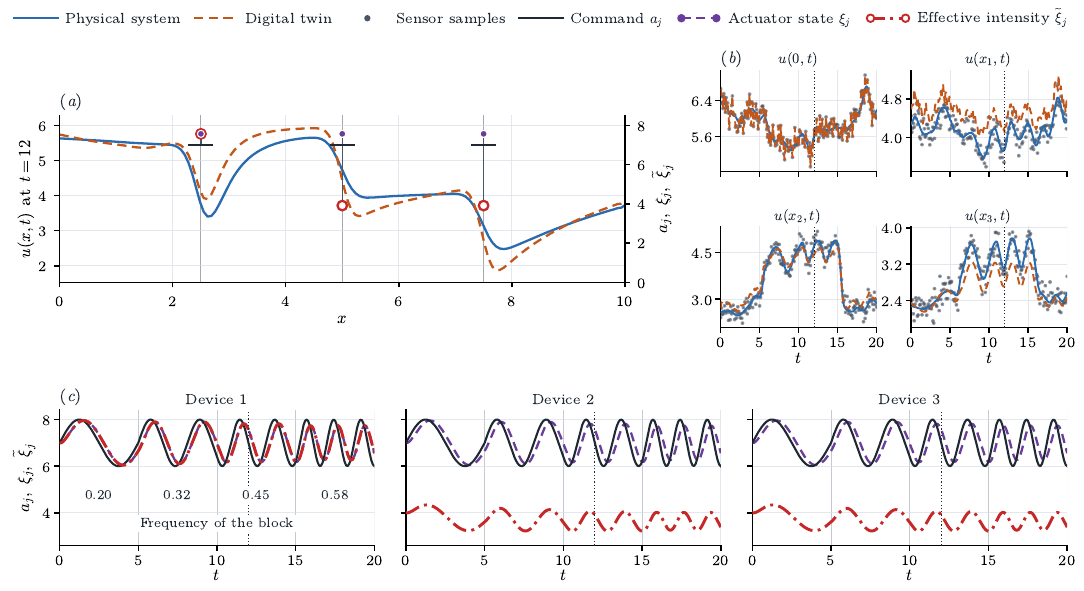}
		\caption{One episode of the information phase under the selected excitation policy, with the devices in phase, a validation episode of the last identification: physical system (full lines) and the digital twin identified in the phase (dashed; second block of Table~\ref{tab:twin_parameters}), in free prediction from the branch.
			(a) Velocity at $t=12$; at each device, on the right-hand axis, the command $a_j$ (black bar), the actuator state $\xi_j$ of \eqref{eq:actuator_dynamics} (filled purple circle) and, for the physical system only, the effective intensity $\widetilde\xi_j$ in \eqref{eq:effective_actuation} (open red circle).
			(b) Velocity at the inlet and at the three devices over the free interval; the dots are the sensor samples \eqref{observation_definition}.
			(c) For each device over the free interval: command $a_j$ (thin black), actuator state $\xi_j$ (purple) and effective intensity $\widetilde\xi_j$ (dashed red). Vertical lines separate the four blocks, whose frequencies are given in the panel of device 1.
			In (b) and (c) the dotted line marks $t=12$. At $t=0$ the twin starts from the analysed state.} %
		\label{fig:excitation_episode}
	\end{figure*}

	\begin{figure}[t]
		\centering
		\includegraphics[width=\columnwidth]{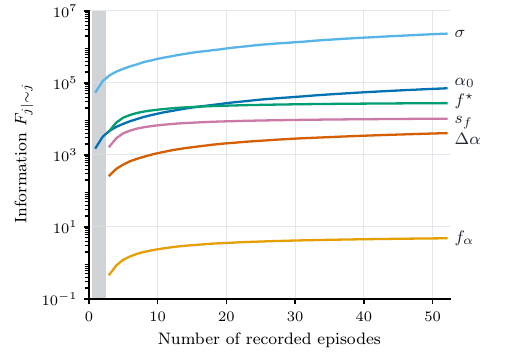}
		\caption{Information coefficient $F_{j\mid\sim j}$ of the recorded episodes on each twin parameter, under the recorded actions and at the identified twin, against the number of recorded episodes. Shaded: the two initial episodes. The four lines that start at the third episode are zero before.}
		\label{fig:information_growth}
	\end{figure}

	The information phase uses excitation policies of the form \eqref{eq:information_policy}, selected by enumeration of the 16 candidates in the library $\Theta_\iota$ using the E-optimal criterion \eqref{eq:rt_e_optimal_policy} (Sections~\ref{sec4p3} and~\ref{sec5p0b}). The phase starts with an empty buffer, since it does not use the episodes of Section~\ref{sec5_baseline}. Two initial episodes ($n_{\mathrm{init}}=2$ in Algorithm~\ref{alg:RT_policy_search}) are therefore recorded under the same fixed excitation policy: the first member of the library, with $\psi_2=\psi_3=0$ and four successive frequency blocks $(0.06,0.30,0.40,0.55)$ over the free interval (see \eqref{eq:information_policy}). These two episodes provide the analysed states and inflow realisations on which the information criterion is first evaluated.

	Three steps are then repeated ten times: (i) the 16 candidates are scored on the two most recent episodes; (ii) the best one is deployed for five episodes; and (iii) the twin is identified from all episodes recorded so far, with the settings of Section~\ref{sec5p0b}. The phase thus counts 50 information episodes after the two initial ones, and ten identifications, after $7,12,\ldots,52$ recorded episodes. In the notation of Algorithm~\ref{alg:RT_policy_search}, $e_n=1$ throughout this phase, with $u_{p,n}=1$ at these ten identifications and $u_{\iota,n}=1$ for $n=3,8,\ldots,48$ (information-policy updates). Neither the model-based reference nor the return surrogate is updated ($u_{\wb\pi,n}=u_{R,n}=0$), and no return enters $\mathcal D_\pi$.

	At each of the ten information-policy updates, the criterion selects the same set of frequencies, $(0.20,0.32,0.45,0.58)$. The phase lags are $(\psi_2,\psi_3)=(\pi,\pi/2)$ at the first update and $(0,0)$, with the devices in phase, at the remaining nine. The first update is performed before any identification, with the information averaged over the $N_\theta=4$ parameter vectors in Section~\ref{sec5p0b}. The corresponding smallest eigenvalue is $\lambda_{\min}=231$, compared with $0$ under the constant command of Section~\ref{sec5_baseline}. At the nine subsequent updates, the information is evaluated at the current twin rather than averaged over the parameter ensemble, so the resulting values are not directly comparable with the first one. At the twin identified at the end of the phase, the other three frequency sets in the library give $\lambda_{\min}=0$, whereas the four choices of phase lag for the selected frequency set give the same value to two digits. The criterion therefore distinguishes the frequency sets but not the phase lags.

As formulated here, each candidate is ranked from the information it is expected to provide on its own, through $\lambda_{\min}[\overline{\bs F}(\bs\theta_\iota)]$. The recorded episodes supply the states and inputs used to evaluate this quantity, but the information already accumulated from those episodes does not enter the ranking explicitly. The criterion can therefore continue to favour an excitation policy that is informative in absolute terms even if much of that information concerns parameter directions that are already well constrained. A sequential experimental design could instead rank candidates from a criterion such as $\lambda_{\min}[\bs F_{\mathcal B}+\overline{\bs F}(\bs\theta_\iota)/N_d]$, where $\bs F_{\mathcal B}$ represents the information already accumulated in the replay buffer. This would favour excitation policies that complement the existing data by improving the parameter directions that remain least informed, and is a natural extension for future work.

	Figure~\ref{fig:excitation_episode} shows one episode under the excitation policy selected among the 16 candidates, with the devices in phase. %
	The command \eqref{eq:information_policy} oscillates between $6$ and $8$, at a frequency that rises from one block to the next. %
	The actuator state follows it through the lag \eqref{eq:actuator_dynamics}, with an amplitude that decreases as the frequency rises. %
	The three devices receive the same command and differ only in their effective intensity \eqref{eq:effective_actuation}, which equals the actuator state at device 1 and is about half of it at devices 2 and 3. We stress that the digital twin cannot represent this effect because of the in-built model error. %

	The velocity downstream responds to the frequency of the block. Averaged over the 52 episodes, the velocity at $x_2$ is below $3$ in the first and last blocks and above $4$ in the second and third, where the frequency is close to $f^\star$ (Figure~\ref{fig:appendix_recovery_gain}). %
	At $x_3$ the rise is smaller. %
	The rise at both sensors comes from the dynamic increment of \eqref{eq:effective_coefficients}, absent under a constant command, through which $\Delta\alpha$, $f^\star$, $s_f$ and $f_\alpha$ now act on the measurements.
	The twin identified in the phase (dashed in Figure~\ref{fig:excitation_episode}) reproduces this recovery at $x_2$. %
	It underestimates the recovery at $x_3$, %
	and at $x_1$ it overestimates the velocity in every block, as under the constant command of Section~\ref{sec5_baseline}. %

	The information phase improves the twin where the excitation policy reveals new dynamics, without degrading its prediction under the constant command. The last identification ends at a validation trajectory loss \eqref{J_o} of $3.95$ (Figure~\ref{fig:identification_curves}a), close to that obtained in Section~\ref{sec5_baseline}, although the two values are computed on different episodes. When the two twins are evaluated on the same constant-command episodes, their losses remain comparable. Under the selected excitation policy, however, the twin identified in Section~\ref{sec5_baseline} gives a loss of $18.6$, almost five times that of the twin identified here. The three other parameter vectors from Section~\ref{sec5_baseline} whose validation loss lay within $5\%$ of the best fit also perform worse under this excitation policy, with losses between $1.6$ and $2.7$ times larger. A small validation loss under routine operation therefore says little about how well a twin predicts the dynamics excited by other policies; that depends on the data used to identify it.

	This improvement is also reflected in the identified parameters, reported in the second block of Table~\ref{tab:twin_parameters}. Two of the four parameters governing the dynamic increment move close to their values in the physical system, with $f^\star=0.399$ and $s_f=0.111$, compared with $0.40$ and $0.10$. The other two do not: $\Delta\alpha$ joins $\alpha_0$ at the upper bound of \eqref{eq:theta_p_box}, while $f_\alpha=0.157$ instead of $0.08$. The fit is thus held at these two bounds, and a wider interval was not tested. The twin therefore continues to use $\alpha_0$ and $\Delta\alpha$ to compensate for the interaction mechanism that is absent from its model structure. Making the parameters of the dynamic increment identifiable improves how the twin represents the excited dynamics, but it does not remove the structural discrepancy; instead, the compensation is redistributed among the available parameters.

The 52 recorded episodes inform five parameter-space directions, while the sixth remains weakly informed: the smallest eigenvalue of the information matrix is more than three orders of magnitude below the others and is associated almost entirely with $f_\alpha$. Figure~\ref{fig:information_growth} shows that the four parameters governing the dynamic increment are uninformed after the two initial episodes and become identifiable only when the excitation frequencies change. Further episodes under the selected excitation continue to add information, but mainly reinforce already informed directions and therefore yield diminishing returns.

	Consistently, the identified twin stabilises early in the information phase. From the first identification, after seven episodes, to the last, the parameters move by at most about $12\%$ of their admissible intervals, while $\alpha_0$ and $\Delta\alpha$ remain on their bounds. Over the same sequence of updates, the validation loss stays within the narrow range $3.78$--$4.04$. Since the successive identifications share much of the same data and start from the same archive, this stability is an indication of convergence rather than an independent validation.

The main limitation concerns $f_\alpha$, which remains the least informed parameter. The dynamic increment enters the observations through $\delta\alpha=\Delta\alpha\,g_\alpha$ in \eqref{eq:effective_coefficients}, with $f_\alpha$ appearing in the factor $1-\exp(-f^2/f_\alpha^2)$. At the two relevant frequencies of the selected excitation policy, this factor is already close to one, leaving little leverage to distinguish $f_\alpha$ from the other parameters. The two initial episodes, recorded at different frequencies, are therefore essential to break this coupling: without them, the information coefficients of $\Delta\alpha$, $f^\star$, $s_f$ and $f_\alpha$ would be several hundred times smaller. If $f_\alpha$ is fixed, the subsequent episodes are sufficient to identify the other three parameters.

	Although selected for information, the excitation policy also improves control performance, which matters because information episodes are run on the physical system. On the reporting inflows (Section~\ref{sec5p_solver}), with the devices in phase, it returns $\mathcal R_{\bb}=66.48$, for $39.14$ under the constant command %
	(Table~\ref{tab:scoreboard}, second row).
	Most of the gain comes from the modulation. A constant command at the mean command of the excitation policy, $a_j=7.00$, returns only $3.4$ more than the command at $7.98$, since the upstream device then leaves more flow to the other two. %
	The modulation adds $23.9$, by raising the mean powers of devices 2 and 3. %

	\subsection{Model-based control}
	\label{sec5_modelbased}

	\begin{figure}[t]
		\centering
		\includegraphics[width=\columnwidth]{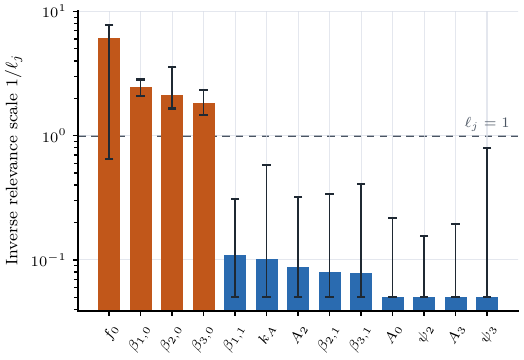}
		\caption{Inverse relevance scales $1/\ell_j$ (Section~\ref{sec3_synth}) of the thirteen policy coordinates \eqref{eq:unified_theta_pi}, fitted to the virtual returns \eqref{eq:operational_twin_return} of 512 policies drawn from a Sobol sequence (Section~\ref{sec5p0b}), on the twin identified in the information phase: median (bars) and range (whiskers) over the 50 fits of Section~\ref{sec5_rt}, one per set of random draws, sorted by the median. Dashed line: $\ell_j=1$, where $\kappa\ell_j=\rho_{\max}$, since $\kappa=\rho_{\max}=0.5$. Above it, the twin narrows the proposal region \eqref{eq:proposal_region} in that coordinate (half-width $\kappa\ell_j<\rho_{\max}$); below it, the half-width is the cap $\rho_{\max}$.} %
		\label{fig:relevance_scales}
	\end{figure}

	\begin{figure}[t]
		\centering
		\includegraphics[width=\columnwidth]{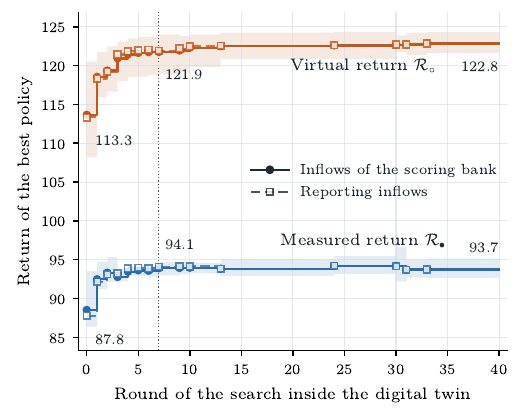}
		\caption{The search for the model-based reference inside the twin (one search, Section~\ref{sec5p0b}): virtual return \eqref{eq:operational_twin_return} (upper curves) and measured return \eqref{eq:ground_truth_dynamic_objective} (lower curves) of the best policy at each round. Full lines with filled circles: means over the eight inflows of the scoring bank, on which the twin selects, with a band from the lowest to the highest inflow. Dashed lines with open squares: means over the eight reporting inflows, with their values printed for the first policy, at round 7 (dotted line) and, at the end, for the reference.} %
		\label{fig:twin_and_physical}
	\end{figure}

	The twin identified in the information phase (second block of Table~\ref{tab:twin_parameters}) is first used alone, before any episode of the policy search. Two computations are performed on it (Section~\ref{sec3_synth}, with the settings of Section~\ref{sec5p0b}): the relevance scales of the virtual return \eqref{eq:operational_twin_return} and the model-based reference \eqref{eq:model_based_policy_synthesis_operator}. These computations require no physical episodes beyond the eight preparatory runs of the scoring bank (Section~\ref{sec5p_solver}), which are not counted in Table~\ref{tab:scoreboard}.

Figure~\ref{fig:relevance_scales} shows the inverse relevance scales $1/\ell_j$ of the thirteen policy coordinates \eqref{eq:unified_theta_pi}, fitted to the virtual returns of 512 policies. Four coordinates have $\ell_j<1$: the upstream frequency $f_0$ in \eqref{action_1} and the three offsets $\beta_{j,0}$ governing the mean allocation in \eqref{eq:learned_quasi_policy}. With $\kappa=0.5$ and $\rho_{\max}=0.5$ (Section~\ref{sec5p0b}), these are precisely the coordinates for which the twin assigns an initial proposal half-width below the cap, since for $g_j=1$,
\(
\rho_j=\min(\kappa\ell_j,\rho_{\max}).
\)
The initial search is therefore more concentrated around the model-based reference in these four directions, with $\rho_j=0.08$ for $f_0$ and between $0.20$ and $0.27$ for the $\beta_{j,0}$. The remaining nine coordinates, including all amplitudes and phase lags of the harmonic part, start at the maximum baseline half-width $\rho_{\max}$.

The twin therefore correctly singles out the offsets $\beta_{j,0}$, which distribute the mean extraction among the three devices, and the frequency $f_0$, which sets the recovery through the gain \eqref{eq:dynamic_induction_gains}, as the policy coordinates that play the dominant role in the control problem. These relevance scales define only the baseline geometry of the proposal region; during the physical search, the expansion factors $g_j$ can enlarge individual half-widths when measured performance indicates that the search is approaching a boundary.

Each run of Section~\ref{sec5_rt} fits its own relevance scales from its own set of 512 policies (Section~\ref{sec5p0b}), with runs sharing the same random draws also sharing this set. In 46 of the 50 sets, the four coordinates with $\ell_j<1$ are exactly those identified above; in the remaining four, only the three $\beta_{j,0}$ satisfy this condition. The resulting scales determine the baseline proposal region of each run and the distance used by its surrogate (Section~\ref{sec_A}).

	Figure~\ref{fig:twin_and_physical} follows the search for the model-based reference inside the twin. Over 40 rounds (Section~\ref{sec5p0b}), the best virtual return on the scoring bank increases from $113.7$ to $122.8$, with fifteen successive incumbent policies. For diagnostic purposes, these policies were also evaluated on the physical system, both on the scoring inflows and on the independent reporting inflows; these evaluations are not used by the model-based strategy itself.

The main gains found by the twin transfer well at the beginning of the search. By round 7, which captures most of the improvement in virtual return, the measured return on the reporting inflows has also increased substantially. Beyond this point, however, further improvements predicted by the twin no longer translate reliably to the physical system: the virtual return continues to rise, while the measured return remains essentially flat and eventually falls slightly below its round-7 value. The same tendency is observed on the scoring inflows on which the twin performs the optimisation.

This loss of transfer is consistent with the discrepancy between the twin and the physical system. Across the successive incumbent policies, the virtual return remains systematically optimistic, and its bias changes by several return units. Once the differences being optimised inside the twin become comparable with this variation, the twin can no longer rank nearby policies reliably. It is therefore an effective guide for the larger improvements found early in the search, but not for resolving the smaller differences encountered near its optimum.

This conclusion should nevertheless be interpreted cautiously. It is based on a single search with one identified twin, and the point at which physical performance is highest is known only retrospectively from evaluations that the model-based strategy does not use. The best round also differs between the scoring and reporting inflows.

	\begin{table*}[t]
		\centering
		\caption{The strategies of the policy search (Section~\ref{sec5_rt}). Each is run 50 times, with $M=500$ physical episodes, of which $n_{\mathrm{init}}=50$ are initial policies, and with the same 50 sets of random draws. UT, FT and CT are each run with the fixed region ($g_j=1$) and with the adaptive region (rule after \eqref{eq:proposal_region}). UG$_{2048}$ draws 2048 candidates at each selection, the other strategies 128. At each twin update, UT moves the centre of its region to the new reference. Trend: basis $\bs\varphi_{\wb}$ of the trend of the surrogate \eqref{eq:calibrated_return_model}. Scales: relevance scales $\ell_j$ of the proposal region and of the distance of the surrogate (Section~\ref{sec3_synth}).}
		\label{tab:arms}
		\footnotesize
		\begin{adjustbox}{max width=\linewidth}
\begin{tabular}{@{}llllll@{}}
			\toprule
			Label & Strategy & Twin updates & Trend & Initial policies and candidates & Scales \\
			\midrule
			UT & updated-twin variant & after episodes 150 and 350 & $[1,\mathcal R_{\wb}^{(m)}]$ & around the reference, moved at updates & from the current twin \\
			FT & fixed-twin variant & none & $[1,\mathcal R_{\wb}^{(0)}]$ & around the initial reference & from the initial twin \\
			CT & constant-trend variant & none & $[1]$ & around the initial reference & from the initial twin \\
			UG & unguided search & no twin & $[1]$ & whole box $\Theta_\pi$ & from the measured returns \\
			UG$_{2048}$ & unguided search, 2048 candidates & no twin & $[1]$ & whole box $\Theta_\pi$ & from the measured returns \\
			\bottomrule
		\end{tabular}
\end{adjustbox}
	\end{table*}

	\begin{figure*}[t]
		\centering
		\includegraphics[width=\textwidth]{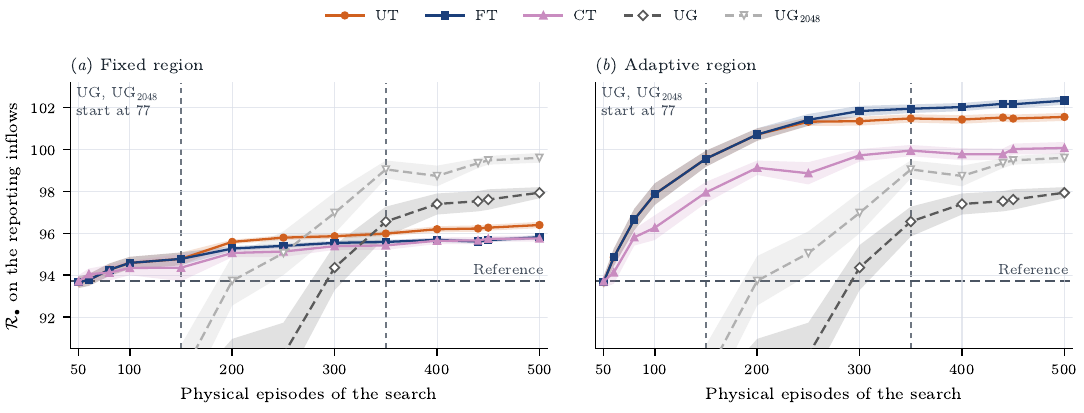}
		\caption{Return on the reporting inflows of the policy of $\mathcal D_\pi$ with the largest posterior mean \eqref{eq:gp_posterior_mean}, against the physical episodes of the search, for the strategies of Table~\ref{tab:arms}: (a) with the fixed region ($g_j=1$) and (b) with the adaptive region; the unguided searches are the same in both. Mean over the 50 runs (lines, with markers at the evaluations) and one standard error (bands); the unguided searches return $77.0$ at 50 episodes and enter the axes after 150 (UG$_{2048}$) and 200 (UG) episodes. Horizontal dashed line: model-based reference (Table~\ref{tab:scoreboard}, third row). Vertical dashed lines: twin updates of UT, which coincides with FT up to the first.} %
		\label{fig:rt_return}
	\end{figure*}

	The final incumbent of the model-based search is taken as the reference policy. On the reporting inflows (Section~\ref{sec5p_solver}), it achieves $\mathcal R_{\bb}=93.71$ (Table~\ref{tab:scoreboard}, third row), improving substantially over the selected excitation policy. Most of this gain comes from increased extraction by the two downstream devices, while the imbalance between device powers changes comparatively little.

Despite identifying a useful policy, the twin remains strongly biased in its absolute prediction: on the same inflows it overpredicts the return by $31\%$. This discrepancy is not primarily caused by state synchronisation. Initialising the twin from the analysed states of the physical episodes (Section~\ref{sec_sync}), rather than from the state reached internally by the twin at the branch (Section~\ref{sec5p0b}), gives almost the same predicted return. As already observed in Section~\ref{sec5_baseline}, the dominant error therefore lies in the model itself rather than in the reconstructed initial state.

	The first reporting inflow shows where the difference arises. On it, the twin predicts a return of $122.2$ and the physical system returns $\mathcal R_{\bb}=93.3$. %
	About $60\%$ of this difference is power that the physical system does not extract, and the rest is a larger balance penalty. %
	The shortfall is at the second device, which extracts almost $40\%$ less than the twin predicts. %
	Its effective intensity \eqref{eq:effective_actuation} is about half the nominal one, a loss that the twin cannot predict (Section~\ref{sec4p1}). %

	The twin thus provides a proposal region and a useful starting point, but not a reliable ranking of the best physical policies.

	\subsection{Reinforcement twinning}
	\label{sec5_rt}

	To quantify the impact of the different forms of twin guidance on the policy-search stage of reinforcement twinning, five strategies are compared, as summarised in Table~\ref{tab:arms}. Three retain progressively fewer contributions from the twin, while two are fully unguided (model-free).

	The updated-twin variant (UT) uses the complete loop, with two scheduled twin updates. The fixed-twin variant (FT) omits these updates but retains the twin identified in Section~\ref{sec5_excitation}, together with its model-based reference, relevance scales and virtual-return prediction. Like UT, it selects its policies from measured returns. Comparing UT with FT therefore isolates the effect of updating the twin during the search.

	The constant-trend variant (CT) removes a second source of guidance. It is identical to FT except that its surrogate models the measured returns without the twin prediction, with the constant trend basis $\bs\varphi_{\wb}(\bs\theta_\pi)=[1]$ in \eqref{eq:calibrated_return_model} instead of $[1,\mathcal R_{\wb}^{(0)}]$. CT nevertheless retains the twin-derived reference, proposal region, initial policies and relevance scales. Comparing FT with CT therefore isolates the contribution of the twin prediction to the return surrogate.

	The two unguided strategies, UG and UG$_{2048}$, use no twin information and are thus fully model-free. Their initial policies and candidates span the whole policy box $\Theta_\pi$, their surrogate has a constant trend, and their relevance scales are learned from measured returns. UG considers $N_{\mathcal Q}=128$ candidates at each selection, as do the guided strategies, whereas UG$_{2048}$ considers 2048. The denser pool tests whether the comparison is affected by sparse coverage of the larger search domain.

The proposal geometry provides a further comparison. Each twin-guided strategy is run with either a fixed region ($g_j=1$) or the adaptive rule of Section~\ref{sec_B}. The fixed region tests the effect of constraining the search to remain close to the model-based reference according to the geometry prescribed by the twin, whereas the adaptive region tests whether physical evidence can progressively relax this guidance when it becomes too restrictive. Figure~\ref{fig:rt_return} tracks how the physical performance of the solution returned by each search evolves as physical episodes are collected, with the fixed-region results in the left panel and the adaptive-region results in the right panel. The comparisons address four questions. The unguided searches test whether twin guidance gives an advantage, at the end of the search and early in it, and the fixed and adaptive regions whether a twin-defined geometry becomes restrictive. FT against CT tests whether the virtual-return trend helps once the region can widen, and UT against FT whether updating the twin improves prediction and control.

Each twin-guided search follows the information phase of Section~\ref{sec5_excitation} and starts from the twin, scoring bank and model-based reference obtained in Section~\ref{sec5_modelbased}. All search episodes are performance episodes ($e_n=0$), and the replay buffer $\mathcal B$ contains only episodes generated during the search. The budget is $M=500$ physical search episodes, including $n_{\mathrm{init}}=50$ initial policies (Section~\ref{sec5p0b}). The information-phase episodes, the eight preparatory runs of the scoring bank and the reporting evaluations are excluded from this budget. UT updates the twin after episodes 150 and 350 ($u_{p,n}=1$, hence $u_{\wb\pi,n}=u_{R,n}=1$; Section~\ref{sec2c}), using the settings of Section~\ref{sec5p0b}.

Each strategy is run 50 times. The runs differ in the random draws used for the relevance-scale fits, initial policies, candidate pools and inflows. The runs guided by the twin share the information phase, the identified twin and the initial reference, and all runs are assessed on the same eight reporting inflows, so their spread measures the variability of the search for this twin, not that of the complete procedure. All strategies use the same 50 sets of draws: paired runs share the inflow of every episode and, for the twin-guided strategies, the initial policies. Comparisons can therefore be made pairwise across runs. Runs based on the same draws coincide until their rules first differ: for example UT and FT are identical until the first twin update at episode 150, and the fixed- and adaptive-region variants until the first widening.

The unguided searches use the same search budget and, run by run, the same inflows as the twin-guided searches. Since they require neither the 52 episodes of the information phase nor the 8 preparatory runs of the scoring bank, they are also compared at equal total physical cost: 500 unguided episodes correspond to 440 search episodes for a twin-guided strategy.

The curves in Figure~\ref{fig:rt_return} report the performance of the current solution returned by each search. At each reporting point, this is taken to be the policy already in $\mathcal D_\pi$ with the largest posterior mean \eqref{eq:gp_posterior_mean}, rather than the policy with the largest single measured return, which may reflect episode-to-episode variability. The selected policy is evaluated independently on the same eight reporting inflows of Section~\ref{sec5p_solver}. Evaluations are performed after 50, 60, 80 and 100 search episodes and then every 50 episodes up to 500, with an additional evaluation after 440 episodes for the equal-total-cost comparison. Their results provide the markers in Figure~\ref{fig:rt_return} and are used only for assessment, entering neither $\mathcal D_\pi$ nor the replay buffer $\mathcal B$.

We first test whether each of the three adaptive twin-guided strategies (UT, FT and CT) outperforms each of the two unguided searches (UG and UG$_{2048}$). The comparisons are made both after 500 search episodes and at equal total physical cost, giving $3\times2\times2=12$ planned comparisons. These comparisons were specified before examining the results of the adaptive runs.

Because the runs are paired, each comparison is based on the return difference between two strategies exposed to the same random draws. Statistical significance is assessed by repeatedly reversing the signs of these paired differences under the hypothesis that neither strategy has a systematic advantage. We use $10^6$ random sign patterns for each test. Since twelve hypotheses are tested together, Holm's correction is applied to keep the probability of making at least one false claim of significance across the full set below $0.05$. The same procedure is applied separately to the fixed-region results.

Two additional questions are examined as secondary analyses: whether widening improves UT and whether UT has an early advantage over the unguided searches. Other comparisons were selected after inspecting the results and are therefore treated as exploratory; they use paired $t$-tests without correction for multiple comparisons. Finally, differences over the complete search history are assessed jointly rather than testing each reporting point in isolation. The test uses the largest absolute $t$-statistic over the 13 reporting points and a paired sign-flip procedure, thereby controlling the probability of detecting a spurious difference anywhere along the search history at $0.05$.

	With the fixed region (left panel of Figure~\ref{fig:rt_return}), the three strategies guided by the twin start from a clear advantage, but eventually level off when the search remains confined to the twin-defined region. After the 50 initial policies, their returns are about $93.7$, close to the model-based reference, whereas both unguided searches return $77.0$. When the reporting points are considered jointly, UT remains ahead of UG through 250 episodes and of UG$_{2048}$ through 150 episodes. The unguided searches then continue to improve and overtake the guided strategies between 250 and 350 episodes.

	At 500 episodes, UT, FT and CT return $96.40$, $95.81$ and $95.75$, respectively, compared with $97.94$ for UG and $99.60$ for UG$_{2048}$ (see Table \ref{tab:scoreboard}). All three guided strategies therefore remain above the model-based reference of $93.71$ on average, but below both unguided searches. The same conclusion holds at equal total physical cost. All twelve final and equal-total-cost differences remain significant after correction for multiple comparisons, in favour of the unguided searches.

\begin{figure}[!tb]
		\centering
		\includegraphics[width=\columnwidth]{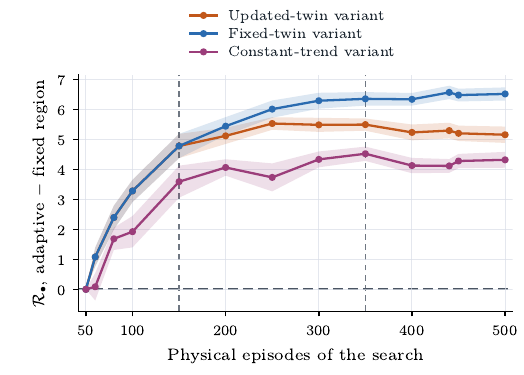}
		\caption{Return on the reporting inflows of each strategy guided by the twin with the adaptive region minus its return with the fixed region, paired over the runs: mean (lines, with markers at the evaluations) and one standard error (bands) over the 50 runs. The two runs of a pair are identical up to the first widening. Vertical dashed lines: twin updates of the updated-twin variant.}
		\label{fig:rule_effect}
	\end{figure}

\begin{figure}[!tb]
		\centering
		\includegraphics[width=\columnwidth]{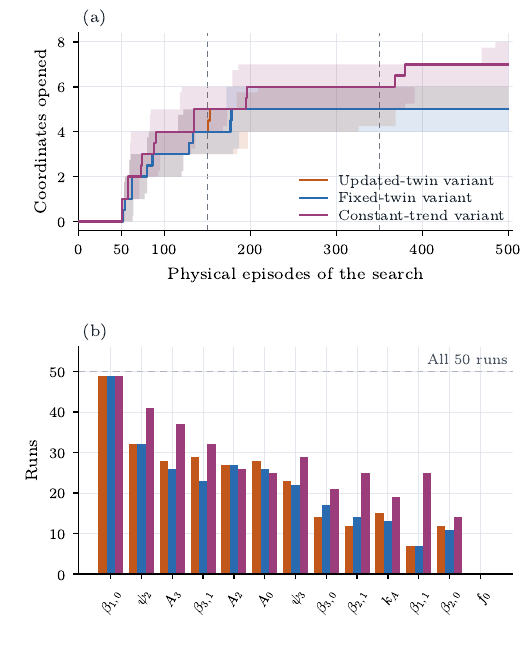}
		\caption{Widening of the proposal region by the rule after \eqref{eq:proposal_region}, for the three strategies guided by the twin with the adaptive region. (a) Number of policy coordinates \eqref{eq:unified_theta_pi} whose half-width has been doubled at least once, against the physical episodes of the search: median (lines) and interquartile range (bands) over the 50 runs. Vertical dashed lines: twin updates of the updated-twin variant. (b) For each coordinate, the number of runs in which its half-width has been doubled at least once.}
		\label{fig:region_opened}
	\end{figure}

	The saturation of the guided searches is consistent with the fixed proposal region becoming restrictive. The right panel of Figure~\ref{fig:rt_return} differs from the left one only in that the measured returns can widen this region; how and where the region widens is shown below with Figures~\ref{fig:rule_effect} and~\ref{fig:region_opened}.

	The adaptive-region search (right panel of Figure~\ref{fig:rt_return}) yields marked improvement in control performance and preserves the early advantage provided by the twin. At 500 episodes, CT, FT and UT return $100.08$, $102.33$ and $101.56$, respectively, about $6$ to $9$ points above the model-based reference (Table~\ref{tab:scoreboard}). UT ends about $3.6$ above UG and $2.0$ above UG$_{2048}$. Essentially the same advantages remain when its return after 440 search episodes is compared with 500 unguided episodes at equal total physical cost.

The advantage of twin guidance is supported by ten of the twelve planned comparisons after correction for multiple comparisons. The two exceptions are the comparisons of CT with UG$_{2048}$, after 500 episodes and at equal total cost. A joint analysis of the complete search histories also supports the advantage of UT over both unguided searches, with a higher mean return at every reporting point from 50 to 500 episodes. The widening rule therefore allows physical evidence to relax the twin-defined geometry where it becomes restrictive.

	Figure~\ref{fig:rule_effect} diagnoses when the adaptive proposal-region mechanism begins to affect performance. For each guided strategy, this figure plots the paired difference between the return obtained with the adaptive and fixed regions against the search episode. The paired runs are identical until the first widening; the lines and bands show the mean difference and one standard error over the 50 runs.

	The adaptive rule is activated early and consistently. Every adaptive run widens at least one coordinate, and half do so within the first three selections after the initial policies. By 500 episodes, widening raises the return by $4.3$ to $6.5$ across the three strategies, with a positive effect already evident from 60 or 80 episodes onwards. Widening is therefore not merely an occasional safeguard: it changes the policies explored and has a sustained effect on their performance.

	Figure~\ref{fig:region_opened} shows how the adaptive mechanism acts coordinate by coordinate. Panel (a) gives, against the search episode, the number of policy coordinates whose half-width has been doubled at least once, as the median and interquartile range over the 50 runs. Panel (b) gives, for each coordinate, the number of runs in which at least one doubling occurs.

	The adaptive mechanism selectively relaxes the twin guidance where the physical returns indicate that its original geometry is too restrictive. The clearest example is $\beta_{1,0}$, the offset of the upstream device, whose half-width is doubled at least once in 49 of the 50 runs of every guided strategy. This is precisely the direction in which the fixed-region solutions press against the region. With the fixed region, the final policy of 41 of the 50 UT runs lies within $\tau\rho_j$ of at least one face of the region that does not coincide with a boundary of $\Theta_\pi$, and in $\beta_{1,0}$ it typically approaches the lower face. Reducing $\beta_{1,0}$ unloads the upstream device and, in the physical system, increases the effective intensity of the downstream devices through \eqref{eq:plant_interaction}--\eqref{eq:effective_actuation}. This interaction is absent from the twin (Section~\ref{sec4p1}), which therefore does not predict the additional downstream benefit. In UT with the adaptive region, the half-width of $\beta_{1,0}$ grows from about $0.20$ to about $0.8$, and the final policy lies well below the lower face of the fixed region. This provides evidence that the twin-defined region can exclude better physical policies.

	The frequency $f_0$ provides the complementary case. Although the twin gives it the narrowest initial region, the physical-evidence rule never doubles its half-width. In UT, $f_0$ can nevertheless broaden because a twin update recomputes the baseline relevance scales: its median half-width changes from $0.08$ initially to $0.50$ at the end. This is distinct from widening by the adaptive rule. FT and CT retain the initial twin and therefore keep the original half-width of $f_0$.

	The right panel of Figure~\ref{fig:rt_return} also isolates the value of the virtual-return prediction by comparing FT and CT. These strategies use the same initial policies, reference, proposal region and relevance scales; only FT retains the calibrated virtual-return trend. Once the region is allowed to adapt, this prediction provides a clear benefit: FT ends $2.3$ above CT ($p<10^{-7}$). With the fixed region in the left panel, by contrast, FT and CT end at essentially the same return.

	The comparison of UT and FT in the same panels isolates the effect of updating the twin. The updates improve the predictive consistency examined below, but they do not improve the final control policy with the adaptive region: UT ends $0.8$ below FT ($p=0.0015$). With the fixed region, the updates instead give the smaller positive difference of $0.6$ ($p=0.0013$). The rest of this section examines what the updates change in the surrogate, the twin and the search.

\begin{figure}[!tb]
		\centering
		\includegraphics[width=\columnwidth]{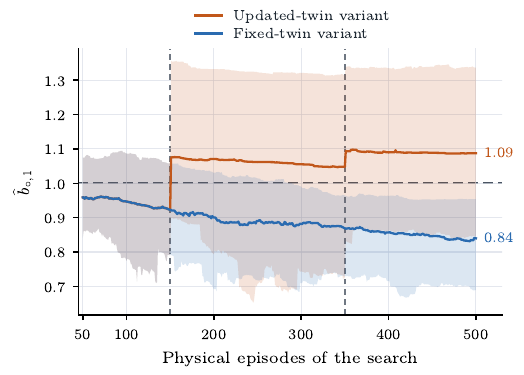}
		\caption{Calibration coefficient $\widehat b_{\wb,1}$ of the trend \eqref{eq:calibrated_return_model}, fitted by \eqref{eq:gls_calibration} after every episode from the end of the initial policies, for the two strategies with the virtual return in the trend, with the adaptive region: median over the 50 runs (lines, with the final value) and range (bands). Horizontal dashed line: $\widehat b_{\wb,1}=1$; vertical dashed lines: twin updates.}
		\label{fig:rt_b1}
	\end{figure}

	Figure~\ref{fig:rt_b1} shows how strongly the surrogate follows the virtual-return prediction in the adaptive-region searches. It plots the fitted coefficient $\widehat b_{\wb,1}$ multiplying $\mathcal R_{\wb}^{(m)}$ in the calibrated trend \eqref{eq:calibrated_return_model}. With $\widehat b_{\wb,1}=1$, a difference of one unit of virtual return between two policies is taken as a difference of approximately one unit of measured return. Values below one downweight the differences predicted by the twin.

	With the twin not updated, the coefficient decreases as measured returns accumulate, reaching a median of $0.84$ at the end of the search. In UT, its median rises from $0.924$ immediately before the first update to $1.076$ immediately afterwards, remains above one, and ends at $1.088$. The surrogate thus progressively scales down the virtual-return differences of the stale fixed twin, while taking those of the updated twin approximately at face value. The fixed-region runs show the same pattern, with final coefficients of $0.89$ and $1.17$.

\begin{figure}[!tb]
		\centering
		\includegraphics[width=\columnwidth]{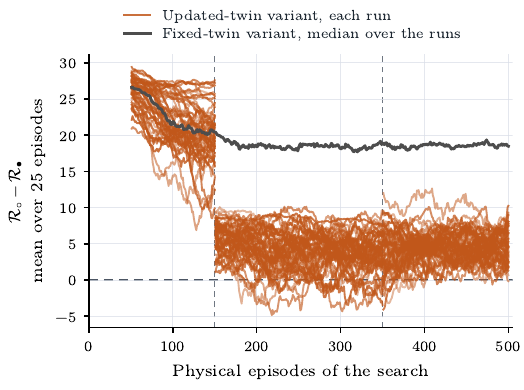}
		\caption{Virtual minus measured return of the policies deployed by the updated-twin variant with the adaptive region, $\mathcal R_{\wb}^{(m)}-\mathcal R_{\bb}$, with the twin in use at each episode: one line per run (50), as a centred moving mean over 25 episodes from the end of the initial policies, taken within each version of the twin. Grey: median over the runs of the fixed-twin variant with the adaptive region, computed in the same way. Vertical dashed lines: twin updates.}
		\label{fig:rt_bias}
	\end{figure}

Figure~\ref{fig:rt_bias} shows how the twin updates affect the error in the predicted return. For each UT run with the adaptive region, it reports the difference $\mathcal R_{\wb}^{(m)}-\mathcal R_{\bb}$ for the deployed policies, using the version of the twin available at each episode. To smooth episode-to-episode fluctuations, the curves show centred moving averages computed separately between successive twin updates; the grey curve shows the corresponding median for FT.

The first update removes most of the systematic bias: its median falls from $+22.5$ to $+4.3$, while the second update produces little further change. In FT, the bias also decreases as the adaptive region shifts the policies being deployed, but only to about $+18$; with the fixed region, it remains around $+26$. This dependence on the policies being explored reflects the structural, policy-dependent discrepancy introduced in Section~\ref{sec4p1}.

	The first update also lowers the total identification loss \eqref{eq:rt_replay_loss} on the validation episodes by $38\%$ in the median. The second changes it little and returns the twin unchanged in 23 of the 50 runs. The two terms of the loss show what is being improved: at the first update, the weighted return term falls in every run, by a factor of about seven in the median, while the trajectory term \eqref{J_o} rises slightly. The update therefore corrects the predicted returns at a small cost in the predicted trajectories. These improvements in model prediction do not translate into a better final control policy, since UT ends below FT in the adaptive-region case.

The third block of Table~\ref{tab:twin_parameters} reports the fitted twin parameters at the end of the adaptive UT searches. Because the twin is structurally misspecified, these values are not estimates of the parameters of the physical system.

In every run, $\alpha_0$ and $\Delta\alpha$ leave the upper bounds at which the initial twin places them and shift towards the physical-system values \eqref{eq:theta_p_physical}. Their final medians are $0.645$ and $5.96$, compared with $0.5$ and $6$ in the physical system. By contrast, $\sigma$ shifts away from its physical-system value in every run, while $f_\alpha$ and, to a lesser extent, $f^\star$ do so in most runs. The fitted correction also varies across runs. As in Section~\ref{sec5_excitation}, the updates therefore redistribute how the twin parameters compensate for missing physics, here in a way that improves the prediction of the recorded returns.

Reducing the bias of the predicted return, however, does not necessarily make the twin more useful for policy search. A nearly uniform offset can be absorbed by the intercept $\widehat b_{\wb,0}$ in \eqref{eq:calibrated_return_model}; what matters for the search is whether the twin ranks candidate policies correctly. This can be assessed directly on the policies visited during the search. For each successive twin, we therefore compare the ranking induced by the virtual returns with that obtained from the measured returns, using Spearman correlation.

The updates improve the ranking over the complete set of 500 visited policies: the median correlation rises from $0.066$ for the initial twin to $0.229$ and $0.234$ after the two updates. In the high-return region, however, the ranking remains poor. Among the 100 policies with the highest measured return, the correlations remain negative, changing only from $-0.189$ to $-0.160$ and $-0.154$. This local ranking is more relevant late in the search, when candidate policies lie close to the best solutions found so far.

Because the 500 visited policies include episodes used to update the twin, we also verify this result out of sample using policies from the paired FT runs after episode 150, which were not used in any twin update. Scoring these policies with the three successive UT twins increases the overall correlation from $-0.116$ to $0.075$ and $0.088$, whereas among the best 100 policies it changes only from $-0.237$ to $-0.209$ and $-0.212$.

The updates therefore improve the ranking over the overall set of policies, but provide little improvement where accurate ranking matters most for optimisation. At the same time, the surrogate gives greater weight to the updated twin and the recomputed relevance scales broaden the baseline region along $f_0$. These effects may help explain the lower return of UT relative to FT with the adaptive region, although the present comparisons do not isolate their individual contributions. A poorer reference is unlikely to be the cause. With the fixed region, the re-synthesised references return slightly more on the physical system than the initial one, with medians of $94.03$ and $94.20$ against $93.71$, and with the adaptive region they move by only about $0.09$ in the normalised coordinates \eqref{eq:policy_whitening}.

	Table~\ref{tab:scoreboard} compares the device-wise mean powers and returns of the final policies. The adaptive-region policies extract less total power than their fixed-region counterparts, but distribute it much more evenly. For UT, the upstream device extracts about a quarter less and the third device more.

	The gain in return therefore comes from a smaller balance penalty, about half that of the fixed-region policy, rather than from greater total extraction. The best policy by return, FT with the adaptive region, extracts $117.6$ in total, about $14\%$ less than the model-based reference. The unguided searches also end with balanced policies.

	Taken together, these results separate the roles of the twin in policy search. The twin supplies a useful starting policy and an informative initial proposal geometry, which give the guided searches their early advantage. If this geometry remains fixed, model mismatch eventually makes it too restrictive. Allowing measured physical returns to widen the region preserves the initial benefit while enabling the search to escape incorrect model guidance. Once this freedom is present, the calibrated virtual-return prediction still improves the search. Online twin updates remove most of the return-prediction bias and improve several model-prediction metrics, but in this benchmark they do not improve the final control return and reduce it relative to the fixed twin. Reducing prediction bias is therefore not sufficient: for control, the decisive property is whether the twin provides a useful local ranking of high-performing policies.

	\section{Conclusions}\label{sec6}

This work has presented reinforcement twinning as a framework for coupling policy search on a physical system with the identification of an imperfect digital twin. The twin supplies a model-based reference, the initial geometry of the policy search, and a prediction of the return calibrated against measured performance within a Gaussian-process surrogate. The physical records, in turn, provide data for parameter identification. An observer first reconstructs the twin state at the branch, and an optimiser fits the parameters by comparing free-replay predictions under the recorded inputs and actions with the measured trajectories and returns. The twin can also synthesise informative excitation policies. Algorithm~\ref{alg:RT_policy_search} brings these operations together while allowing their computationally expensive components to be scheduled separately from physical interaction.

The Burgers-flow benchmark illustrates why both directions of this coupling matter. Routine operation under a steady controller left the parameters governing dynamic recovery unidentifiable, regardless of how many such episodes were recorded. The excitation policy synthesised on the twin, together with the two initial episodes recorded at other frequencies, exposed these dynamics and improved their prediction, although one parameter direction remained weakly informed. The resulting twin provided a useful control policy despite a persistent structural discrepancy. These observations distinguish insufficient excitation, which can be addressed by changing the data collected, from missing model structure, which parameter fitting alone cannot remove.

For policy search, the central finding is that useful model guidance must leave room for correction by physical evidence. The twin provided a favourable starting point, but a fixed proposal region eventually constrained the search and allowed unguided alternatives to overtake it. Allowing measured performance to expand this region preserved the early advantage of reinforcement twinning and led to better final policies than either unguided search tested here, including when preparatory interactions were counted. Comparisons with variants where individual contributions of the twin were selectively removed also showed a benefit from retaining the calibrated twin prediction in the surrogate once the region could expand. The twin was therefore useful both in directing initial exploration and guiding subsequent choices, provided its preferred region did not become a permanent restriction.

The benefit of updating the twin was less direct. Scheduled parameter updates substantially reduced return-prediction bias for visited policies, but the improvement in the replay return term could come at the expense of the trajectory fit. The fitted parameters continued to compensate for the missing interaction mechanism, and the twin remained unreliable at ranking the best policies. With an adaptive proposal region, the fixed-twin variant achieved a higher final return than the updated-twin variant. Reducing prediction bias therefore did not by itself ensure better control. Since an update also changes the reference, the relevance scales, and the surrogate, the present comparisons do not isolate the cause of this outcome.

Several extensions could follow from these findings. Excitation-policy synthesis could account for the information already accumulated, rather than score each candidate in isolation \cite{GoodwinPayne1977,WalterPronzato1997}, and information episodes could be scheduled within policy search rather than confined to a preceding phase. After a twin update, the proposal region could also be re-centred on the evaluated policy with the largest posterior mean, following the principle of trust-region Bayesian optimisation \cite{Eriksson2019}. The present benchmark demonstrates the computational framework under sparse sensing and deliberate model misspecification, while the recovery law and balance penalty adopted here determine the magnitude of the observed gains. The twin also shares the recovery-law structure the search must exploit and lacks only the hidden interaction between the devices; a twin whose error affects this dominant mechanism was not tested. Assessing reinforcement twinning for dynamic-induction control will require validation on richer flow models and experiments, building on the large-eddy simulations \cite{Munters2017,Munters2018} and wind-tunnel studies \cite{Frederik2020Periodic} that motivate this example. Experimental proof-of-concept studies of reinforcement twinning on other controlled systems are in preparation.

More broadly, these results point towards reinforcement twinning as a route to control systems in which models, experiments, and control policies are adapted together as operational evidence accumulates.

	\section*{CRediT authorship contribution statement}
	\textbf{Miguel A. Mendez:} Conceptualization, Methodology, Software, Formal analysis, Investigation, Data curation, Visualization, Writing -- original draft, Writing -- review \& editing, Project administration, Funding acquisition. \textbf{Lorenzo Schena:} Conceptualization, Methodology, Software, Validation, Writing -- review \& editing. \textbf{Romain Poletti:} Conceptualization, Methodology, Software, Validation, Writing -- review \& editing. \textbf{Yannick Lecomte:} Conceptualization, Methodology, Software, Validation, Writing -- review \& editing. \textbf{Sebastiano Randino:} Conceptualization, Methodology, Software, Validation, Writing -- review \& editing. \textbf{Jan van den Berghe:} Conceptualization, Methodology, Software, Validation, Writing -- review \& editing.

	\section*{Declaration of competing interest}
	The authors declare that they have no known competing financial interests or personal relationships that could have appeared to influence the work reported in this paper.

	\section*{Declaration of generative AI and AI-assisted technologies in the manuscript preparation process}
During the preparation of this work, the authors used Claude (Anthropic)
and ChatGPT (OpenAI) to assist in the conceptual development of the
algorithm, the drafting and refinement of the manuscript based on
the authors' results and instructions, and the development of scripts
for generating figures and tables. The authors reviewed and, where
necessary, revised all AI-assisted content and take full responsibility
for the published article.

	\section*{Data availability}
	The code and the data that support the findings of this study will be made available in an open repository upon publication.

	\section*{Acknowledgements}
	This work was funded by the European Research Council (ERC, grant agreement No 101165479 RE-TWIST StG), which also supports R. Poletti, Y. Lecomte and J. van den Berghe. Views and opinions expressed are however those of the authors only and do not necessarily reflect those of the European Union or the European Research Council. Neither the European Union nor the granting authority can be held responsible for them. L. Schena is supported by the Fonds Wetenschappelijk Onderzoek (FWO), project number 1S67925N. S. Randino is funded by an FNRS FRIA Fellowship (fellowship number 40029825).

	\appendix
	\setcounter{table}{0}
	\setcounter{figure}{0}
	\renewcommand{\theHtable}{A.\arabic{table}}
	\renewcommand{\theHfigure}{A.\arabic{figure}}
	\section{Technical details of the Burgers benchmark}\label{app:benchmark_details}

	\begin{table}[t]
		\centering
		\caption{Remaining constants of the Burgers benchmark: recovery law \eqref{eq:effective_coefficients}, interaction weights and policy. The main constants are in Table~\ref{tab:burgers_settings}; the values of the twin parameters in the physical system and the box $\Theta_p$ are given in \eqref{eq:theta_p_physical} and \eqref{eq:theta_p_box}.}
		\label{tab:burgers_full_settings}
		\footnotesize
		\begin{tabularx}{\columnwidth}{@{}>{\raggedright\arraybackslash}Xll@{}}
			\toprule
			Quantity & Symbol & Value \\
			\midrule
		Amplitude/frequency exponents & $p_A,p_f$ & $4,\;4$ \\
			Recovery centre and width & $x_{\mathrm{rec}},\sigma_{\mathrm{rec}}$ & $2.90,\;1.00$ \\
			Recovery transition width & $\ell_{\mathrm{rec}}$ & $0.2$ \\
			Recovery modulation depth & $\mu_\alpha$ & $0.50$ \\
			Recovery phase lag & $\phi_\alpha$ & $\pi/2$ \\
			Interaction weights & $w_{12},w_{13},w_{23}$ & $1,\;0.3287,\;0.6713$ \\
			Amplitude normalisation & $A_{\mathrm{ref}}$ & $1.0$ \\
			Policy reference inflow & $u_{\mathrm{ref}}$ & $6.0$ \\
			Policy inflow scale & $s_{\mathrm{in}}$ & $1.0$ \\
			Mean-allocation coefficient bounds & $\beta_{j,0},\beta_{j,1}$ & $[-6,6]$ \\
			Inlet-filter time constant & $\tau_{\mathrm f}$ & $1.0$ \\
			Harmonic frequency bounds & $f_0$ & $[0,2]$ \\
			Upstream base-amplitude bounds & $A_0$ & $[0,1]$ \\
			Upstream inlet-gain bounds & $k_A$ & $[0,80]$ \\
			Upstream harmonic limit & $d_{1,\max}$ & $1.0$ \\
			Downstream amplitude and phase-lag bounds & $A_j,\ \psi_j$ & $[0,1],\ [0,2\pi]$ \\
			\bottomrule
		\end{tabularx}
	\end{table}

	The interaction weights in \eqref{eq:plant_interaction} act only from upstream to downstream. For each $j>1$ and $i<j$,
	\begin{equation}
		\widetilde w_{ij}=\exp\!\left[-\frac{x_j-x_i}{\ell_w}\right],
		\qquad
		w_{ij}=\frac{\widetilde w_{ij}}{\sum_{l<j}\widetilde w_{lj}},
	\end{equation}
	and all other entries are zero; with $\ell_w$ of Table~\ref{tab:burgers_settings}, this gives the weights in Table~\ref{tab:burgers_full_settings}. The states $\zeta_j$ are clipped to $[0,1]$ after each numerical step.

	The twin parameters \eqref{eq:unified_theta_p} take the values
	\begin{equation}
		\label{eq:theta_p_physical}
		\bs\theta_p^{\bb}
		=
		[0.50,\,6.0,\,0.40,\,0.10,\,0.08,\,0.20]^\top
	\end{equation}
	in the physical system and are identified in the box $\Theta_p=[\bs\theta_p^{\min},\bs\theta_p^{\max}]$, with
	\begin{equation}
		\label{eq:theta_p_box}
		\begin{aligned}
		\bs\theta_p^{\min}
		&=
		[0.20,\,1.00,\,0.20,\,0.04,\,0.02,\,0.10]^\top,
		\\
		\bs\theta_p^{\max}
		&=
		[0.80,\,10.00,\,0.70,\,0.25,\,0.20,\,0.40]^\top.
		\end{aligned}
	\end{equation}

	\begin{figure}[!ht]
		\centering
		\includegraphics[width=0.78\columnwidth]{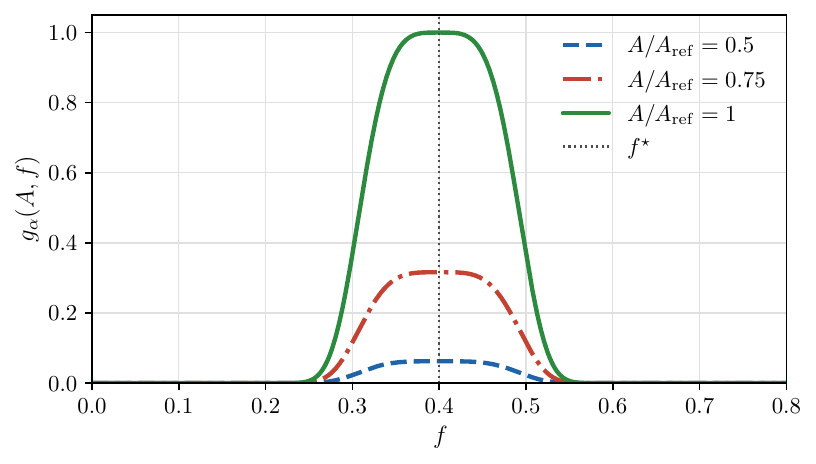}
		\caption{Recovery gain $g_\alpha(A,f)$ \eqref{eq:dynamic_induction_gains} as a function of the frequency $f$ of the upstream harmonic, for three normalised amplitudes. The dotted line marks $f^\star$, the centre of the frequency band.}
		\label{fig:appendix_recovery_gain}
	\end{figure}

	The recovery mask in \eqref{eq:effective_coefficients} is
	\begin{equation}
		\begin{split}
		q_{\mathrm{rec}}(x)
		={}&
		\exp\!\left[
		-\frac{1}{2}
		\left(\frac{x-x_{\mathrm{rec}}}{\sigma_{\mathrm{rec}}}\right)^2
		\right]
		\\
		&\times
		\frac{1}{2}
		\left[
		1+
		\tanh\!\left(\frac{x-x_1}{\ell_{\mathrm{rec}}}\right)
		\right],
		\end{split}
	\end{equation}
	a Gaussian bump just downstream of the first device, smoothly cut off at the location $x_1$ of that device. The recovery gain is
	\begin{equation}
		\label{eq:dynamic_induction_gains}
		\begin{split}
		g_\alpha(A,f)
		={}&
		\left[
		\frac{\min(A,A_{\mathrm{ref}})}{A_{\mathrm{ref}}}
		\right]^{p_A}
		\left[
		1-\exp\!\left(-\frac{f^2}{f_\alpha^2}\right)
		\right]
		\\
		&\times
		\exp\!\left[
		-\left|\frac{f-f^\star}{s_f}\right|^{p_f}
		\right].
		\end{split}
	\end{equation}

	The phase-locked factor in \eqref{eq:effective_coefficients} keeps the increment between $0.5$ and $1.5$ times its mean. The fourth powers (Table~\ref{tab:burgers_full_settings}) confine the gain to amplitudes near $A_{\mathrm{ref}}$ and frequencies near $f^\star$: half the amplitude gives $1/16$ of the gain and, with the value $s_f=0.10$ of the physical system, $|f-f^\star|=0.2$ a factor $\exp(-16)$. Figure~\ref{fig:appendix_recovery_gain} shows the frequency dependence for three amplitudes: at the values of the physical system, the gain exceeds 1\% of its peak only for frequencies between about $0.25$ and $0.55$.

	\bibliographystyle{preprint_unsrt}
	\bibliography{bibliography,cmame_additions}

\begin{thebibliography}{130}
\providecommand{\natexlab}[1]{#1}
\providecommand{\url}[1]{\texttt{#1}}
\expandafter\ifx\csname urlstyle\endcsname\relax
  \providecommand{\doi}[1]{doi: #1}\else
  \providecommand{\doi}{doi: \begingroup \urlstyle{rm}\Url}\fi

\bibitem[Jones et~al.(2020)Jones, Snider, Nassehi, Yon, and
  Hicks]{Jones2020DTReview}
D.~Jones, C.~Snider, A.~Nassehi, J.~Yon, and B.~Hicks.
\newblock Characterising the digital twin: A systematic literature review.
\newblock \emph{CIRP Journal of Manufacturing Science and Technology},
  29:\penalty0 36--52, 2020.
\newblock \doi{10.1016/j.cirpj.2020.02.002}.

\bibitem[Bokhtiar Al~Zami et~al.(2025)Bokhtiar Al~Zami, Shaon, Khanh~Quy, and
  Nguyen]{BokhtiarAlZami2025}
M.~Bokhtiar Al~Zami, S.~Shaon, V.~Khanh~Quy, and D.~C. Nguyen.
\newblock Digital twin in industries: A comprehensive survey.
\newblock \emph{IEEE Access}, 13:\penalty0 47291--47336, 2025.
\newblock \doi{10.1109/access.2025.3551532}.

\bibitem[Lee(2008)]{Lee2008CPS}
E.~A. Lee.
\newblock Cyber physical systems: Design challenges.
\newblock In \emph{2008 11th IEEE International Symposium on Object and
  Component-Oriented Real-Time Distributed Computing}, pages 363--369, 2008.
\newblock \doi{10.1109/ISORC.2008.25}.

\bibitem[Lee and Seshia(2017)]{Lee2015CPS}
E.~A. Lee and S.~A. Seshia.
\newblock \emph{Introduction to Embedded Systems: A Cyber-Physical Systems
  Approach}.
\newblock MIT Press, second edition, 2017.

\bibitem[Kagermann et~al.(2013)Kagermann, Wahlster, and
  Helbig]{Kagermann2013Industry4}
H.~Kagermann, W.~Wahlster, and J.~Helbig.
\newblock Recommendations for implementing the strategic initiative {Industrie
  4.0}.
\newblock Final report of the {Industrie 4.0} working group, National Academy
  of Science and Engineering, 2013.

\bibitem[Ficili et~al.(2025)Ficili, Giacobbe, Tricomi, and
  Puliafito]{Ficili2025}
I.~Ficili, M.~Giacobbe, G.~Tricomi, and A.~Puliafito.
\newblock From sensors to data intelligence: Leveraging {IoT}, cloud, and edge
  computing with {AI}.
\newblock \emph{Sensors}, 25\penalty0 (6):\penalty0 1763, March 2025.
\newblock \doi{10.3390/s25061763}.

\bibitem[Korenhof et~al.(2021)Korenhof, Blok, and Kloppenburg]{Korenhof2021}
P.~Korenhof, V.~Blok, and S.~Kloppenburg.
\newblock Steering representations---towards a critical understanding of
  digital twins.
\newblock \emph{Philosophy \& Technology}, 34\penalty0 (4):\penalty0
  1751--1773, October 2021.
\newblock \doi{10.1007/s13347-021-00484-1}.

\bibitem[Wagg et~al.(2025)Wagg, Burr, Shepherd, Xuereb~Conti, Enzer, and
  Niederer]{Wagg2025}
D.~J. Wagg, C.~Burr, J.~Shepherd, Z.~Xuereb~Conti, M.~Enzer, and S.~Niederer.
\newblock The philosophical foundations of digital twinning.
\newblock \emph{Data-Centric Engineering}, 6:\penalty0 e12, 2025.
\newblock \doi{10.1017/dce.2025.4}.

\bibitem[Torzoni et~al.(2024)Torzoni, Tezzele, Mariani, Manzoni, and
  Willcox]{Torzoni2024digital}
M.~Torzoni, M.~Tezzele, S.~Mariani, A.~Manzoni, and K.~E. Willcox.
\newblock A digital twin framework for civil engineering structures.
\newblock \emph{Computer Methods in Applied Mechanics and Engineering},
  418:\penalty0 116584, 2024.
\newblock \doi{10.1016/j.cma.2023.116584}.

\bibitem[Zaki and Wang(2021)]{ZakiWang2021}
T.~A. Zaki and M.~Wang.
\newblock From limited observations to the state of turbulence: Fundamental
  difficulties of flow reconstruction.
\newblock \emph{Physical Review Fluids}, 6\penalty0 (10):\penalty0 100501,
  2021.
\newblock \doi{10.1103/PhysRevFluids.6.100501}.

\bibitem[Zaki(2025)]{Zaki2025}
T.~A. Zaki.
\newblock Turbulence from an observer perspective.
\newblock \emph{Annual Review of Fluid Mechanics}, 57\penalty0 (1):\penalty0
  311--334, 2025.
\newblock \doi{10.1146/annurev-fluid-030424-114735}.

\bibitem[H{\oe}pffner et~al.(2005)H{\oe}pffner, Chevalier, Bewley, and
  Henningson]{Hoepffner2005}
J.~H{\oe}pffner, M.~Chevalier, T.~R. Bewley, and D.~S. Henningson.
\newblock State estimation in wall-bounded flow systems. {P}art 1. {P}erturbed
  laminar flows.
\newblock \emph{Journal of Fluid Mechanics}, 534:\penalty0 263--294, 2005.
\newblock \doi{10.1017/S0022112005004210}.

\bibitem[Colburn et~al.(2011)Colburn, Cessna, and Bewley]{Colburn2011}
C.~H. Colburn, J.~B. Cessna, and T.~R. Bewley.
\newblock State estimation in wall-bounded flow systems. {P}art 3. {T}he
  ensemble {K}alman filter.
\newblock \emph{Journal of Fluid Mechanics}, 682:\penalty0 289--303, 2011.
\newblock \doi{10.1017/jfm.2011.222}.

\bibitem[Mons et~al.(2016)Mons, Chassaing, Gomez, and Sagaut]{Mons2016}
V.~Mons, J.-C. Chassaing, T.~Gomez, and P.~Sagaut.
\newblock Reconstruction of unsteady viscous flows using data assimilation
  schemes.
\newblock \emph{Journal of Computational Physics}, 316:\penalty0 255--280,
  2016.
\newblock \doi{10.1016/j.jcp.2016.04.022}.

\bibitem[Manohar et~al.(2018)Manohar, Brunton, Kutz, and Brunton]{Manohar2018}
K.~Manohar, B.~W. Brunton, J.~N. Kutz, and S.~L. Brunton.
\newblock Data-driven sparse sensor placement for reconstruction: Demonstrating
  the benefits of exploiting known patterns.
\newblock \emph{IEEE Control Systems Magazine}, 38\penalty0 (3):\penalty0
  63--86, 2018.
\newblock \doi{10.1109/MCS.2018.2810460}.

\bibitem[Fukami et~al.(2021)Fukami, Maulik, Ramachandra, Fukagata, and
  Taira]{Fukami2021}
K.~Fukami, R.~Maulik, N.~Ramachandra, K.~Fukagata, and K.~Taira.
\newblock Global field reconstruction from sparse sensors with {V}oronoi
  tessellation-assisted deep learning.
\newblock \emph{Nature Machine Intelligence}, 3\penalty0 (11):\penalty0
  945--951, 2021.
\newblock \doi{10.1038/s42256-021-00402-2}.

\bibitem[{\"O}zalp et~al.(2026){\"O}zalp, N{\'o}voa, and Magri]{Ozalp2026real}
E.~{\"O}zalp, A.~N{\'o}voa, and L.~Magri.
\newblock Real-time forecasting of chaotic dynamics from sparse data and
  autoencoders.
\newblock \emph{Computer Methods in Applied Mechanics and Engineering},
  450:\penalty0 118600, 2026.
\newblock \doi{10.1016/j.cma.2025.118600}.

\bibitem[Asch et~al.(2016)Asch, Bocquet, and Nodet]{AschBocquetNodet2016}
M.~Asch, M.~Bocquet, and M.~Nodet.
\newblock \emph{Data Assimilation: Methods, Algorithms, and Applications}.
\newblock Fundamentals of Algorithms. Society for Industrial and Applied
  Mathematics, Philadelphia, PA, 2016.
\newblock \doi{10.1137/1.9781611974546}.

\bibitem[Moireau et~al.(2008)Moireau, Chapelle, and
  Le~Tallec]{Moireau2008joint}
P.~Moireau, D.~Chapelle, and P.~Le~Tallec.
\newblock Joint state and parameter estimation for distributed mechanical
  systems.
\newblock \emph{Computer Methods in Applied Mechanics and Engineering},
  197\penalty0 (6--8):\penalty0 659--677, 2008.
\newblock \doi{10.1016/j.cma.2007.08.021}.

\bibitem[Smith and Gottwald(2025)]{Smith2025}
L.~D. Smith and G.~A. Gottwald.
\newblock Data assimilation for networks of coupled oscillators: inferring
  unknown model parameters from partial observations.
\newblock \emph{Proceedings of the Royal Society A: Mathematical, Physical and
  Engineering Sciences}, 481\penalty0 (2313):\penalty0 20240813, May 2025.
\newblock \doi{10.1098/rspa.2024.0813}.

\bibitem[Marques et~al.(2024)Marques, Ahizi, and Mendez]{Marques2024}
P.~A. Marques, S.~Ahizi, and M.~A. Mendez.
\newblock Real-time data assimilation for the thermodynamic modeling of
  cryogenic storage tanks.
\newblock \emph{Energy}, 302:\penalty0 131739, September 2024.
\newblock \doi{10.1016/j.energy.2024.131739}.

\bibitem[Ozan et~al.(2025)Ozan, Nóvoa, Rigas, and Magri]{Ozan2025}
D.~E. Ozan, A.~Nóvoa, G.~Rigas, and L.~Magri.
\newblock Data-assimilated model-informed reinforcement learning.
\newblock \emph{Proceedings of the Royal Society A: Mathematical, Physical and
  Engineering Sciences}, 481\penalty0 (2327):\penalty0 20250476, December 2025.
\newblock \doi{10.1098/rspa.2025.0476}.

\bibitem[N{\'o}voa et~al.(2024)N{\'o}voa, Racca, and Magri]{Novoa2024inferring}
A.~N{\'o}voa, A.~Racca, and L.~Magri.
\newblock Inferring unknown unknowns: Regularized bias-aware ensemble {K}alman
  filter.
\newblock \emph{Computer Methods in Applied Mechanics and Engineering},
  418:\penalty0 116502, 2024.
\newblock \doi{10.1016/j.cma.2023.116502}.

\bibitem[Goodwin and Payne(1977)]{GoodwinPayne1977}
G.~C. Goodwin and R.~L. Payne.
\newblock \emph{Dynamic System Identification: Experiment Design and Data
  Analysis}.
\newblock Academic Press, 1977.

\bibitem[Ljung(1999)]{Ljung1999}
L.~Ljung.
\newblock \emph{System Identification: Theory for the User}.
\newblock Prentice Hall, 2nd edition, 1999.

\bibitem[Mehra(1974)]{Mehra1974}
R.~K. Mehra.
\newblock Optimal input signals for parameter estimation in dynamic
  systems--survey and new results.
\newblock \emph{IEEE Transactions on Automatic Control}, 19\penalty0
  (6):\penalty0 753--768, 1974.
\newblock \doi{10.1109/TAC.1974.1100701}.

\bibitem[Walter and Pronzato(1997)]{WalterPronzato1997}
{\'E}.~Walter and L.~Pronzato.
\newblock \emph{Identification of Parametric Models from Experimental Data}.
\newblock Communications and Control Engineering. Springer, London, UK, 1997.

\bibitem[Gevers(2005)]{Gevers2005}
M.~Gevers.
\newblock Identification for control: From the early achievements to the
  revival of experiment design.
\newblock \emph{European Journal of Control}, 11\penalty0 (4--5):\penalty0
  335--352, 2005.
\newblock \doi{10.3166/ejc.11.335-352}.

\bibitem[Hjalmarsson(2005)]{Hjalmarsson2005}
H.~Hjalmarsson.
\newblock From experiment design to closed-loop control.
\newblock \emph{Automatica}, 41\penalty0 (3):\penalty0 393--438, 2005.
\newblock \doi{10.1016/j.automatica.2004.11.021}.

\bibitem[Chiuso and Pillonetto(2019)]{Chiuso2019}
A.~Chiuso and G.~Pillonetto.
\newblock System identification: A machine learning perspective.
\newblock \emph{Annual Review of Control, Robotics, and Autonomous Systems},
  2\penalty0 (1):\penalty0 281--304, 2019.
\newblock \doi{10.1146/annurev-control-053018-023744}.

\bibitem[Bhattacharya et~al.(2026)Bhattacharya, Cao, and
  Stuart]{Bhattacharya2026optimal}
K.~Bhattacharya, L.~Cao, and A.~M. Stuart.
\newblock Optimal experimental design for reliable learning of
  history-dependent constitutive laws.
\newblock \emph{Computer Methods in Applied Mechanics and Engineering},
  457:\penalty0 119022, 2026.
\newblock \doi{10.1016/j.cma.2026.119022}.

\bibitem[Yang et~al.(2025)Yang, Chen, and Wu]{Yang2025active}
H.~Yang, C.~Chen, and J.-L. Wu.
\newblock Active learning of model discrepancy with {B}ayesian experimental
  design.
\newblock \emph{Computer Methods in Applied Mechanics and Engineering},
  446:\penalty0 118198, 2025.
\newblock \doi{10.1016/j.cma.2025.118198}.

\bibitem[Huan and Marzouk(2016)]{HuanMarzouk2016}
X.~Huan and Y.~M. Marzouk.
\newblock Sequential {B}ayesian optimal experimental design via approximate
  dynamic programming.
\newblock arXiv preprint arXiv:1604.08320, 2016.

\bibitem[Shen and Huan(2023)]{ShenHuan2023}
W.~Shen and X.~Huan.
\newblock Bayesian sequential optimal experimental design for nonlinear models
  using policy gradient reinforcement learning.
\newblock \emph{Computer Methods in Applied Mechanics and Engineering},
  416:\penalty0 116304, 2023.
\newblock \doi{10.1016/j.cma.2023.116304}.

\bibitem[Shen et~al.(2025)Shen, Dong, and Huan]{Shen2025variational}
W.~Shen, J.~Dong, and X.~Huan.
\newblock Variational sequential optimal experimental design using
  reinforcement learning.
\newblock \emph{Computer Methods in Applied Mechanics and Engineering},
  444:\penalty0 118068, 2025.
\newblock \doi{10.1016/j.cma.2025.118068}.

\bibitem[Foster et~al.(2021)Foster, Ivanova, Malik, and
  Rainforth]{Foster2021DAD}
A.~Foster, D.~R. Ivanova, I.~Malik, and T.~Rainforth.
\newblock Deep adaptive design: Amortizing sequential {B}ayesian experimental
  design.
\newblock In \emph{Proceedings of the 38th International Conference on Machine
  Learning (ICML)}, volume 139 of \emph{Proceedings of Machine Learning
  Research}, pages 3384--3395. PMLR, 2021.

\bibitem[Blau et~al.(2022)Blau, Bonilla, Chades, and Dezfouli]{Blau2022}
T.~Blau, E.~V. Bonilla, I.~Chades, and A.~Dezfouli.
\newblock Optimizing sequential experimental design with deep reinforcement
  learning.
\newblock In \emph{Proceedings of the 39th International Conference on Machine
  Learning (ICML)}, volume 162 of \emph{Proceedings of Machine Learning
  Research}, pages 2107--2128. PMLR, 2022.

\bibitem[Rainforth et~al.(2024)Rainforth, Foster, Ivanova, and
  Bickford~Smith]{Rainforth2024}
T.~Rainforth, A.~Foster, D.~R. Ivanova, and F.~Bickford~Smith.
\newblock Modern {B}ayesian experimental design.
\newblock \emph{Statistical Science}, 39\penalty0 (1):\penalty0 100--114, 2024.
\newblock \doi{10.1214/23-STS915}.

\bibitem[Wagenmaker and Jamieson(2020)]{WagenmakerJamieson2020}
A.~Wagenmaker and K.~Jamieson.
\newblock Active learning for identification of linear dynamical systems.
\newblock In \emph{Proceedings of the 33rd Conference on Learning Theory
  (COLT)}, volume 125 of \emph{Proceedings of Machine Learning Research}, pages
  3487--3582, 2020.

\bibitem[Mania et~al.(2022)Mania, Jordan, and Recht]{Mania2022}
H.~Mania, M.~I. Jordan, and B.~Recht.
\newblock Active learning for nonlinear system identification with guarantees.
\newblock \emph{Journal of Machine Learning Research}, 23\penalty0
  (32):\penalty0 1--30, 2022.

\bibitem[Buisson-Fenet et~al.(2020)Buisson-Fenet, Solowjow, and
  Trimpe]{BuissonFenet2020}
M.~Buisson-Fenet, F.~Solowjow, and S.~Trimpe.
\newblock Actively learning {G}aussian process dynamics.
\newblock In \emph{Proceedings of the 2nd Conference on Learning for Dynamics
  and Control (L4DC)}, volume 120 of \emph{Proceedings of Machine Learning
  Research}, pages 5--15. PMLR, 2020.

\bibitem[Sukhija et~al.(2023)Sukhija, Treven, Sancaktar, Blaes, Coros, and
  Krause]{Treven2023}
B.~Sukhija, L.~Treven, C.~Sancaktar, S.~Blaes, S.~Coros, and A.~Krause.
\newblock Optimistic active exploration of dynamical systems.
\newblock In \emph{Advances in Neural Information Processing Systems 36}, pages
  38122--38153, 2023.
\newblock \doi{10.52202/075280-1656}.

\bibitem[Mehta et~al.(2022)Mehta, Paria, Schneider, Ermon, and
  Neiswanger]{Mehta2022}
V.~Mehta, B.~Paria, J.~Schneider, S.~Ermon, and W.~Neiswanger.
\newblock An experimental design perspective on model-based reinforcement
  learning.
\newblock In \emph{International Conference on Learning Representations
  (ICLR)}, 2022.

\bibitem[Memmel et~al.(2024)Memmel, Wagenmaker, Zhu, Fox, and
  Gupta]{Memmel2024ASID}
M.~Memmel, A.~Wagenmaker, C.~Zhu, D.~Fox, and A.~Gupta.
\newblock {ASID}: Active exploration for system identification in robotic
  manipulation.
\newblock In \emph{International Conference on Learning Representations
  (ICLR)}, 2024.

\bibitem[Ogbodo et~al.(2026)Ogbodo, Rogers, Dal~Borgo, and Wagg]{Ogbodo2026}
C.~O. Ogbodo, T.~J. Rogers, M.~Dal~Borgo, and D.~J. Wagg.
\newblock Adaptive sensor steering strategy using deep reinforcement learning
  for dynamic data acquisition in digital twins.
\newblock \emph{Proceedings of the Royal Society A: Mathematical, Physical and
  Engineering Sciences}, 482\penalty0 (2329):\penalty0 20250326, January 2026.
\newblock \doi{10.1098/rspa.2025.0326}.

\bibitem[Stengel(1994)]{stengel1994optimal}
R.~F. Stengel.
\newblock \emph{Optimal Control and Estimation}.
\newblock Dover Publications, 1994.

\bibitem[Bertsekas(2017)]{Bertsekas2012}
D.~P. Bertsekas.
\newblock \emph{Dynamic Programming and Optimal Control, Vol. I}.
\newblock Athena Scientific, 4 edition, 2017.

\bibitem[Rawlings et~al.(2017)Rawlings, Mayne, and Diehl]{RawlingsMayne2017}
J.~B. Rawlings, D.~Q. Mayne, and M.~M. Diehl.
\newblock \emph{Model Predictive Control: Theory, Computation, and Design}.
\newblock Nob Hill Publishing, 2nd edition, 2017.

\bibitem[Gunzburger(2002)]{gunzburger2002perspectives}
M.~D. Gunzburger.
\newblock \emph{Perspectives in Flow Control and Optimization}.
\newblock SIAM, 2002.

\bibitem[{\AA}str{\"o}m and Wittenmark(1973)]{AstromWittenmark1973}
K.~J. {\AA}str{\"o}m and B.~Wittenmark.
\newblock On self-tuning regulators.
\newblock \emph{Automatica}, 9\penalty0 (2):\penalty0 185--199, 1973.
\newblock \doi{10.1016/0005-1098(73)90073-3}.

\bibitem[{\AA}str{\"o}m and Wittenmark(2008)]{astrom_adaptive_2008}
K.~J. {\AA}str{\"o}m and B.~Wittenmark.
\newblock \emph{Adaptive Control}.
\newblock Dover Publications, 2008.

\bibitem[Annaswamy and Fradkov(2021)]{Annaswamy2021}
A.~M. Annaswamy and A.~L. Fradkov.
\newblock A historical perspective of adaptive control and learning.
\newblock \emph{Annual Reviews in Control}, 52:\penalty0 18--41, 2021.
\newblock \doi{10.1016/j.arcontrol.2021.10.014}.

\bibitem[Matni et~al.(2019)Matni, Proutiere, Rantzer, and Tu]{Matni2019}
N.~Matni, A.~Proutiere, A.~Rantzer, and S.~Tu.
\newblock From self-tuning regulators to reinforcement learning and back again.
\newblock In \emph{2019 IEEE 58th Conference on Decision and Control (CDC)},
  pages 3724--3740. IEEE, December 2019.
\newblock \doi{10.1109/cdc40024.2019.9029916}.

\bibitem[Hjalmarsson et~al.(1998)Hjalmarsson, Gevers, Gunnarsson, and
  Lequin]{Hjalmarsson1998IFT}
H.~Hjalmarsson, M.~Gevers, S.~Gunnarsson, and O.~Lequin.
\newblock Iterative feedback tuning: theory and applications.
\newblock \emph{IEEE Control Systems Magazine}, 18\penalty0 (4):\penalty0
  26--41, 1998.
\newblock \doi{10.1109/37.710876}.

\bibitem[Sutton and Barto(2018)]{SuttonBarto2018}
R.~S. Sutton and A.~G. Barto.
\newblock \emph{Reinforcement Learning: An Introduction}.
\newblock MIT Press, 2 edition, 2018.

\bibitem[Bucci et~al.(2019)Bucci, Semeraro, Allauzen, Wisniewski, Cordier, and
  Mathelin]{Bucci2019}
M.~A. Bucci, O.~Semeraro, A.~Allauzen, G.~Wisniewski, L.~Cordier, and
  L.~Mathelin.
\newblock Control of chaotic systems by deep reinforcement learning.
\newblock \emph{Proceedings of the Royal Society A: Mathematical, Physical and
  Engineering Sciences}, 475\penalty0 (2231):\penalty0 20190351, November 2019.
\newblock \doi{10.1098/rspa.2019.0351}.

\bibitem[Pino et~al.(2023)Pino, Schena, Rabault, and Mendez]{Pino2023}
F.~Pino, L.~Schena, J.~Rabault, and M.~A. Mendez.
\newblock Comparative analysis of machine learning methods for active flow
  control.
\newblock \emph{Journal of Fluid Mechanics}, 958:\penalty0 A39, March 2023.
\newblock \doi{10.1017/jfm.2023.76}.

\bibitem[Pino et~al.(2026)Pino, Fracchia, Scheid, and Mendez]{Pino2026}
F.~Pino, E.~Fracchia, B.~Scheid, and M.~A. Mendez.
\newblock Integral modeling and reinforcement learning control of
  three-dimensional liquid metal coating on a moving substrate.
\newblock \emph{Physical Review Fluids}, 11\penalty0 (4):\penalty0 044003,
  April 2026.
\newblock \doi{10.1103/77zf-ks57}.

\bibitem[Park and Law(2016)]{ParkLaw2016}
J.~Park and K.~H. Law.
\newblock {B}ayesian ascent: A data-driven optimization scheme for real-time
  control with application to wind farm power maximization.
\newblock \emph{IEEE Transactions on Control Systems Technology}, 24\penalty0
  (5):\penalty0 1655--1668, 2016.
\newblock \doi{10.1109/TCST.2015.2508007}.

\bibitem[Sutton(1991)]{sutton1991dyna}
R.~S. Sutton.
\newblock Dyna, an integrated architecture for learning, planning, and
  reacting.
\newblock \emph{ACM SIGART Bulletin}, 2\penalty0 (4):\penalty0 160--163, 1991.
\newblock \doi{10.1145/122344.122377}.

\bibitem[Deisenroth and Rasmussen(2011)]{Deisenroth2011}
M.~P. Deisenroth and C.~E. Rasmussen.
\newblock {PILCO}: A model-based and data-efficient approach to policy search.
\newblock In \emph{Proceedings of the 28th International Conference on Machine
  Learning (ICML)}, pages 465--472, 2011.

\bibitem[Nagabandi et~al.(2018)Nagabandi, Kahn, Fearing, and
  Levine]{nagabandi_neural_2018}
A.~Nagabandi, G.~Kahn, R.~S. Fearing, and S.~Levine.
\newblock Neural network dynamics for model-based deep reinforcement learning
  with model-free fine-tuning.
\newblock In \emph{2018 IEEE International Conference on Robotics and
  Automation (ICRA)}, pages 7559--7566. IEEE, May 2018.
\newblock \doi{10.1109/icra.2018.8463189}.

\bibitem[Moerland et~al.(2023)Moerland, Broekens, Plaat, and
  Jonker]{moerland_model-based_2022}
T.~M. Moerland, J.~Broekens, A.~Plaat, and C.~M. Jonker.
\newblock Model-based reinforcement learning: A survey.
\newblock \emph{Foundations and Trends in Machine Learning}, 16\penalty0
  (1):\penalty0 1--118, 2023.
\newblock \doi{10.1561/2200000086}.

\bibitem[Bansal et~al.(2017)Bansal, Calandra, Chua, Levine, and
  Tomlin]{bansal_mbmf_2017}
S.~Bansal, R.~Calandra, K.~Chua, S.~Levine, and C.~Tomlin.
\newblock {MBMF}: Model-based priors for model-free reinforcement learning.
\newblock arXiv preprint arXiv:1709.03153, 2017.

\bibitem[Johannink et~al.(2019)Johannink, Bahl, Nair, Luo, Kumar, Loskyll,
  Ojea, Solowjow, and Levine]{Johannink2019}
T.~Johannink, S.~Bahl, A.~Nair, J.~Luo, A.~Kumar, M.~Loskyll, J.~A. Ojea,
  E.~Solowjow, and S.~Levine.
\newblock Residual reinforcement learning for robot control.
\newblock In \emph{2019 International Conference on Robotics and Automation
  (ICRA)}, pages 6023--6029. IEEE, May 2019.
\newblock \doi{10.1109/icra.2019.8794127}.

\bibitem[Liu and Wang(2021)]{Liu2021}
X.-Y. Liu and J.-X. Wang.
\newblock Physics-informed dyna-style model-based deep reinforcement learning
  for dynamic control.
\newblock \emph{Proceedings of the Royal Society A: Mathematical, Physical and
  Engineering Sciences}, 477\penalty0 (2255):\penalty0 20210618, November 2021.
\newblock \doi{10.1098/rspa.2021.0618}.

\bibitem[Tomasetto et~al.(2025)Tomasetto, Braghin, and
  Manzoni]{Tomasetto2025latent}
M.~Tomasetto, F.~Braghin, and A.~Manzoni.
\newblock Latent feedback control of distributed systems in multiple scenarios
  through deep learning-based reduced order models.
\newblock \emph{Computer Methods in Applied Mechanics and Engineering},
  442:\penalty0 118030, 2025.
\newblock \doi{10.1016/j.cma.2025.118030}.

\bibitem[Werner and Peitz(2024)]{Werner2024}
S.~Werner and S.~Peitz.
\newblock Numerical evidence for sample efficiency of model-based over
  model-free reinforcement learning control of partial differential equations.
\newblock In \emph{2024 European Control Conference (ECC)}, pages 2965--2971.
  IEEE, June 2024.
\newblock \doi{10.23919/ecc64448.2024.10590945}.

\bibitem[Werner and Peherstorfer(2025)]{Werner2025system}
S.~W.~R. Werner and B.~Peherstorfer.
\newblock System stabilization with policy optimization on unstable latent
  manifolds.
\newblock \emph{Computer Methods in Applied Mechanics and Engineering},
  433:\penalty0 117483, 2025.
\newblock \doi{10.1016/j.cma.2024.117483}.

\bibitem[Rai et~al.(2019)Rai, Antonova, Meier, and
  Atkeson]{Rai2019SimulationBO}
A.~Rai, R.~Antonova, F.~Meier, and C.~G. Atkeson.
\newblock Using simulation to improve sample-efficiency of {B}ayesian
  optimization for bipedal robots.
\newblock \emph{Journal of Machine Learning Research}, 20\penalty0
  (49):\penalty0 1--24, 2019.

\bibitem[Nobar et~al.(2025)Nobar, Keller, Rupenyan, Khosravi, and
  Lygeros]{Nobar2025}
M.~Nobar, J.~Keller, A.~Rupenyan, M.~Khosravi, and J.~Lygeros.
\newblock Guided {B}ayesian optimization: Data-efficient controller tuning with
  digital twin.
\newblock \emph{IEEE Transactions on Automation Science and Engineering},
  22:\penalty0 11304--11317, 2025.
\newblock \doi{10.1109/tase.2024.3454176}.

\bibitem[Nobar et~al.(2026)Nobar, Keller, Forino, Lygeros, and
  Rupenyan]{Nobar2026}
M.~Nobar, J.~Keller, A.~Forino, J.~Lygeros, and A.~Rupenyan.
\newblock Guided multi-fidelity {B}ayesian optimization for data-driven
  controller tuning with digital twins.
\newblock \emph{IEEE Robotics and Automation Letters}, 11\penalty0
  (5):\penalty0 5294--5301, 2026.
\newblock \doi{10.1109/LRA.2026.3671557}.

\bibitem[Feldbaum(1960)]{Feldbaum1960}
A.~A. Feldbaum.
\newblock Dual control theory. {I}.
\newblock \emph{Automation and Remote Control}, 21\penalty0 (9):\penalty0
  874--880, 1960.

\bibitem[Feldbaum(1963)]{feldbaum_dual_1963}
A.~A. Feldbaum.
\newblock Dual control theory problems.
\newblock \emph{IFAC Proceedings Volumes}, 1\penalty0 (2):\penalty0 541--550,
  1963.
\newblock \doi{10.1016/S1474-6670(17)69687-3}.

\bibitem[Rummery and Niranjan(1994)]{RummeryNiranjan1994}
G.~A. Rummery and M.~Niranjan.
\newblock On-line {Q}-learning using connectionist systems.
\newblock Technical Report CUED/F-INFENG/TR 166, Department of Engineering,
  University of Cambridge, 1994.

\bibitem[Mnih et~al.(2015)Mnih, Kavukcuoglu, Silver, Rusu, Veness, Bellemare,
  Graves, Riedmiller, Fidjeland, Ostrovski, Petersen, Beattie, Sadik,
  Antonoglou, King, Kumaran, Wierstra, Legg, and Hassabis]{Mnih2015}
V.~Mnih, K.~Kavukcuoglu, D.~Silver, A.~A. Rusu, J.~Veness, M.~G. Bellemare,
  A.~Graves, M.~Riedmiller, A.~K. Fidjeland, G.~Ostrovski, S.~Petersen,
  C.~Beattie, A.~Sadik, I.~Antonoglou, H.~King, D.~Kumaran, D.~Wierstra,
  S.~Legg, and D.~Hassabis.
\newblock Human-level control through deep reinforcement learning.
\newblock \emph{Nature}, 518\penalty0 (7540):\penalty0 529--533, 2015.
\newblock \doi{10.1038/nature14236}.

\bibitem[Lillicrap et~al.(2016)Lillicrap, Hunt, Pritzel, Heess, Erez, Tassa,
  Silver, and Wierstra]{LillicrapEtAl2016}
T.~P. Lillicrap, J.~J. Hunt, A.~Pritzel, N.~Heess, T.~Erez, Y.~Tassa,
  D.~Silver, and D.~Wierstra.
\newblock Continuous control with deep reinforcement learning.
\newblock In \emph{International Conference on Learning Representations
  (ICLR)}, 2016.

\bibitem[Haarnoja et~al.(2018)Haarnoja, Zhou, Abbeel, and
  Levine]{HaarnojaEtAl2018}
T.~Haarnoja, A.~Zhou, P.~Abbeel, and S.~Levine.
\newblock Soft actor-critic: Off-policy maximum entropy deep reinforcement
  learning with a stochastic actor.
\newblock In \emph{Proceedings of the 35th International Conference on Machine
  Learning (ICML)}, pages 1861--1870, 2018.

\bibitem[Bar-Shalom and Tse(1974)]{BarShalomTse1974}
Y.~Bar-Shalom and E.~Tse.
\newblock Dual effect, certainty equivalence, and separation in stochastic
  control.
\newblock \emph{IEEE Transactions on Automatic Control}, 19\penalty0
  (5):\penalty0 494--500, 1974.
\newblock \doi{10.1109/TAC.1974.1100635}.

\bibitem[Bernhardsson(1989)]{bernhardsson_dual_1989}
B.~Bernhardsson.
\newblock Dual control of a first-order system with two possible gains.
\newblock \emph{International Journal of Adaptive Control and Signal
  Processing}, 3\penalty0 (1):\penalty0 15--22, 1989.
\newblock \doi{10.1002/acs.4480030103}.

\bibitem[Åström and Helmersson(1986)]{astrom_dual_1986}
K.~J. Åström and A.~Helmersson.
\newblock Dual control of an integrator with unknown gain.
\newblock \emph{Computers \& Mathematics with Applications}, 12\penalty0 (6,
  Part A):\penalty0 653--662, 1986.
\newblock \doi{10.1016/0898-1221(86)90052-0}.

\bibitem[Filatov and Unbehauen(2000)]{Filatov2000}
N.~M. Filatov and H.~Unbehauen.
\newblock Survey of adaptive dual control methods.
\newblock \emph{IEE Proceedings -- Control Theory and Applications},
  147\penalty0 (1):\penalty0 118--128, 2000.
\newblock \doi{10.1049/ip-cta:20000107}.

\bibitem[Wittenmark(1995)]{Wittenmark1995}
B.~Wittenmark.
\newblock Adaptive dual control methods: An overview.
\newblock \emph{IFAC Proceedings Volumes}, 28\penalty0 (13):\penalty0 67--72,
  1995.
\newblock \doi{10.1016/S1474-6670(17)45327-4}.

\bibitem[Mesbah(2018)]{Mesbah2018}
A.~Mesbah.
\newblock Stochastic model predictive control with active uncertainty learning:
  A survey on dual control.
\newblock \emph{Annual Reviews in Control}, 45:\penalty0 107--117, 2018.
\newblock \doi{10.1016/j.arcontrol.2017.11.001}.

\bibitem[Klenske and Hennig(2016)]{Klenske2016}
E.~D. Klenske and P.~Hennig.
\newblock Dual control for approximate {B}ayesian reinforcement learning.
\newblock \emph{Journal of Machine Learning Research}, 17\penalty0
  (127):\penalty0 1--30, 2016.

\bibitem[Arcari et~al.(2020)Arcari, Hewing, Schlichting, and
  Zeilinger]{Arcari2020}
E.~Arcari, L.~Hewing, M.~Schlichting, and M.~N. Zeilinger.
\newblock Dual stochastic {MPC} for systems with parametric and structural
  uncertainty.
\newblock In \emph{Proceedings of the 2nd Conference on Learning for Dynamics
  and Control (L4DC)}, volume 120 of \emph{Proceedings of Machine Learning
  Research}, pages 894--903, 2020.

\bibitem[Chen et~al.(2021)Chen, Rhodes, and Liu]{chen_dual_2021}
W.-H. Chen, C.~Rhodes, and C.~Liu.
\newblock Dual control for exploitation and exploration ({DCEE}) in autonomous
  search.
\newblock \emph{Automatica}, 133:\penalty0 109851, 2021.
\newblock \doi{10.1016/j.automatica.2021.109851}.

\bibitem[Schena et~al.(2024)Schena, Marques, Poletti, Ahizi, Van~den Berghe,
  and Mendez]{Schena2024}
L.~Schena, P.~A. Marques, R.~Poletti, S.~Ahizi, J.~Van~den Berghe, and M.~A.
  Mendez.
\newblock Reinforcement twinning: From digital twins to model-based
  reinforcement learning.
\newblock \emph{Journal of Computational Science}, 82:\penalty0 102421, October
  2024.
\newblock \doi{10.1016/j.jocs.2024.102421}.

\bibitem[Poletti et~al.(2025)Poletti, Schena, Koloszar, Degroote, and
  Mendez]{Poletti2025}
R.~Poletti, L.~Schena, L.~Koloszar, J.~Degroote, and M.~A. Mendez.
\newblock Reinforcement twinning for hybrid control of flapping-wing drones.
\newblock arXiv preprint arXiv:2505.18201, 2025.

\bibitem[Kennedy and O'Hagan(2000)]{Kennedy2000}
M.~C. Kennedy and A.~O'Hagan.
\newblock Predicting the output from a complex computer code when fast
  approximations are available.
\newblock \emph{Biometrika}, 87\penalty0 (1):\penalty0 1--13, 2000.
\newblock \doi{10.1093/biomet/87.1.1}.

\bibitem[Kennedy and O'Hagan(2001)]{Kennedy2001}
M.~C. Kennedy and A.~O'Hagan.
\newblock Bayesian calibration of computer models.
\newblock \emph{Journal of the Royal Statistical Society: Series B},
  63\penalty0 (3):\penalty0 425--464, 2001.
\newblock \doi{10.1111/1467-9868.00294}.

\bibitem[Guillas et~al.(2014)Guillas, Glover, and
  Malki-Epshtein]{Guillas2014bayesian}
S.~Guillas, N.~Glover, and L.~Malki-Epshtein.
\newblock {B}ayesian calibration of the constants of the $k$--$\varepsilon$
  turbulence model for a {CFD} model of street canyon flow.
\newblock \emph{Computer Methods in Applied Mechanics and Engineering},
  279:\penalty0 536--553, 2014.
\newblock \doi{10.1016/j.cma.2014.06.008}.

\bibitem[Girolami et~al.(2021)Girolami, Febrianto, Yin, and
  Cirak]{Girolami2021statistical}
M.~Girolami, E.~Febrianto, G.~Yin, and F.~Cirak.
\newblock The statistical finite element method ({statFEM}) for coherent
  synthesis of observation data and model predictions.
\newblock \emph{Computer Methods in Applied Mechanics and Engineering},
  375:\penalty0 113533, 2021.
\newblock \doi{10.1016/j.cma.2020.113533}.

\bibitem[Forrester et~al.(2007)Forrester, S{\'o}bester, and
  Keane]{Forrester2007}
A.~I.~J. Forrester, A.~S{\'o}bester, and A.~J. Keane.
\newblock Multi-fidelity optimization via surrogate modelling.
\newblock \emph{Proceedings of the Royal Society A: Mathematical, Physical and
  Engineering Sciences}, 463\penalty0 (2088):\penalty0 3251--3269, 2007.
\newblock \doi{10.1098/rspa.2007.1900}.

\bibitem[Le~Gratiet and Garnier(2014)]{LeGratiet2014}
L.~Le~Gratiet and J.~Garnier.
\newblock Recursive co-kriging model for design of computer experiments with
  multiple levels of fidelity.
\newblock \emph{International Journal for Uncertainty Quantification},
  4\penalty0 (5):\penalty0 365--386, 2014.
\newblock \doi{10.1615/Int.J.UncertaintyQuantification.2014006914}.

\bibitem[Bonfiglio et~al.(2018)Bonfiglio, Perdikaris, Brizzolara, and
  Karniadakis]{Bonfiglio2018multi}
L.~Bonfiglio, P.~Perdikaris, S.~Brizzolara, and G.~E. Karniadakis.
\newblock Multi-fidelity optimization of super-cavitating hydrofoils.
\newblock \emph{Computer Methods in Applied Mechanics and Engineering},
  332:\penalty0 63--85, 2018.
\newblock \doi{10.1016/j.cma.2017.12.009}.

\bibitem[Peherstorfer et~al.(2018)Peherstorfer, Willcox, and
  Gunzburger]{Peherstorfer2018}
B.~Peherstorfer, K.~Willcox, and M.~Gunzburger.
\newblock Survey of multifidelity methods in uncertainty propagation,
  inference, and optimization.
\newblock \emph{SIAM Review}, 60\penalty0 (3):\penalty0 550--591, 2018.
\newblock \doi{10.1137/16M1082469}.

\bibitem[Foumani et~al.(2023)Foumani, Shishehbor, Yousefpour, and
  Bostanabad]{Foumani2023multi}
Z.~Z. Foumani, M.~Shishehbor, A.~Yousefpour, and R.~Bostanabad.
\newblock Multi-fidelity cost-aware {B}ayesian optimization.
\newblock \emph{Computer Methods in Applied Mechanics and Engineering},
  407:\penalty0 115937, 2023.
\newblock \doi{10.1016/j.cma.2023.115937}.

\bibitem[Fiore et~al.(2026)Fiore, Bresciani, Mendez, and van
  Beeck]{Fiore2026predicting}
M.~Fiore, A.~Bresciani, M.~A. Mendez, and J.~van Beeck.
\newblock Predicting wind loads on container ships in harbour environments
  through multi-fidelity modelling.
\newblock \emph{Ocean Engineering}, 364:\penalty0 126928, 2026.
\newblock \doi{10.1016/j.oceaneng.2026.126928}.

\bibitem[Hart and van Bloemen~Waanders(2023)]{Hart2023hyper}
J.~Hart and B.~van Bloemen~Waanders.
\newblock Hyper-differential sensitivity analysis with respect to model
  discrepancy: Optimal solution updating.
\newblock \emph{Computer Methods in Applied Mechanics and Engineering},
  412:\penalty0 116082, 2023.
\newblock \doi{10.1016/j.cma.2023.116082}.

\bibitem[Marchetti et~al.(2009)Marchetti, Chachuat, and Bonvin]{Marchetti2009}
A.~Marchetti, B.~Chachuat, and D.~Bonvin.
\newblock Modifier-adaptation methodology for real-time optimization.
\newblock \emph{Industrial \& Engineering Chemistry Research}, 48\penalty0
  (13):\penalty0 6022--6033, 2009.
\newblock \doi{10.1021/ie801352x}.

\bibitem[del Rio~Chanona et~al.(2019)del Rio~Chanona, Alves~Graciano, Bradford,
  and Chachuat]{delRioChanona2019}
E.~A. del Rio~Chanona, J.~E. Alves~Graciano, E.~Bradford, and B.~Chachuat.
\newblock Modifier-adaptation schemes employing {G}aussian processes and trust
  regions for real-time optimization.
\newblock \emph{IFAC-PapersOnLine}, 52\penalty0 (1):\penalty0 52--57, 2019.
\newblock \doi{10.1016/j.ifacol.2019.06.036}.

\bibitem[del Rio~Chanona et~al.(2021)del Rio~Chanona, Petsagkourakis, Bradford,
  Alves~Graciano, and Chachuat]{delRioChanona2021}
E.~A. del Rio~Chanona, P.~Petsagkourakis, E.~Bradford, J.~E. Alves~Graciano,
  and B.~Chachuat.
\newblock Real-time optimization meets {B}ayesian optimization and
  derivative-free optimization: A tale of modifier adaptation.
\newblock \emph{Computers \& Chemical Engineering}, 147:\penalty0 107249, 2021.
\newblock \doi{10.1016/j.compchemeng.2021.107249}.

\bibitem[Marco et~al.(2017)Marco, Berkenkamp, Hennig, Schoellig, Krause,
  Schaal, and Trimpe]{Marco2017}
A.~Marco, F.~Berkenkamp, P.~Hennig, A.~P. Schoellig, A.~Krause, S.~Schaal, and
  S.~Trimpe.
\newblock Virtual vs. real: Trading off simulations and physical experiments in
  reinforcement learning with {B}ayesian optimization.
\newblock In \emph{2017 IEEE International Conference on Robotics and
  Automation (ICRA)}, pages 1557--1563, 2017.
\newblock \doi{10.1109/ICRA.2017.7989186}.

\bibitem[Munters and Meyers(2017)]{Munters2017}
W.~Munters and J.~Meyers.
\newblock An optimal control framework for dynamic induction control of wind
  farms and their interaction with the atmospheric boundary layer.
\newblock \emph{Philosophical Transactions of the Royal Society A:
  Mathematical, Physical and Engineering Sciences}, 375\penalty0
  (2091):\penalty0 20160100, March 2017.
\newblock \doi{10.1098/rsta.2016.0100}.

\bibitem[Munters and Meyers(2018)]{Munters2018}
W.~Munters and J.~Meyers.
\newblock Towards practical dynamic induction control of wind farms: analysis
  of optimally controlled wind-farm boundary layers and sinusoidal induction
  control of first-row turbines.
\newblock \emph{Wind Energy Science}, 3\penalty0 (1):\penalty0 409--425, June
  2018.
\newblock \doi{10.5194/wes-3-409-2018}.

\bibitem[Gillijns and De~Moor(2007)]{GillijnsDeMoor2007}
S.~Gillijns and B.~De~Moor.
\newblock Unbiased minimum-variance input and state estimation for linear
  discrete-time systems.
\newblock \emph{Automatica}, 43\penalty0 (1):\penalty0 111--116, 2007.
\newblock \doi{10.1016/j.automatica.2006.08.002}.

\bibitem[Mo{\v{c}}kus(1975)]{Bo_1}
J.~Mo{\v{c}}kus.
\newblock On {B}ayesian methods for seeking the extremum.
\newblock In G.~I. Marchuk, editor, \emph{Optimization Techniques IFIP
  Technical Conference Novosibirsk, July 1--7, 1974}, pages 400--404, Berlin,
  Heidelberg, 1975. Springer Berlin Heidelberg.

\bibitem[Jones et~al.(1998)Jones, Schonlau, and Welch]{Jones1998}
D.~R. Jones, M.~Schonlau, and W.~J. Welch.
\newblock Efficient global optimization of expensive black-box functions.
\newblock \emph{Journal of Global Optimization}, 13\penalty0 (4):\penalty0
  455--492, 1998.
\newblock \doi{10.1023/A:1008306431147}.

\bibitem[Calandra et~al.(2016)Calandra, Seyfarth, Peters, and
  Deisenroth]{Calandra2016bayesian}
R.~Calandra, A.~Seyfarth, J.~Peters, and M.~P. Deisenroth.
\newblock Bayesian optimization for learning gaits under uncertainty: An
  experimental comparison on a dynamic bipedal walker.
\newblock \emph{Annals of Mathematics and Artificial Intelligence}, 76\penalty0
  (1--2):\penalty0 5--23, 2016.
\newblock \doi{10.1007/s10472-015-9463-9}.

\bibitem[Ng and Jordan(2000)]{NgJordan2000}
A.~Y. Ng and M.~I. Jordan.
\newblock {PEGASUS}: A policy search method for large {MDP}s and {POMDP}s.
\newblock In \emph{Proceedings of the Sixteenth Conference on Uncertainty in
  Artificial Intelligence (UAI)}, pages 406--415, 2000.

\bibitem[Abbeel et~al.(2006)Abbeel, Quigley, and Ng]{Abbeel2006}
P.~Abbeel, M.~Quigley, and A.~Y. Ng.
\newblock Using inaccurate models in reinforcement learning.
\newblock In \emph{Proceedings of the 23rd International Conference on Machine
  Learning (ICML)}, pages 1--8, 2006.
\newblock \doi{10.1145/1143844.1143845}.

\bibitem[Deisenroth et~al.(2013)Deisenroth, Neumann, and
  Peters]{Deisenroth2013}
M.~P. Deisenroth, G.~Neumann, and J.~Peters.
\newblock A survey on policy search for robotics.
\newblock \emph{Foundations and Trends in Robotics}, 2\penalty0
  (1--2):\penalty0 1--142, 2013.
\newblock \doi{10.1561/2300000021}.

\bibitem[Gao et~al.(2020)Gao, Yu, and Li]{GaoYuLi2020BOwLS}
Y.~Gao, T.~Yu, and J.~Li.
\newblock {B}ayesian optimization with local search.
\newblock In \emph{Machine Learning, Optimization, and Data Science (LOD
  2020)}, volume 12566 of \emph{Lecture Notes in Computer Science}, pages
  350--361, Cham, 2020. Springer.
\newblock \doi{10.1007/978-3-030-64580-9_30}.

\bibitem[Kyriakidis et~al.(2024)Kyriakidis, Mendez, and
  B{\"a}hr]{KyriakidisMendezBahr2024BOIPOPT}
L.~Kyriakidis, M.~A. Mendez, and M.~B{\"a}hr.
\newblock A hybrid algorithm based on {B}ayesian optimization and {Interior
  Point OPTimizer} for optimal operation of energy conversion systems.
\newblock \emph{Energy}, 312:\penalty0 133416, 2024.
\newblock \doi{10.1016/j.energy.2024.133416}.

\bibitem[Reddi et~al.(2018)Reddi, Kale, and Kumar]{ReddiKaleKumar2018}
S.~J. Reddi, S.~Kale, and S.~Kumar.
\newblock On the convergence of {A}dam and beyond.
\newblock In \emph{International Conference on Learning Representations
  (ICLR)}, 2018.

\bibitem[Griewank and Walther(2008)]{GriewankWalther2008}
A.~Griewank and A.~Walther.
\newblock \emph{Evaluating Derivatives: Principles and Techniques of
  Algorithmic Differentiation}.
\newblock SIAM, Philadelphia, PA, 2nd edition, 2008.
\newblock \doi{10.1137/1.9780898717761}.

\bibitem[O'Hagan(1978)]{OHagan1978}
A.~O'Hagan.
\newblock Curve fitting and optimal design for prediction.
\newblock \emph{Journal of the Royal Statistical Society: Series B},
  40\penalty0 (1):\penalty0 1--24, 1978.
\newblock \doi{10.1111/j.2517-6161.1978.tb01643.x}.

\bibitem[Santner et~al.(2018)Santner, Williams, and Notz]{Santner2018}
T.~J. Santner, B.~J. Williams, and W.~I. Notz.
\newblock \emph{The Design and Analysis of Computer Experiments}.
\newblock Springer, New York, 2nd edition, 2018.
\newblock \doi{10.1007/978-1-4939-8847-1}.

\bibitem[Poloczek et~al.(2017)Poloczek, Wang, and Frazier]{Poloczek2017}
M.~Poloczek, J.~Wang, and P.~I. Frazier.
\newblock Multi-information source optimization.
\newblock In \emph{Advances in Neural Information Processing Systems},
  volume~30, pages 4288--4298, 2017.

\bibitem[Rasmussen and Williams(2006)]{Rasmussen2006}
C.~E. Rasmussen and C.~K.~I. Williams.
\newblock \emph{Gaussian Processes for Machine Learning}.
\newblock The MIT Press, 2006.
\newblock \doi{10.7551/mitpress/3206.001.0001}.

\bibitem[Cao et~al.(2015)Cao, Brubaker, Fleet, and Hertzmann]{Cao2015}
Y.~Cao, M.~A. Brubaker, D.~J. Fleet, and A.~Hertzmann.
\newblock Efficient optimization for sparse {G}aussian process regression.
\newblock \emph{IEEE Transactions on Pattern Analysis and Machine
  Intelligence}, 37\penalty0 (12):\penalty0 2415--2427, December 2015.
\newblock \doi{10.1109/tpami.2015.2424873}.

\bibitem[Picheny et~al.(2013)Picheny, Wagner, and Ginsbourger]{Picheny2013}
V.~Picheny, T.~Wagner, and D.~Ginsbourger.
\newblock A benchmark of kriging-based infill criteria for noisy optimization.
\newblock \emph{Structural and Multidisciplinary Optimization}, 48\penalty0
  (3):\penalty0 607--626, 2013.
\newblock \doi{10.1007/s00158-013-0919-4}.

\bibitem[Eriksson et~al.(2019)Eriksson, Pearce, Gardner, Turner, and
  Poloczek]{Eriksson2019}
D.~Eriksson, M.~Pearce, J.~R. Gardner, R.~Turner, and M.~Poloczek.
\newblock Scalable global optimization via local {B}ayesian optimization.
\newblock In \emph{Advances in Neural Information Processing Systems 32
  (NeurIPS 2019)}, 2019.

\bibitem[Y{\i}lmaz and Meyers(2018)]{Yilmaz2018}
A.~E. Y{\i}lmaz and J.~Meyers.
\newblock Optimal dynamic induction control of a pair of inline wind turbines.
\newblock \emph{Physics of Fluids}, 30\penalty0 (8):\penalty0 085106, 2018.
\newblock \doi{10.1063/1.5038600}.

\bibitem[Frederik et~al.(2020{\natexlab{a}})Frederik, Weber, Cacciola,
  Campagnolo, Croce, Bottasso, and van Wingerden]{Frederik2020Periodic}
J.~A. Frederik, R.~Weber, S.~Cacciola, F.~Campagnolo, A.~Croce, C.~Bottasso,
  and J.-W. van Wingerden.
\newblock Periodic dynamic induction control of wind farms: proving the
  potential in simulations and wind tunnel experiments.
\newblock \emph{Wind Energy Science}, 5\penalty0 (1):\penalty0 245--257,
  2020{\natexlab{a}}.
\newblock \doi{10.5194/wes-5-245-2020}.

\bibitem[Frederik et~al.(2020{\natexlab{b}})Frederik, Doekemeijer, Mulders, and
  van Wingerden]{Frederik2020Helix}
J.~A. Frederik, B.~M. Doekemeijer, S.~P. Mulders, and J.-W. van Wingerden.
\newblock The helix approach: Using dynamic individual pitch control to enhance
  wake mixing in wind farms.
\newblock \emph{Wind Energy}, 23\penalty0 (8):\penalty0 1739--1751,
  2020{\natexlab{b}}.
\newblock \doi{10.1002/we.2513}.

\bibitem[Clark Di~Leoni et~al.(2020)Clark Di~Leoni, Mazzino, and
  Biferale]{ClarkDiLeoni2020}
P.~Clark Di~Leoni, A.~Mazzino, and L.~Biferale.
\newblock Synchronization to big data: nudging the {N}avier--{S}tokes equations
  for data assimilation of turbulent flows.
\newblock \emph{Physical Review X}, 10\penalty0 (1):\penalty0 011023, 2020.
\newblock \doi{10.1103/PhysRevX.10.011023}.

\bibitem[Zauner et~al.(2022)Zauner, Mons, Marquet, and Leclaire]{Zauner2022}
M.~Zauner, V.~Mons, O.~Marquet, and B.~Leclaire.
\newblock Nudging-based data assimilation of the turbulent flow around a square
  cylinder.
\newblock \emph{Journal of Fluid Mechanics}, 937:\penalty0 A38, 2022.
\newblock \doi{10.1017/jfm.2022.133}.

\bibitem[Brynjarsd{\'o}ttir and O'Hagan(2014)]{Brynjarsdottir2014}
J.~Brynjarsd{\'o}ttir and A.~O'Hagan.
\newblock Learning about physical parameters: the importance of model
  discrepancy.
\newblock \emph{Inverse Problems}, 30\penalty0 (11):\penalty0 114007, 2014.
\newblock \doi{10.1088/0266-5611/30/11/114007}.

\end{thebibliography}

\end{document}